\documentclass[a4paper,11pt]{article}
\usepackage[utf8]{inputenc}
\usepackage{graphicx}
\graphicspath{ {Images/} }
\usepackage[margin=10pt,font=small,labelfont=bf]{caption}
\usepackage{subcaption}
\usepackage{setspace}
\usepackage{mciteplus}
\usepackage{enumitem}
\setlist{  
  listparindent=\parindent,
  parsep=0pt,
}

\newcommand\mainmatter{%
    \cleardoublepage
  \pagenumbering{arabic}}

\newcommand\backmatter{%
  \if@openright
    \cleardoublepage
  \else
    \clearpage
  \fi
   }

\makeatother

\usepackage{fancyhdr}
\usepackage[english]{babel}
\usepackage{physics}
\usepackage{tensor}
\usepackage{amsthm}
\usepackage{amsmath}
\usepackage{amsfonts}
\usepackage{amssymb}
\usepackage{dsfont}
\usepackage{multicol}
\usepackage{xcolor}
\usepackage{hyperref}
\usepackage{url}

\definecolor{UGentblue}{RGB}{30,100,200}
\definecolor{OIST}{HTML}{C80019} 
\hypersetup{colorlinks,linkcolor=OIST,citecolor=blue, urlcolor=UGentblue, anchorcolor=sepia, linktocpage}
\usepackage{CJKutf8}

\numberwithin{equation}{section}
\numberwithin{figure}{section}

\newcommand{\xplus}{X^+}
\newcommand{\xminus}{X^-}
\newcommand{\GN}{G_{\text{N}}}
\newcommand{\lads}{l_{\text{AdS}}}

\newcommand{\ads}{AdS$_2$}
\newcommand{\eps}{\varepsilon}
\newcommand{\normord}[1]{ :\mathrel{#1}: }
\newcommand{\schw}[2]{\{\mathrel{#1}, \mathrel{#2} \}}

\author{Julian De Vuyst\footnote{\href{mailto:julian.devuyst@oist.jp}{julian.devuyst@oist.jp}}, 
~Thomas G. Mertens\footnote{\href{mailto:thomas.mertens@ugent.be}{thomas.mertens@ugent.be}}}
\date{$^{*, \dagger}$\emph{Department of Physics \& Astronomy,\\
Ghent University, Krijgslaan, S9, 9000 Ghent, Belgium}\\
$^*$\emph{Qubits and Spacetime Unit,\\ Okinawa Institute of Science and Technology 
(\begin{CJK}{UTF8}{min}沖縄科学技術大学院大学\end{CJK}),\\
1919-1 Tancha, Onna-son, Kunigami-gun, Okinawa 904-0495, Japan}}
\title{ {\huge Operational islands and black hole dissipation in JT gravity }\vspace{0.5cm}}

\begin{document}

\maketitle
\begin{abstract}
In this work, we revisit the problem of finding entanglement islands in 2d Jackiw-Teitelboim (JT) gravity. We implement the following adjustments to the traditional setup: (1) we do not explicitly couple to a non-gravitating system, instead we implement only pure absorption into a fiducial detector, (2) we utilise the operationally defined renormalised matter entanglement entropy, as defined by the boundary observer's worldline. We show that this leads to a unitary Page curve that we explicitly compute, with an island outside of the event horizon. Next, we extend the analysis to a charged and/or supersymmetric black hole. We find that in a certain regime the charged black hole grows first as it emits superradiation before eventually dissipating. We obtain similar results when embedding the system in a supersymmetric setting.
\end{abstract}
\thispagestyle{empty}

\mainmatter

\newpage
\tableofcontents
\hrulefill
\vspace{0.5cm}

\section{Introduction and summary}
In the past couple of years, significant attention has been given to the black hole information paradox which is still a key problen in quantum gravity arising from the discovery of Hawking radiation \cite{Hawking1971, Hawking1975, Hawking1976}. One of the recent proposed resolutions for spacetimes with a holographic dual \cite{Maldacena1999} is the \textbf{Island Conjecture} \cite{Almheiri2020a, Almheiri2020c}, closely tied to the appearance of replica wormholes in the gravitational path integral \cite{Penington2019, Almheiri2020, Engelhardt2021, Goto2020}.\footnote{See \cite{Almheiri2020c} for a review.} 
Within this framework, the generalised entropy associated to the nontrivial island becomes the minimal one and dominates the entropy, leading to the decreasing part of the Page curve \cite{Page1993a, Page2013}. Moreover, the AMPS paradox \cite{Almheiri2013} can be resolved by noting that the island only describes 
a part of the black hole interior, namely the part within its causal diamond -- the entanglement wedge \cite{Czech2012, Wall2014, Headrick2014a, Almheiri2020c}. This region starts on a null line a scrambling time to the past and tells us when we can decode an object after throwing it into the black hole according 
to the Hayden--Preskill protocol \cite{Hayden2007, Sekino2008}. Since this island is associated to the radiation 
which already left the black hole a long time ago, both regions should be attributed to the same Hilbert space. Maximal entanglement now only happens between two states: one on $\mathcal{R} \cup \mathcal{I}$ and one on its complement $\mathcal{BH}$ associated to the fine-grained entropy of the black hole. 
This identification can be viewed as a manifestation of $\text{ER} = \text{EPR}$ \cite{Maldacena2013, vanRaamsdonk2010}.\\
\indent The key idea is the following formula for the generalised radiation entropy:
\begin{equation}
    S(\mathcal{R}) = \underset{\mathcal{I}}{\text{Min}}~\underset{\mathcal{I}}{\text{Ext}} \left[ \frac{A(\mathcal{\partial \mathcal{I}})}{4 \GN}  
    + S_{\text{QFT}}(\mathcal{R} \cup \mathcal{I})  \right],
    \label{island}
\end{equation}
where $\mathcal{R}$ is the radiation region, $\mathcal{I}$ is the island and $S_{\text{QFT}}(\mathcal{R} \cup \mathcal{I})$ is the entanglement entropy of the quantum fields 
in this bulk region of spacetime. \\

\noindent Whilst the island formula computes the fine-grained entropy of the radiation, there is an analogous formula for the fine-grained entropy associated to the black hole.\footnote{This formula was actually used prior to the island formula.}
Instead of an island, we look for a \textbf{Quantum Extremal Surface} $\mathcal{X}$ \cite{Engelhardt2015} from which one computes the generalised entropy as:
\begin{equation}
    S(\mathcal{X}) = \underset{\mathcal{X}}{\text{Min}}~\underset{\mathcal{X}}{\text{Ext}} \left[ \frac{A(\mathcal{X})}{4\GN} + S_{\text{QFT}}(\mathcal{X})\right].
    \label{QES}
\end{equation}
This formula can be viewed as a natural extension of the RT/HRT prescription \cite{Takayanagi2017, Hubeny2007}, improving further on the JLMS extension \cite{Jafferis:2015del} where the classical Bekenstein--Hawking contribution \cite{Bekenstein1972, Bekenstein1973} is augmented by the bulk matter contribution of quanta surrounding the black hole. It is understood that bringing this semiclassical contribution inside the extremisation procedure leads to an expression which is valid at all orders of $\GN$. 
The local divergences in the QFT entanglement entropy computation can be absorbed into a renormalisation of Newton's constant, in 4d of the form: $1/G_{\text{N,ren}} \equiv 1/G_{\text{N,bare}} + \#/\epsilon^2$ in the standard framework of renormalisation making the entire quantity \eqref{QES} finite. If the black hole was formed from an initially pure state, we expect these two formulas to coincide $S(\mathcal{X}) = S(\mathcal{R})$.  \\
\indent Considering purely the matter or radiation contributions $S_{\text{QFT}}$ in either computation, one can make sense of its UV-divergence on its own by considering the so-called \emph{renormalised entanglement entropy}, studied extensively in the somewhat older literature \cite{Holzhey1994,Fiola1994}. For this quantity, one subtracts the matter entanglement entropy of the same region in a reference matter state: 
\begin{equation}
\label{renS}
S_{\text{ren}} \equiv S_{\text{bare}} - S_{\text{ref}}.
\end{equation}
For the asymptotically flat CGHS/RST model, this definition of the radiation entropy was explored in \cite{Almheiri:2013wka}, whereas for the JT model this was done in \cite{Mertens:2019bvy}. One of the appealing benefits is that the renormalised matter entropy can be defined operationally and independently of the gravitational piece of the entropy.
\\
\indent The above QES \eqref{QES} and island \eqref{island} prescriptions have been tested extensively, starting initially with the JT gravity model. In this work, guided by the above considerations, we make two adjustments to the typical analysis:
\begin{itemize}

\item 
Following the above arguments, we will utilise the renormalised entanglement entropy \eqref{renS} for the matter sector, and insert this for $S_{\text{QFT}}$ in the QES formula \eqref{QES}. The apparent ambiguity of choice of reference state is naturally addressed in the JT gravity model due to the preferred choice of boundary coordinate $t$ as time flows. This leads to an operational definition of the island points, which we will call \emph{operational islands}. Our reinterpretation of the renormalisation procedure of \eqref{QES} can be viewed as an alternative proposal, the main benefits of which will be made apparent in the main text below.

\item 
In most of the literature, a flat heat bath is added to the boundary of spacetime to accommodate for the radiation region in the island formula \cite{Almheiri2019, Almheiri2019a, Almheiri2020, Almheiri2020a, Geng:2020fxl, Goto2020, Chen2020, ZheChen2020, Hollowood2020, Hollowood2020a, Geng:2021hlu, Alishahiha:2020qza, Azarnia:2021uch, Omidi:2021opl}.\footnote{For calculations with a gravitating bath, see \cite{Anderson2021}.} One may wonder whether the physics of this heat bath does indeed change the time evolution of the entropy. For example, it was found that the location of the island depends on the initial temperature of the heat bath \cite{ZheChen2020}. Yet, gluing such a heat bath still seems like an arbitrary process, that has nothing to do with the internal dynamics of the evaporating black hole itself. In this paper, we will, following \cite{Engelsoey2016, Mertens:2019bvy}, not include any explicit heat bath. Instead, we implement solely the absorbing boundary conditions of the boundary detector, and consider the resulting purely dissipative dynamics. 
\end{itemize}

The result of these adjustments is the island structure of Fig.~\ref{sect5:figislandTrajectory} and unitary Page curve in Fig.~\ref{sect5:figPageCurve} that we will show later on. This provides for an alternative renormalisation scheme of the island formula than the one utilised in the past couple of years. \\

Next to this, we will also generalise the relevant dissipative system to include charge and supercharge dissipation through the holographic boundary, and present a solution for the energetics and radiation entropy during evaporation of such more general black holes. In particular, we will see that the black hole can actually grow during the initial stage of evaporation by superradiant mode emission, to eventually dissipate as expected. The matter entanglement entropy profile as a function of time is also drawn for these black holes, and has a qualitatively similar Page curve as expected. \\

\noindent In the remainder of the paper, we provide the details of these calculations. The paper is structured as follows: we begin with a review of JT gravity in Section \ref{sect:JT} where emphasis is put on the perspective of the boundary observer. Next, in Section \ref{sect:EBH} we describe the evaporating black hole in this model, and the entropy associated to the matter sector. Of importance will be how we deal with the UV cutoffs. 
Subsequently, in Section \ref{sect:archipelago} we present the QES calculations in our model, with the final Page curve in \ref{sect:conclusion}. In the second half of the paper, we consider more general dissipative systems than the purely energetic dissipation of the uncharged black hole. In particular, we focus on adding charged dissipation and then solve the coupled equations of motion. As a further argument in favour of our results, we also embed and generalise these calculations in $\mathcal{N}=2$ and higher supersymmetric black hole dissipation. This is done in Sections \ref{sect:charged} and \ref{sect:chargedEvap}. 

Section \ref{sect:archipelago} on the one hand, and \ref{sect:charged} and \ref{sect:chargedEvap} on the other, can be read independently, and the reader only interested in one of the two can safely skip the other parts.

The appendices contain some of the more technical details, in particular Appendix \ref{sect:SCFTEE} contains a discussion on the 2d CFT entanglement entropy generalising the pure frame dependence to include gauge and superframe dependence. \\

\noindent For convenience, we present a short \textbf{Glossary} for the different kinds of entropy: \\
${} \quad S_{\text{ren}, \mathcal{R}}(t) = $ renormalised outgoing radiation entropy, \\
${} \quad S_{\text{BH}}(t) = $ semiclassical Bekenstein--Hawking entropy, \\
${} \quad S_{\text{pre}}(t) = $ generalised entropy before the Page time, \\
${} \quad S_{\text{post}}(t) = $ generalised entropy after the Page time.

\section{A short review of JT gravity}
\label{sect:JT}
We first present a concise review of the classical dynamics of JT gravity.
\subsection{Action and dynamics}
We begin by writing down the action for 2d JT gravity with metric $g$ and dilaton $\Phi$ coupled to conformal matter $\phi$ ($\lads = 1$) \cite{jackiw, teitelboim, Almheiri2015, Jensen2016, Maldacena2016a, Engelsoey2016}
\begin{subequations}
    \begin{align}
        S[g, \Phi, \phi] &= S_{\text{top}}[g] + S_{\text{JT}}[g, \Phi] + S_m[g, \phi], \\
        S_{\text{top}}[g] &= \frac{\Phi_0}{16 \pi \GN} \left[\int_\mathcal{M} \sqrt{-g} R + 2 \int_{\partial \mathcal{M}} \sqrt{- \gamma} K   \right], \\
        S_{\text{JT}}[g, \Phi] &= \frac{1}{16 \pi \GN} \left[\int_\mathcal{M} \sqrt{-g} \Phi (R + 2) + 2 \int_{\partial \mathcal{M}} \sqrt{- \gamma} \Phi_b (K - 1) \right].
    \end{align}
\end{subequations}
\indent The first action $S_{\text{top}}[g]$ is purely topological and adds a constant contribution $\sim \Phi_0 \chi$ through the Gauss--Bonnet theorem where $\chi$ is the Euler characteristic of the corresponding manifold $\mathcal{M}$. 
The second action $S_{\text{JT}}[g, \Phi]$ captures the leading deviation from extremality of higher-dimensional black holes where $\Phi_0 + \Phi$ measures the area of the transverse space in the parent theory. 
Moreover, we added the usual Gibbons--Hawking--York boundary term with boundary metric $\gamma$, 
curvature $K$, and boundary value $\phi_b$ for the dilaton. There is an additional holographic counterterm required which explains the $(K-1)$ combination. 
The last action $S_m[g, \phi]$  is simply a 2d CFT action for a matter field which does not couple to the dilaton directly.\\
\\
Varying the total action w.r.t.~$\Phi$ imposes the constant curvature constraint $R+2 =0$.
In other words, the geometry is everywhere locally AdS$_2$ with a constant negative curvature. In Poincar\'e coordinates $(F, Z)$ this is 
\begin{equation}
    ds^2 = \frac{-dF^2 + dZ^2}{Z^2} = -\frac{4}{(\xplus - \xminus)^2} d\xplus d\xminus ,
    \label{sect2:poincare}
\end{equation}
with $X^\pm = F \pm Z$ lightcone coordinates, future and past horizon at $X^\pm = \pm \infty$, and the boundary located at $\xplus = \xminus$ as per usual in the Poincar\'e patch $\xminus \leq \xplus$.

\indent The other dynamics can be found from varying with respect to the metric and yields the EoM for the dilaton field sourced by the conserved energy-momentum of the matter CFT sector \cite{Almheiri2015}.

\subsection{Boundary particle}
\label{sect:boundaryPart}
The classical dynamics of JT gravity can be conveniently described in terms of a dynamical \textbf{boundary particle}. This description originated from the SYK model, and was developed in parallel in  \cite{Maldacena2016a, Jensen2016, Engelsoey2016,Kitaev:2017awl}. 
From this viewpoint, the boundary time $t$ is a preferred coordinate from which we can define a dynamical variable at the \ads~boundary: the time reparametrisation $F(t)$ of a fixed reference time $F$, the Poincar\'e time. This preferred boundary time $t$, associated to a boundary particle/observer, naturally describes the time evolution along the \ads~boundary $u = v \equiv t$. By requiring that these coordinates $(u, v)$ near the boundary coincide with the \ads~boundary in the Poincar\'e patch itself $\xplus(u) = \xminus(v)$, we acquire the dynamical/holographic boundary curve
\begin{equation}
    \xplus(t) = \xminus(t) \equiv F(t),
\end{equation}

\indent After introducing a regulator for the \ads~boundary $\varepsilon$ which moves the boundary slightly inwards, the boundary observer's coordinate frame $(t, z)$ 
can be related to that of a bulk observer through 
\begin{subequations}
    \begin{align}
        \label{sect2:boundary1}
        \frac{\xplus(t + \eps) + \xminus(t-\eps)}{2} &= F(t),\\
        \label{sect2:boundary2}
        \frac{\xplus(t + \eps) - \xminus(t-\eps)}{2} &= \eps F'(t),
    \end{align}
\end{subequations}
where $\eps F'(t)$ is equal to the distance between the holographic boundary curve and the true boundary.\\
\indent There is a quite natural way to extend this coordinate frame into the entire bulk: a boundary observer can construct a bulk frame by shooting in and collecting light rays $u = t + z,~v = t - z$. 
They shoot in a light ray at $v = t_1$ as measured on their clock and collect it back at $u = t_2$. Now, taking this procedure to the Poincar\'e patch 
where the boundary observer sends at $F_1 = F(t_1)$ and receives at $F_2 = F(t_2)$, they can associate the coordinates $\xminus = F_1,~\xplus = F_2$ to every bulk point.  
From this they can construct a unique bulk frame and metric -- the radar definition of the bulk \cite{Blommaert2019, Mertens:2019bvy}
\begin{subequations}
    \begin{alignat}{2}
        \label{sect2:xplusminusB}
        &\xplus(u) = F(u), \hspace{1cm} \xminus(v) = F(v&), \\
        &ds^2(F) = \frac{F'(u) F'(v)}{[F(u) - F(v)]^2} (dz^2 - dt^2&).
    \end{alignat}
\end{subequations}
This procedure is boundary-intrinsic and is constructed via local operations from the boundary observer perspective. Allowing quantum fluctuations, the boundary observer experiences a fixed bulk location but fluctuating metric whilst a Poincar\'e observer would see a fuzzy bulk location and a fixed metric.\\
\indent The boundary particle dynamics can be derived by imposing  specific boundary conditions as \cite{Almheiri2015, Maldacena2016a, Engelsoey2016}
\begin{equation}
    \eval{g_{tt}}_{\partial \mathcal{M}} = \frac{1}{\eps^2}, \hspace{1,5cm} \eval{\Phi}_{\partial \mathcal{M}} = \Phi_b = \frac{\phi_r}{\eps},
\end{equation}
where $\phi_b$ is large and $\phi_r$ is a fixed constant. In this approach, the JT action reduces to a boundary term and the theory becomes a Schwarzian theory in 0+1 dimensions
\begin{align}
    S_{\text{JT}} \to -\frac{\phi_r}{8 \pi \GN} \int \frac{dt}{\eps} \frac{K - 1}{\eps} = -C \int dt \{F(t), t \}, \quad \{F(t), t \} = \frac{F'''}{F'} - \frac{3}{2} \left(\frac{F''}{F'} \right)^2,
    \label{sect2:schwarzianAction}
\end{align}
where $C \equiv \frac{\phi_r}{8 \pi \GN}$, with induced metric $dt / \eps$ and extrinsic curvature $K = 1 + \eps^2 \{F(t), t \} + \hdots $.

\indent With the aforementioned result, the total energy can be computed directly by relating the action (\ref{sect2:schwarzianAction}) to the boundary Hamiltonian
\begin{equation}
    E(t) = - \frac{\phi_r}{8 \pi \GN} \{F(t), t \}.
    \label{sect2:ADM}
\end{equation}

\noindent When allowing for backreaction in a semiclassical setup, the rate at which the total energy changes is merely equal to a net flux of the energy flow at the holographic boundary:
\begin{equation}
    \dv{E}{t}(t) = \left\langle T_{vv}(t) \right\rangle - \left\langle T_{uu}(t) \right\rangle,
\end{equation}
where the stress tensors showing up are the covariant ones \cite{Davies:1976ei, Christensen:1977jc} (see also the book \cite{Fabbri2005} for a thorough pedagogical treatment):
\begin{subequations}
    \begin{align}
        \label{sect2:tensor1}
        T_{uu} &= -\frac{c}{12 \pi} \left[(\partial_u \omega)^2 - \partial_u^2 \omega  \right] + \normord{T_{uu}}, \\
        \label{sect2:tensor2}
        T_{vv} &= -\frac{c}{12 \pi} \left[(\partial_v \omega)^2 - \partial_v^2 \omega  \right] + \normord{T_{vv}}, \\
        T_{uv} &= -\frac{c}{12 \pi} \partial_u \partial_v \omega,
    \end{align}
\end{subequations}
for a spacetime $ds^2 = -e^{2 \omega(u, v)} du dv$. The second term is the normal-ordered stress tensor which is chirally conserved and frame-dependent since normal-ordering is always with respect to a certain vacuum. This normal-ordered term is exactly what one would measure via a detector calibrated to their vacuum. However, this term transforms non-covariantly under a general conformal transformation $(U(u) , V(v))$ due to the conformal anomaly:\footnote{Observe how for a M\"obius transformation $U \to \frac{aU+b}{cU+d}$ in PSL$(2,\mathbb{R})$ (and independently for $V$), the anomalous term disappears. This is directly related to the fact that these transformations leave the vacuum state invariant.}
\begin{equation}
    \normord{T_{uu}}~= \left(\frac{d U}{du}  \right)^2 \normord{T_{UU}} - \frac{c}{24 \pi} \{U, u\}, \hspace{1cm} \normord{T_{vv}}~= \left(\frac{d V}{dv}  \right)^2 \normord{T_{VV}} - \frac{c}{24 \pi} \{V, v\},
    \label{sect2:stress}
\end{equation}

\indent Only the total sum is covariant in the sense that $\nabla_\mu T^{\mu \nu} = 0$ and transforms as a tensor. These properties are important to be consistent with the Einstein equation. By giving an operational meaning to this equation, it resolves a paradox related to the experience of accelerated observers: the Minkowski vacuum for an accelerated observer looks like a thermal Rindler state, but if this thermal stress-energy backreacts on the spacetime, it would deform the flat space we started with and invalidate the entire setup. However, what the accelerated observer measures, as stated, is the normal-ordered piece and not the total covariant stress tensor. So by adding the vacuum/Casimir piece -- the first term -- to their measurement, the effect cancels out and no net backreaction is happening. \\
\indent The case for AdS$_2$ gravity is special because:
\begin{equation}
    -\frac{c}{12 \pi} \left[(\partial_u \omega)^2 - \partial_u^2 \omega  \right] = \frac{c}{24 \pi} \{ F(u), u \},
\end{equation}
and analogously for $v$, which means the Casimir piece of both stress tensors (\ref{sect2:tensor1}, \ref{sect2:tensor2}) are equal to the expression above 
\begin{equation}
    T_{uu}(u) = \frac{c}{24 \pi} \{ F(u), u \} ~+ \normord{T_{uu}(u)}, \hspace{1cm} T_{vv}(v) = \frac{c}{24 \pi} \{ F(v), v \} ~+ \normord{T_{vv}(v)} ,
    \label{sect2:cov}
\end{equation}
such that for the difference at the boundary $u=v=t$, they cancel each other out: 
\begin{equation}
    \dv{E}{t} = \left\langle T_{vv}(t) \right\rangle - \left\langle T_{uu}(t)\right\rangle =~ \left\langle \normord{T_{vv}(t)} \right\rangle  - \left\langle \normord{T_{uu}(t)} \right\rangle. 
    \label{sect2:flux}
\end{equation}
This equation of motion should be viewed as the JT version of the semiclassical Einstein equation:
\begin{equation}
    G_{\mu\nu} = 8 \pi G \left\langle T_{\mu\nu}\right\rangle .
\end{equation}

\subsection{Static black hole solution}
For cases in which the matter stress tensor vanishes and with metric (\ref{sect2:poincare}), we obtain the static black hole background with mass $M=E$ and dilaton profile 
\begin{equation}
    \Phi = 2 \phi_r \frac{1 - (\pi T)^2\xplus \xminus}{\xplus - \xminus},
    \label{sect2:dilatonStatic}
\end{equation}
at Hawking temperature $T$ in \ads~\cite{Almheiri2015}. The time reparametrisation in this case $F(t)$ becomes 
\begin{equation}
    F(t) = \frac{1}{\pi T} \tanh(\pi T t),
    \label{sect2:thermal}
\end{equation}
which is a solution for a constant Schwarzian
\begin{equation}
    \{F(t), t \} = - \frac{2\pi^2}{\beta^2}.
\end{equation}
From this follows the metric and dilaton in the new coordinates $(u, v)$ through $(\xplus(u), \xminus(v))$
\begin{equation}
    ds^2 = - \frac{4 (\pi T)^2}{\sinh^2[\pi T (u - v)]} du dv, \qquad         \Phi = 2 \phi_r \pi T \coth[\pi T (u - v)].
		\label{staticmetric}
\end{equation}
Both expressions are periodic in Euclidean time with period $\beta$.\\
\indent This spacetime has a future and past horizon where $u = +\infty,~v = -\infty$ which translates in Poincar\'e coordinates to 
\begin{equation}
    X^\pm = \pm \frac{1}{\pi T}.
    \label{sect2:bifurcate}
\end{equation}
These two points are exactly where the boundary particle meets the \ads~boundary for $X^\pm = F(t \to \pm \infty)$ at the two points respectively. The black hole is characterised by the following thermodynamical relations above extremality:
\begin{equation}
    E = \frac{\pi \phi_r}{4 \GN}T^2, \qquad S =  \frac{2 \pi \phi_r}{4 \GN} T.
    \label{sect2:staticEnergy}
\end{equation}
The energy is found by plugging the time reparametrisation (\ref{sect2:thermal}) into (\ref{sect2:ADM}). The entropy above extremality is found by either using the thermodynamical definition, the RT formula or the Bekenstein--Hawking result in combination with (\ref{sect2:staticEnergy}). Notice that for the latter two, the minimal surface $\partial_\pm \phi = 0$ and the horizon $\eval{g_{tt}}_h = 0$ are both codimension-2 surfaces which in our spacetime translate to a point. This is in correspondence to the dilaton being related to the area of the transverse space of the parent theory via 
dimensional reduction. 

\section{Evaporating black holes}
\label{sect:EBH}
We now implement evaporation in this model.

\subsection{Setup}
\begin{figure}[h!]
    \centering
    \begin{subfigure}[b]{0.45\textwidth}
        \centering
				    \includegraphics[width=0.9\linewidth]{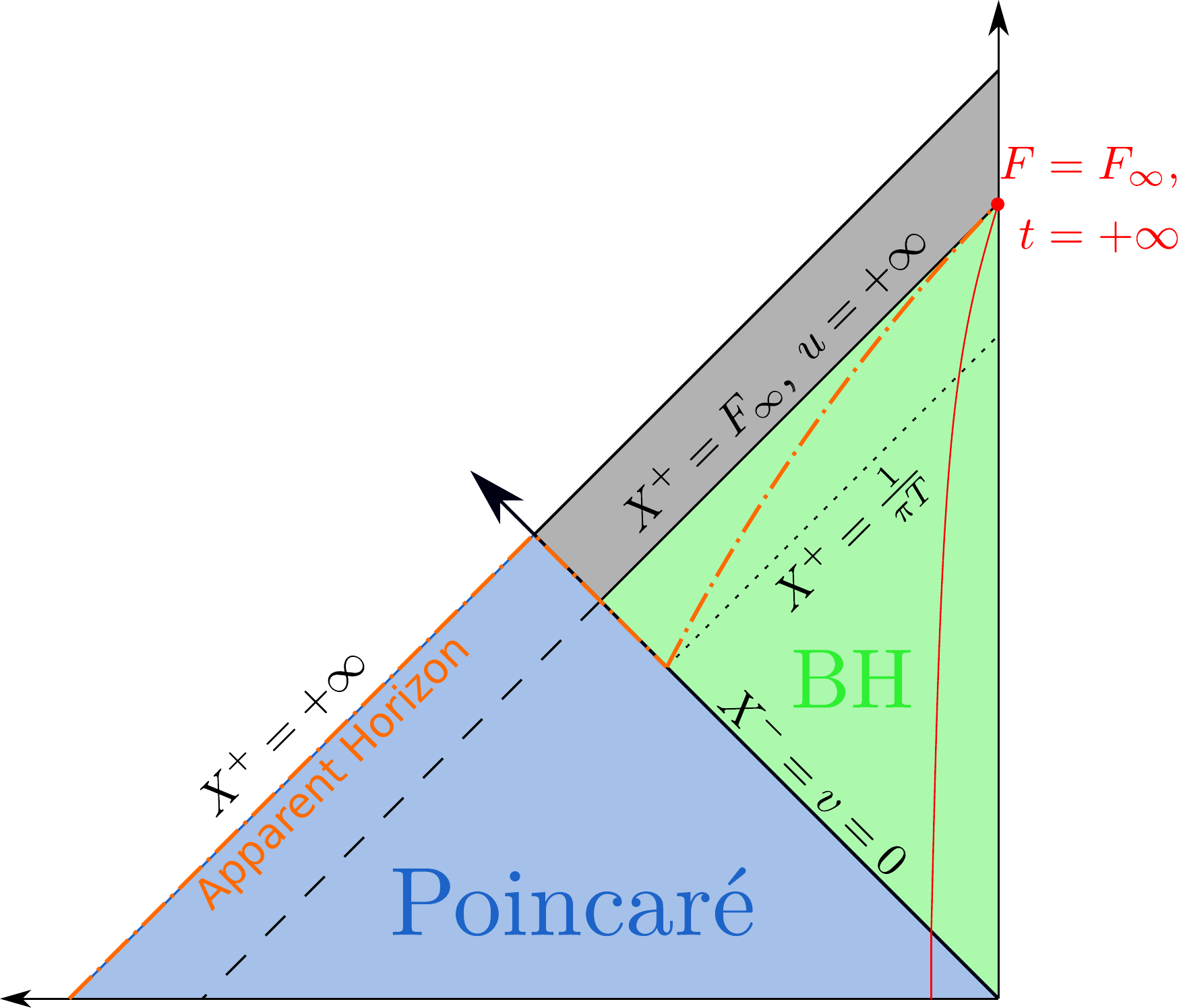}
               \caption{}
        \label{sect3:figEBH}
    \end{subfigure}
    \hfill
    \begin{subfigure}[b]{0.53\textwidth}
        \centering
				 \includegraphics[width=\textwidth]{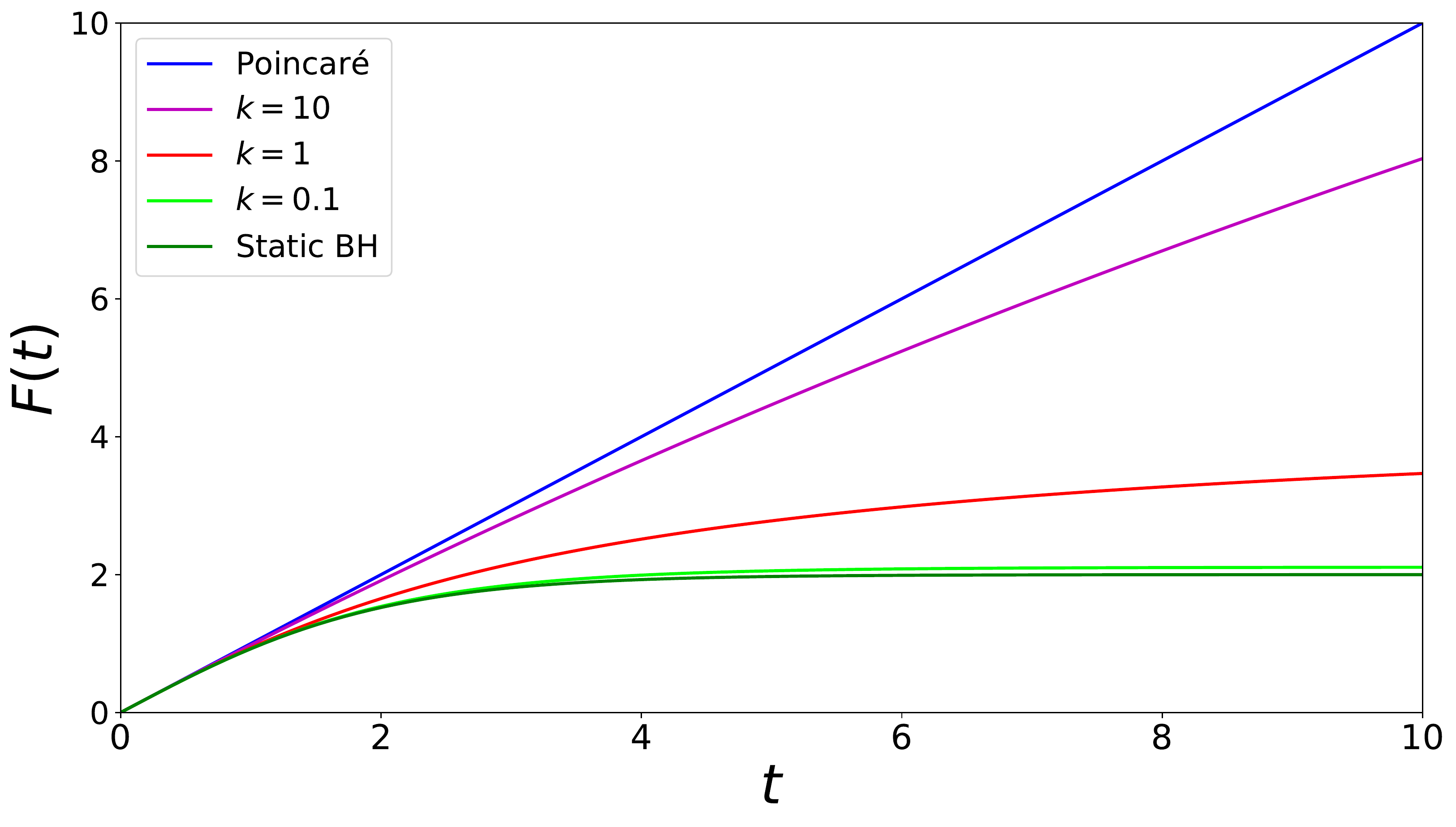}
        \caption{}
        \label{sect3:figReparam}
    \end{subfigure}
    \caption{\textbf{(a)} Evaporating black hole setup. Starting in the Poincar\'e patch, we send in a pulse at $t=0$ to form a black hole. If the black hole were static, the $(u, v)$ coordinates could only describe up to 
    the would-be horizon $\xplus = \frac{1}{\pi T}$. In the evaporating case, these coordinates describe the patch up to $F_\infty$. The red wiggly curve represents the holographic boundary curve. The apparent horizon is denoted in dashed orange. \textbf{(b)} Reparametrisation profiles for the different scenarios and $T = \frac{1}{2\pi}$. The time reparametrisation of the evaporating black hole interpolates between the eternal black hole and the Poincar\'e patch for 
    increasing evaporation rate $k$.}
\end{figure}

We start with the extremal black hole (the Poincar\'e patch) and form a non-extremal black hole by sending in a classical matter pulse at $t=0$ with fixed energy $E_0$
\begin{equation}
    \dv{E}{t} = E_0 \delta(t) + \left\langle \normord{T_{vv}(t)} \right\rangle - \left\langle \normord{T_{uu}(t)} \right\rangle. 
    \label{sect3:pulse}
\end{equation}
The quantum state in which we evaluate the matter fluxes is hence always the initial Poincar\'e frame. In this Poincar\'e patch, we have ~$\left\langle \normord{T_{++}(t)} \right\rangle ~= 0 =~ \left\langle \normord{T_{--} (t)} \right\rangle$ for all times $t$.\\
\indent Before the pulse $t < 0$ there is nothing ingoing or outgoing such that $\left\langle T_{uu} \right\rangle  = 0 = \left\langle T_{vv} \right\rangle$. So, the equations to solve 
(\ref{sect2:ADM}, \ref{sect3:pulse}) become 
\begin{equation}
    \{ F(t), t \} = 0, \hspace{1,5cm} \dv{}{t} \{ F(t), t \} = 0, 
\end{equation}
which has the trivial Poincar\'e solution $F(t) = t$. A boundary observer would also measure the 
Poincar\'e vacuum $\left\langle \normord{T_{vv}} \right\rangle ~= 0 =~ \left\langle \normord{T_{uu}(t)} \right\rangle$.\\  
\indent After we have sent in the pulse $t > 0$, we do not throw anything in anymore: $\left\langle T_{vv}\right\rangle = 0 = \left\langle T_{uu} \right\rangle$. By imposing reflecting BC, the Hawking radiation sent out by the 
black hole eventually falls back into it again and feeds it such that it remains at a constant energy. The Poincar\'e vacuum is now described in the black hole frame $(u, v)$ which has a nonzero Schwarzian contribution (\ref{sect2:cov}):
\begin{equation}
    \left\langle \normord{T_{uu}} \right\rangle ~=~ \left\langle \normord{T_{vv}} \right\rangle~= -\frac{c}{24 \pi} \schw{F(t)}{t}.
\end{equation}
Equations (\ref{sect2:ADM}, \ref{sect3:pulse}) now dictate that the energy stays constant at $E_0$
\begin{equation}
    \{ F(t), t \} = E_0, \hspace{1,5cm} \dv{}{t} \{ F(t), t \} = 0. 
\end{equation}
These two conditions lead to the static black hole solution with reparametrisation (\ref{sect2:thermal}) and Hawking temperature ({\ref{sect2:staticEnergy}}).
Hence, as viewed from the black hole frame, we obtain an eternal Unruh heat bath \cite{Spradlin:1999bn}
\begin{equation}
   \left\langle \normord{T_{uu}} \right\rangle ~=~ \left\langle \normord{T_{vv}} \right\rangle ~= \frac{\pi c}{12 \beta^2}.
\end{equation}
\indent To let our black hole evaporate, we take away the outgoing Hawking radiation. We can imagine a boundary observer moving along the holographic boundary curve 
and who holds a detector perfectly absorbing all emitted Hawking radiation; hence $ \left\langle \normord{T_{vv}} \right\rangle ~= 0$ -- there is no ingoing flux feeding the black hole. 
The outgoing Hawking radiation is nonzero and measured at the boundary as:
\begin{equation}
    \left\langle \normord{T_{uu}} \right\rangle ~= -\frac{c}{24 \pi} \schw{F(t)}{t}.
\end{equation}
We now have a nonzero energy rate (\ref{sect3:pulse}) $\dv{E}{t} = - \left\langle \normord{T_{uu}} \right\rangle $ and together with the total energy (\ref{sect2:ADM}) we have to solve the following differential equation
\begin{equation}
    \frac{d}{dt} \schw{F(t)}{t} = - k \schw{F(t)}{t}, \quad t>0, 
    \label{sect3:ODE}
\end{equation}
or in terms of the energy profile:
\begin{equation}
    \dv{E}{t} = - k E \,\, \Rightarrow \,\,  E(t) = E_0 e^{-kt},
    \label{sect3:ECondition}
\end{equation}
after implementation of the initial value $E(0) = E_0$ and where we defined the evaporation rate $k$ as 
\begin{equation}
    k \equiv \frac{c \GN}{3 \phi_r} = \frac{c}{24 \pi C},
    \label{sect3:k}
\end{equation}
with units $[k] = [\text{energy}] = [\text{time}]^{-1}$. We obtain an exponentially decaying energy and with this we can find the reparametrisation $F(t)$ by solving 
\begin{subequations}
    \begin{align}
        \label{sect3:EBHSch}
        &\schw{F(t)}{t} = -2 (\pi T)^2 e^{-kt},\\
        &F(0) = 0, \hspace{.5cm} F'(0) = 1, \hspace{.5cm} F''(0) = 0,
        \label{sect3:BC}
    \end{align}
\end{subequations}
with the BC stemming from gluing along the pulse at $t = 0$ which also determines $E_0$ in terms of the would-be temperature via (\ref{sect2:staticEnergy}).
The solution is pretty complicated and contains the modified Bessel functions of the first and second kind \cite{Engelsoey2016}
\begin{equation}
    F(t) = \frac{1}{\pi T} \frac{I_0(\alpha) K_0\left(\alpha e^{-\frac{kt}{2}}\right) - K_0(\alpha) I_0\left(\alpha e^{-\frac{kt}{2}}\right)}{I_1(\alpha) K_0\left(\alpha e^{-\frac{kt}{2}}\right) + K_1(\alpha) I_0\left(\alpha e^{-\frac{kt}{2}}\right)}, 
    \hspace{1,5cm} \alpha = \frac{2 \pi T}{k}.
    \label{sect3:reparam}
\end{equation}
The reparametrisation $F(t)$ increases monotonically and asymptotes to a fixed value $F_\infty$ for $t \to +\infty$ beyond the horizon of the would-be static black hole. However, it does not reach the original Poincar\'e horizon as shown in Fig.~\ref{sect3:figEBH}.\footnote{ Indeed, we have
\begin{equation}
    F_\infty \equiv F(t \to +\infty) = \frac{1}{\pi T} \frac{I_0(\alpha)}{I_1(\alpha)},
    \label{sect3:horizon}
\end{equation}
and Poincar\'e time stops flowing as $F'(t \to +\infty) \to 0$ \cite{Engelsoey2016}.} Different profiles are plotted in Fig.~\ref{sect3:figReparam}.

\indent  For a macroscopic black hole $k/T \ll 1$, the reparametrisation and horizon value can be approximated by the following expressions \cite{ZheChen2020}\footnote{This just boils down to expanding the modified Bessel functions for large arguments \cite{NIST:DLMF}
\begin{equation*}
    I_\nu(z) = \frac{e^z}{\sqrt{2\pi z}} \left(1 - \frac{4\nu^2 - 1}{8z} \right) + \mathcal{O}(z^ {-2}), \hspace{1cm} K_\nu(z) = \sqrt{\frac{\pi}{2z}} e^{-z} \left(1 + \frac{4\nu^2 - 1}{8z} \right) + \mathcal{O}(z^ {-2}).
\end{equation*}} 
\begin{equation}
    F(t) \approx F_\infty \tanh \left[ \frac{2\pi T}{k}\left(1- e^{-\frac{k}{2}t} \right) \right], \hspace{1,5cm} F_\infty = \frac{1}{\pi T} + \mathcal{O}(k).
    \label{sect3:reparamApprox}
\end{equation}
This can be compared to the static black hole solution \eqref{sect2:thermal}.

\subsection{Matter sector: recalibrating the entropy}
\label{Entropy}

To find islands through the QES formula, we require an expression for the entanglement entropy of the bulk fields across the spacelike interval between the island and the boundary. 
The fact that this interval is anchored at one end on the holographic boundary provides us with an entropy evolving in boundary time $t$. 
Since the matter sector does not couple to the dilaton directly, we can treat it as a QFT on a fixed AdS$_2$ background.
Moreover, since information flow in a CFT is preserved along null lines, we will be able to interpret our setup as a boundary observer with a detector calibrated according to their vacuum in boundary coordinates $(u, v)$, who measures the entanglement between the radiation already captured (early time) and the radiation yet to come out of the remaining black hole (late time). \\

\noindent Let us start in flat space in lightcone coordinates $ds^2 = - d\xplus d\xminus$ and measure the entanglement entropy across a single interval for a free, massless scalar with respect to the Minkowski vacuum $\ket{0_+}$ \cite{Holzhey1994, Fiola1994, Mertens:2019bvy}. For a free massless scalar, the CFT decouples in a right-moving and left-moving part, thus we can calculate the contribution to the entropy from these modes separately across intervals $[\xplus_1, \xplus_2]$ and $[\xminus_1, \xminus_2]$
\begin{equation}
    S = \frac{c}{12} \ln \frac{(\xplus_1 - \xplus_2)^2}{\delta_1 \delta_2} + (\xplus_i \to \xminus_i),
    \label{sect3:entropyX}
\end{equation}
where $\delta_i$ are the cutoffs in the points $(\xplus_i, \xminus_i)$ as measured by an observer in this frame.\\
\indent Alternatively, we can look at the entropy with respect to another vacuum $\ket{0_u}$ related to the original vacuum by a conformal transformation $u(\xplus)$, similarly for $v(\xminus)$
\begin{equation}
    S = \frac{c}{12} \ln \frac{(u_1 - u_2)^2}{\hat{\delta}_1 \hat{\delta}_2} + (u_i \to v_i),
    \label{sect3:Sintermed}
\end{equation}
with cutoffs $\hat{\delta}_i$ as measured by an observer in the $(u ,v)$ frame. \\
\indent The cutoffs $\delta_i$ and $\hat{\delta}_i$ are related by how the clocks tick for the observers in the different frames $\hat{\delta}_i = u_i' \delta_i$, since they have their detectors 
calibrated to a different vacuum and hence to a different time coordinate with respect to which they define positive frequency modes. Plugging this relation into (\ref{sect3:Sintermed}) leads to 
\begin{equation}
    S = \frac{c}{12} \ln \frac{(u_1 - u_2)^2}{u_1' u_2' \delta_1 \delta_2} + (u_i \to v_i).
\end{equation}

In these new coordinates, the metric is $ds^2 = - \partial_u \xplus \partial_v \xminus du dv = -e^{2 \omega} du dv$ and the entropy can alternatively be written as 
\begin{equation}
    S = \frac{c}{6}(\omega_1 + \omega_2) + \frac{c}{12} \ln \frac{(u_1 - u_2)^2}{\delta_1 \delta_2} + (u_i \to v_i).
    \label{sect3:entropyST}
\end{equation}
In \cite{Fiola1994} it was argued that this is the correct analogue for a generic curved spacetime with conformal factor $e^{2 \omega}$. Moreover, this equation is directly derived using a standard twist fields approach using the replica trick \cite{Almheiri2019}.\\
\indent We can now apply this procedure to our setup in \ads
\begin{equation}
    ds^2 = - \frac{4}{(\xplus - \xminus)^2} d\xplus d\xminus = - \frac{4 \partial_u \xplus \partial_v \xminus}{(\xplus - \xminus)^2} du dv.
\end{equation}

\noindent We obtain for the vacuum $\ket{0_+}$:
\begin{align}
    S_{\text{bare}} = \frac{c}{6} \left( \ln \frac{2}{\xplus_1 - \xminus_1} + \ln \frac{2}{\xplus_2 - \xminus_2} \right) 
    + \frac{c}{12} \ln \frac{(\xplus_1 - \xplus_2)^2}{\delta_1 \delta_2} + (\xplus_i \to \xminus_i).
    \label{sect3:entropyBare}
\end{align}
However, a boundary observer in the coordinate frame $(u , v)$ would calibrate their detector with respect to their vacuum $\ket{0_u}$, so we should subtract 
a reference entropy 
\begin{align}
    S_{\text{ref}} = \frac{c}{6} \left( \ln \frac{2\partial_u \xplus_1 \partial_v \xminus_1}{\xplus_1 - \xminus_1} + \ln \frac{2\partial_u \xplus_2 \partial_v \xminus_2}{\xplus_2 - \xminus_2} \right) 
    + \frac{c}{12} \ln \frac{(u_1- u_2)^2}{\delta_1 \delta_2} + (\xplus_i \to \xminus_i, u_i \to v_i),
\end{align}
from the bare entropy to obtain a renormalised quantity 
\begin{align}
    S_{\text{ren}} &\equiv S_{\text{bare}} - S_{\text{ref}} = \frac{c}{12} \ln \frac{(\xplus_1 - \xplus_2)^2}{\partial_u \xplus_1 \partial_u \xplus_2 (u_1 - u_2)^2} + (\xplus_i \to \xminus_i, u_i \to v_i).
    \label{sect3:entropyRen}
\end{align}
\indent This is a finite quantity that a boundary observer would measure.\footnote{One might think that we are working in a boundary CFT, and we should hence add the Affleck--Ludwig boundary entropy $\ln g$, and to the two-interval case a term $\ln G(\eta)$ related to the OPE coefficient of the two-point function \cite{Affleck1994}, and where $\eta$ is the crossratio. However, by setting absorbing BC, the boundary can be seen as ``transparent'', and we are then effectively considering a single interval in a space without boundaries.}
Notice that this expression is also equal to the renormalised entropy in flat space, since we can rewrite it as \cite{Mertens:2019bvy}:
\begin{align}
    S_{\text{ren}} &\equiv S_{\text{bare}} - S_{\text{ref}} = \frac{c}{12} \ln \frac{(\xplus_1 - \xplus_2)^2}{\partial_u \xplus_1 \partial_u \xplus_2 \delta_1\delta_2} - \frac{c}{12} \ln \frac{(u_1-u_2)^2}{\delta_1\delta_2}+ (\xplus_i \to \xminus_i, u_i \to v_i).
    \label{renisflat}
\end{align}
This is quite convenient since we know a lot about expressions of this kind.

\subsection{Boundary observer perspective}
\label{sect:perspective}
In the previous section we introduced our model for an evaporating black hole in JT gravity, without specifying a heat bath. 
We took the perspective of a boundary observer who follows the holographic boundary curve with a detecting apparatus in hand, absorbing all the outgoing radiation from the black hole.
This effectively turns our CFT into a chiral CFT since only one set of movers remains. \\

\begin{figure}[h!]
    \centering
    \begin{subfigure}[b]{0.4\textwidth}
        \centering
            \includegraphics[width=\linewidth]{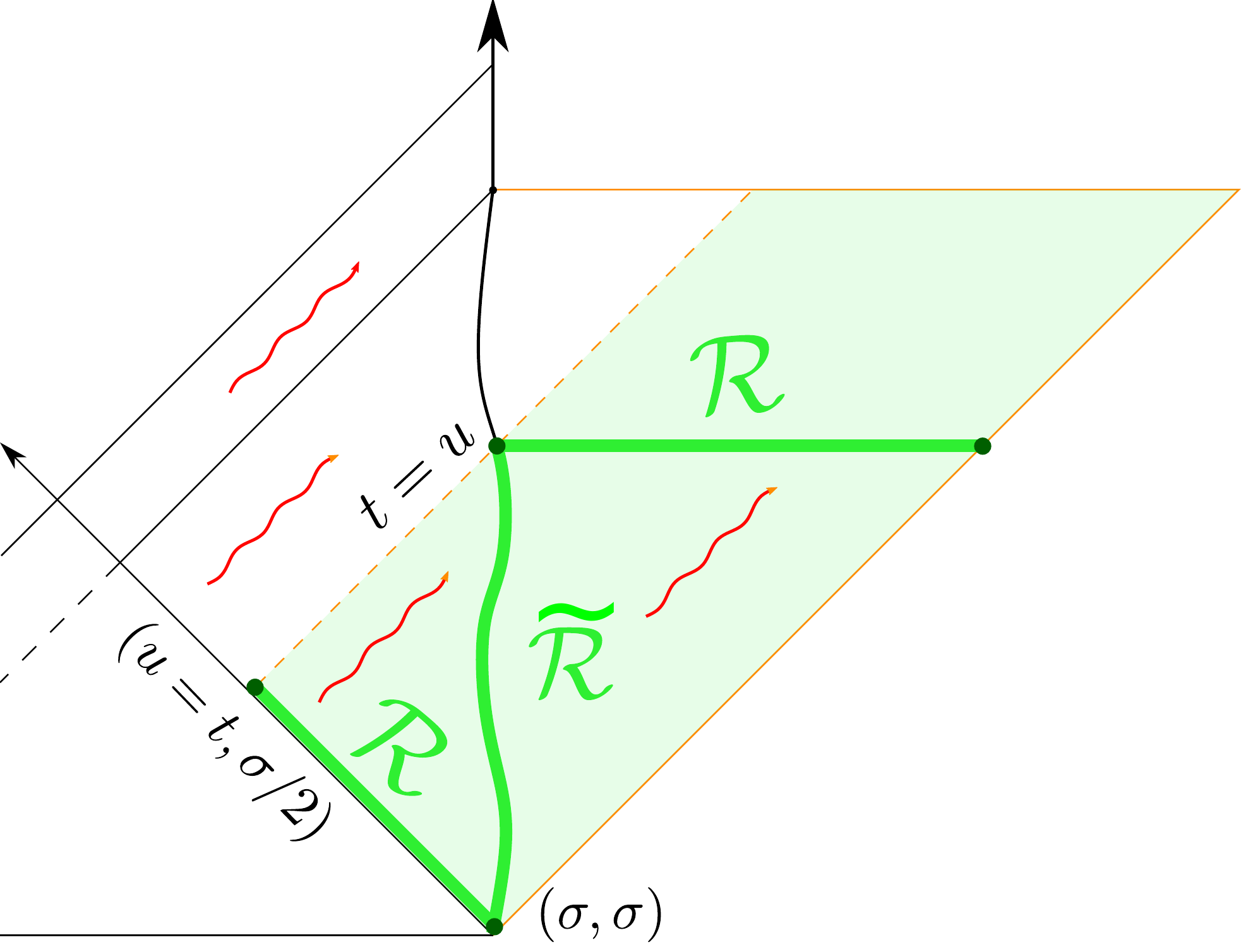}
    \caption{}
    \label{sect4:figNull}
    \end{subfigure}
    \hfill
    \begin{subfigure}[b]{0.58\textwidth}
        \centering
				 \includegraphics[width=0.95\textwidth]{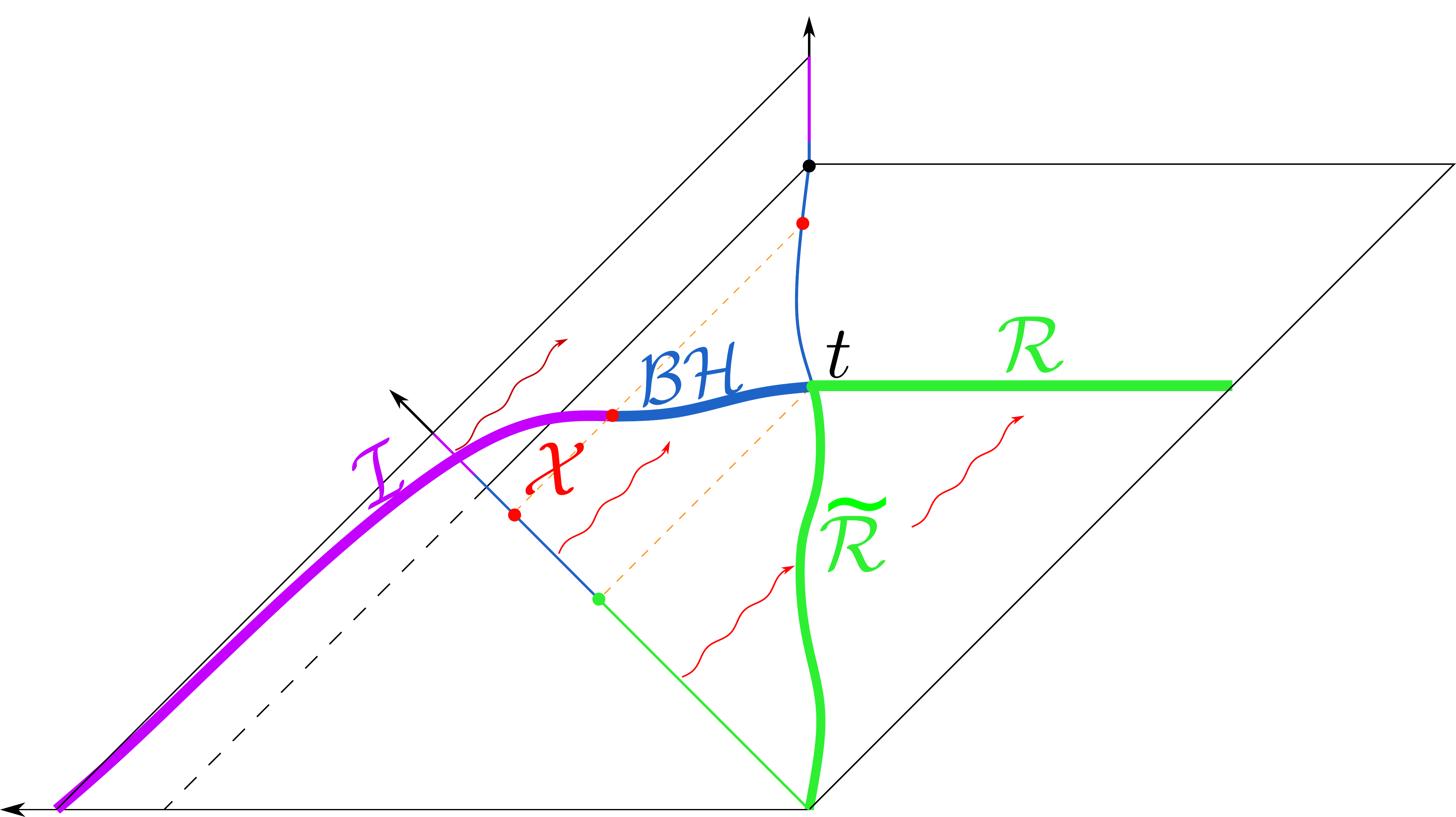}
        \caption{}
        \label{sect4:figOutside}
    \end{subfigure}
    \caption{\textbf{(a)} The radiation region $\mathcal{R}$ up to time $t=u$ is shown as the (nearly) null surface in green. We can add an internal detector system where the radiation caught by the observer is stored. Every interval beginning at the left null ray and ending on the right one constitutes a correct radiation region. We can even map it into the timelike boundary interval $\widetilde{\mathcal{R}}$. \textbf{(b)} Full Cauchy slice $\Sigma = \mathcal{I} \cup \mathcal{BH} \cup \mathcal{R}$, for the case of an island outside the horizon, as we will construct. The union $\mathcal{I} \cup \mathcal{BH} \cup \tilde{\mathcal{R}}$ is not spacelike everywhere, but it serves the same goal for purely right-moving CFT matter flows as is the case here.}
\end{figure}

\noindent For the radiation region $\mathcal{R}$, we take it to be the interval stretching from $(u=\sigma, \, v=\sigma)$ up until $(u = t,\, v=\sigma / 2)$ along the pulse \cite{Mertens:2019bvy}, where $\sigma \ll 1$ is infinitesimal.
We can imagine that as soon as the detector starts absorbing radiation, a heat bath develops behind the boundary which grows as more radiation is getting absorbed -- some kind of internal spacetime for the detector. 
In this heat bath, a horizontal slice is equivalent to the radiation region as the latter can be mapped to the interval by moving along an outgoing null ray. This view is illustrated in Fig.~\ref{sect4:figNull}. 

This links our setup with the usual setup in which a flat space heat bath is glued to the boundary. However, we need not specify any particular dynamics or geometry for this region, as this region is fictitious in our case. The only property it needs is for it to act as an absorber for the radiation. \\
\indent This information preservation along null lines also give rises to a nice interpretation of the island. Given that all radiation can be seen as originating from the pulse 
$v=0$, to each interval along the pulse we can associate an interval $\widetilde{\mathcal{R}}$ which lies along the holographic boundary curve by shifting it over an outgoing null line (Fig.~\ref{sect4:figNull}). An observer starting at $t = 0$ moving along the boundary will start collecting radiation and its associated interval along the pulse will grow. The entropy the observer measures is then a measure for the entanglement between the early radiation -- the collected radiation $\widetilde{\mathcal{R}}$ and the island region $\mathcal{I}$ -- and the late radiation -- the radiation yet to escape $\mathcal{BH}$. This total slice of the spacetime $\mathcal{I} \cup \mathcal{BH} \cup \tilde{\mathcal{R}}$ is \emph{not} spacelike, but can be mapped to a spacelike slice without losing information as discussed above. A full Cauchy slice of this extended spacetime is the union of these three intervals: $\mathcal{I} \cup \mathcal{BH} \cup \mathcal{R}$, as shown in Fig.~\ref{sect4:figOutside}. 

\indent With our operational definition, we will search for and find an island region outside of the horizon. One could ask whether the presence of such an island violates causality. The first appearance of such an island was in \cite{Almheiri2019a} for an eternal AdS$_2$ black hole; causality was restored by means of the quantum focussing conjecture \cite{Bousso2016}. Other cases also report a possible island extending to the outside \cite{Chen2020a, Rozali2020, Gautason2020, Hartman2020, Hashimoto2020, ZheChen2020, Almheiri:2019psy, Geng:2020qvw}. Considering quantum corrections to the event horizon, it was suggested that the island may be inside the stretched horizon \cite{Susskind1993} but outside the classical horizon \cite{Gautason2020}. In higher-dimensional systems this effect can be explained in terms of entanglement wedge nesting \cite{Headrick2014, Wall2014, Chen2020b, Chen2020a}.

\begin{figure}[h!]
    \centering
    \includegraphics[width=.45\linewidth]{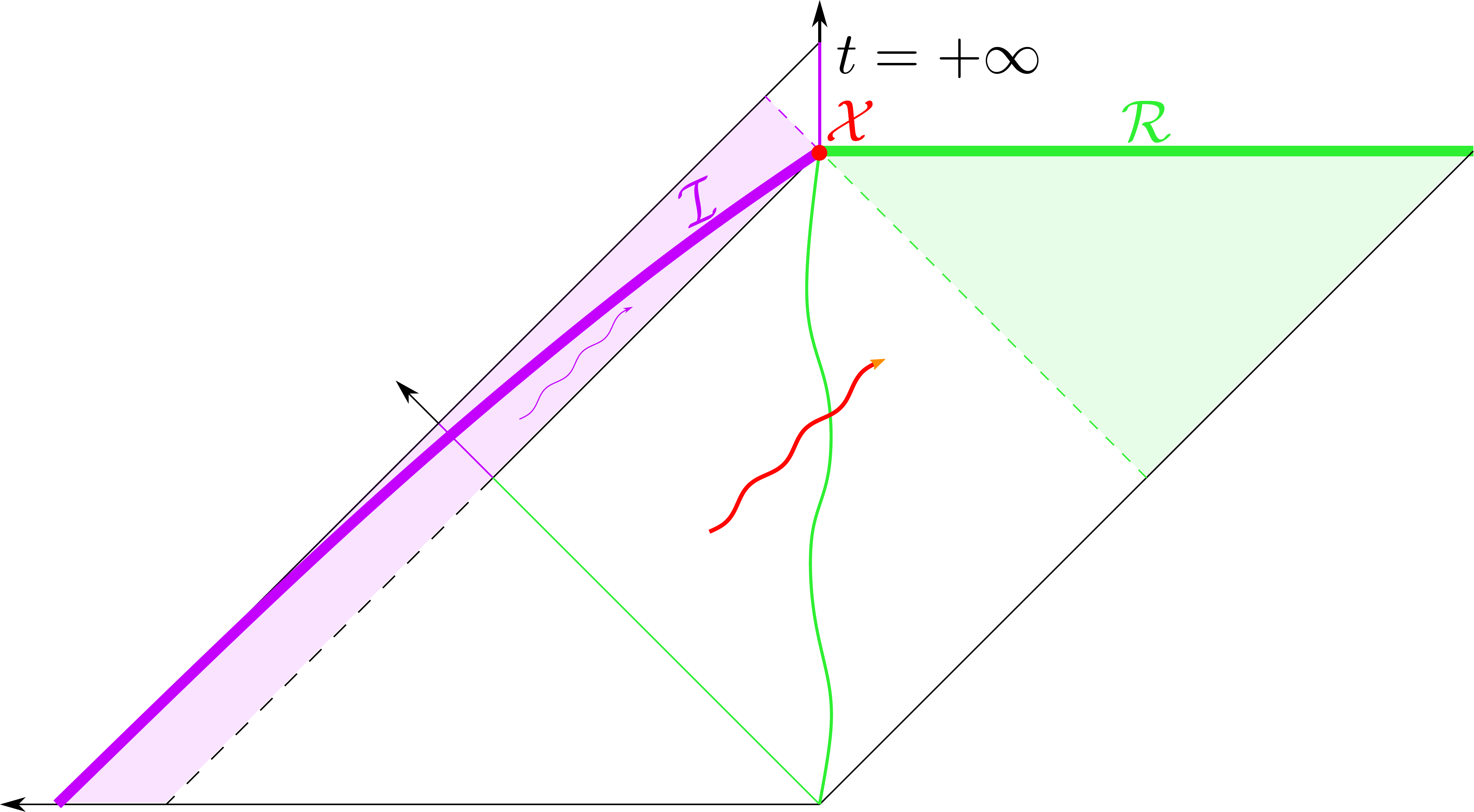}
    \caption{The final island. The island eventually ends up at the same location as the observer on the boundary. There is no $\mathcal{BH}$ anymore, a full Cauchy slice is given by $\mathcal{I} \cup \mathcal{R}$. 
    At this point, the observer has caught all the outgoing radiation and can construct the interior (in purple).}
    \label{sect4:figFinal}
\end{figure}

\indent As the boundary observers moves to $t \to +\infty$ and ends at $F_\infty$, the detector will have collected all the outgoing radiation from the black hole resulting in full knowledge and a zero entropy. Eventually, the region $\mathcal{BH}$ disappears and the full Cauchy slice is now $\Sigma = \mathcal{I} \cup \mathcal{R}$ (see Fig.~\ref{sect4:figFinal}).
If the quantum state was pure on this Cauchy slice, then surely $S(\mathcal{BH}) = S(\mathcal{I} \cup \mathcal{R}) = S(\Sigma) = 0$. Therefore, we expect that the island eventually recedes to this boundary point for $t \to +\infty$. The captured radiation can be used to almost fully reconstruct the black hole interior -- the part lying within the entanglement wedge of the island and the only region relevant for the observer.
This is a clear difference between information loss and preservation when incorporating the existence of such an island. At this point, the scrambling time $t_\text{scr} = t - v_\mathcal{X}$ is zero since the island and the boundary observer coincide; all the information is immediately available. \\

\begin{figure}[h!]
    \centering
    \begin{subfigure}[b]{0.49\textwidth}
        \centering
        \includegraphics[width=0.72\textwidth]{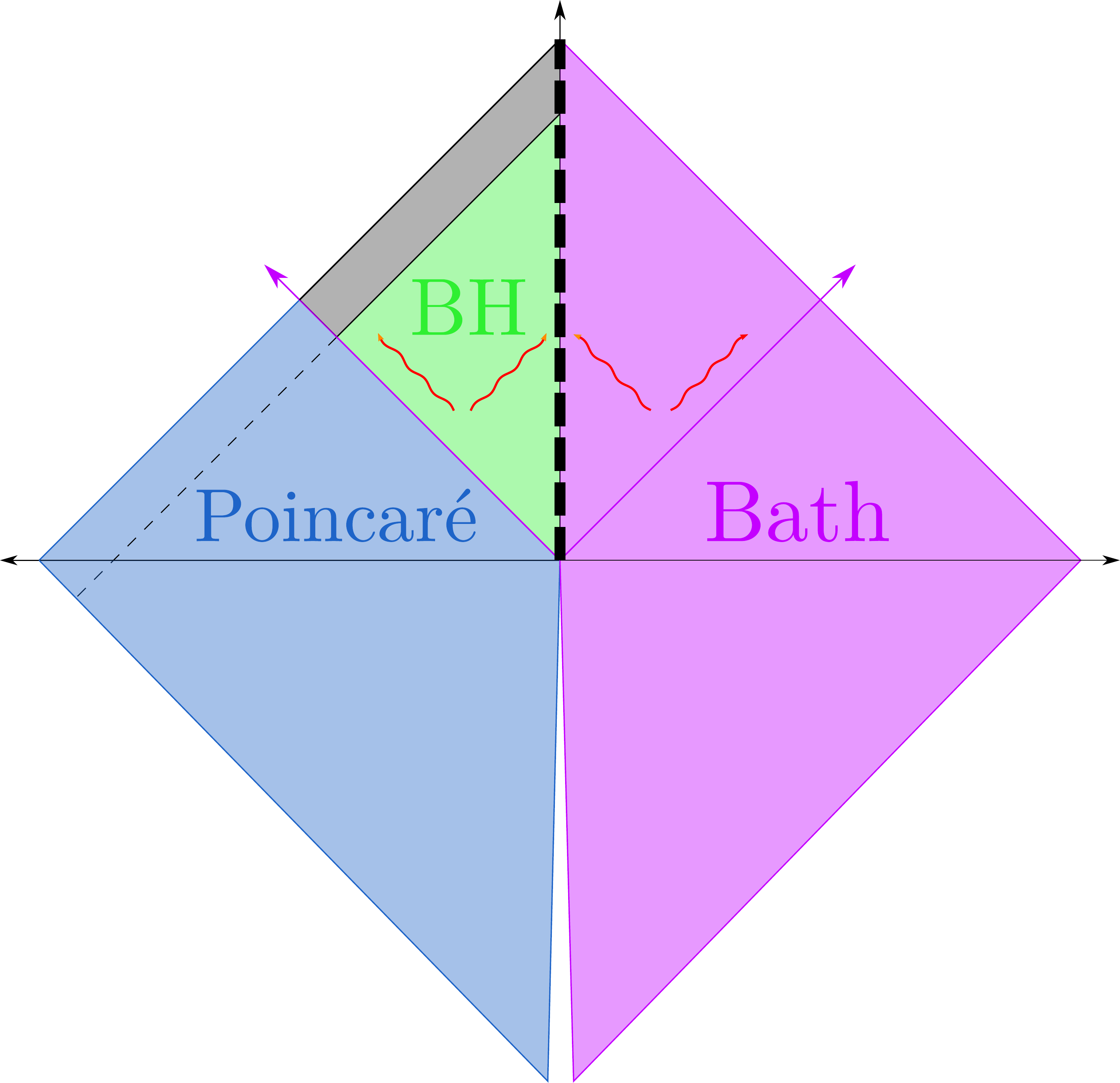}
        \caption{}
        \label{sect4:figLiterature}
    \end{subfigure}
    \hfill
    \begin{subfigure}[b]{0.49\textwidth}
        \centering
        \includegraphics[width=0.36\textwidth]{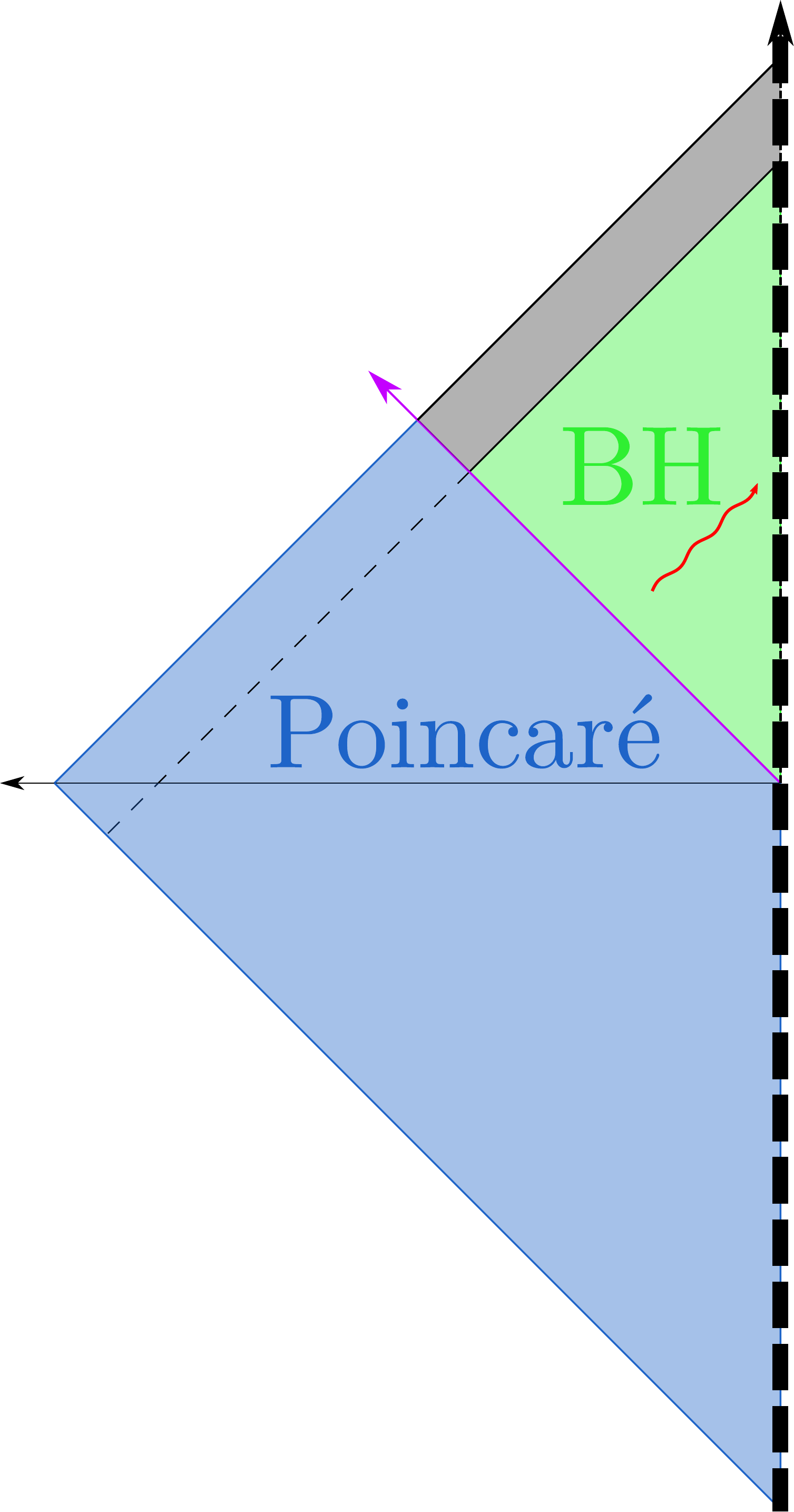}
        \caption{}
        \label{sect4:figEBH}
    \end{subfigure}
    \caption{ \textbf{(a)} The setup in most of the literature. Due to the coupling at $t = 0$ a pulse is sent in forming the black hole. By setting transparent BC, evaporation is achieved and we have radiation in both directions. The dashed vertical line denotes the transparant BC.
    \textbf{(b)} Our setup of an evaporating black hole. A classical pulse is sent in and outgoing radiation is absorbed by the boundary observer. Nothing is ever reflected back in.}
\end{figure}

\noindent The setup we described here varies from the usual setup in most recent papers \cite{Almheiri2019, Almheiri2019a, Almheiri2020, Almheiri2020a, Geng:2020fxl, Goto2020, Chen2020, ZheChen2020, Hollowood2020, Hollowood2020a, Geng:2021hlu, Alishahiha:2020qza, Azarnia:2021uch, Omidi:2021opl}. So we juxtapose both models depicted in Fig.~\ref{sect4:figLiterature} and \ref{sect4:figEBH}.\\
\indent The first model glues a flat region to the holographic boundary. This extra region of spacetime acts as a heat bath collecting the Hawking radiation but where gravity is effectively turned off. Before the pulse is sent in, these two regions of spacetime do not interact with each other and the boundary is assumed to be perfectly reflecting, whereas after the coupling it becomes transparent. This results in a pulse that forms the black hole, reminiscent of a quench procedure (see Appendix B of \cite{Almheiri2018}). \\
\indent In our model, we do not glue any spacetime to the boundary but take the perspective of a boundary observer. This boundary observer moves along the holographic boundary associated to the Poincar\'e patch for $t < 0$. Evaporation for $t > 0$ is achieved by the observer collecting the outgoing Hawking radiation with a detector, without making any assumptions about the dynamics of the absorbing medium, and aiming at purely modelling the dissipation of the black hole system.

\subsection{Radiation and Bekenstein--Hawking entropy}
\label{sect2:sectRadiation}
We consider the radiation region $\mathcal{R}$ to lie along the pulse; this is the interval between $(\sigma, \sigma)$ and $(u = t,~\sigma / 2)$ where $\sigma$ is very small and only there to keep it spacelike (Fig.~\ref{sect4:figNull}). Without loss of generality we set $\phi_0 = 0$ and work in units of $2 \phi_r / 4 \GN$ for all coming computations. 

\begin{figure}[h!]
    \centering
    \begin{subfigure}[b]{0.49\textwidth}
        \centering
        \includegraphics[width=\textwidth]{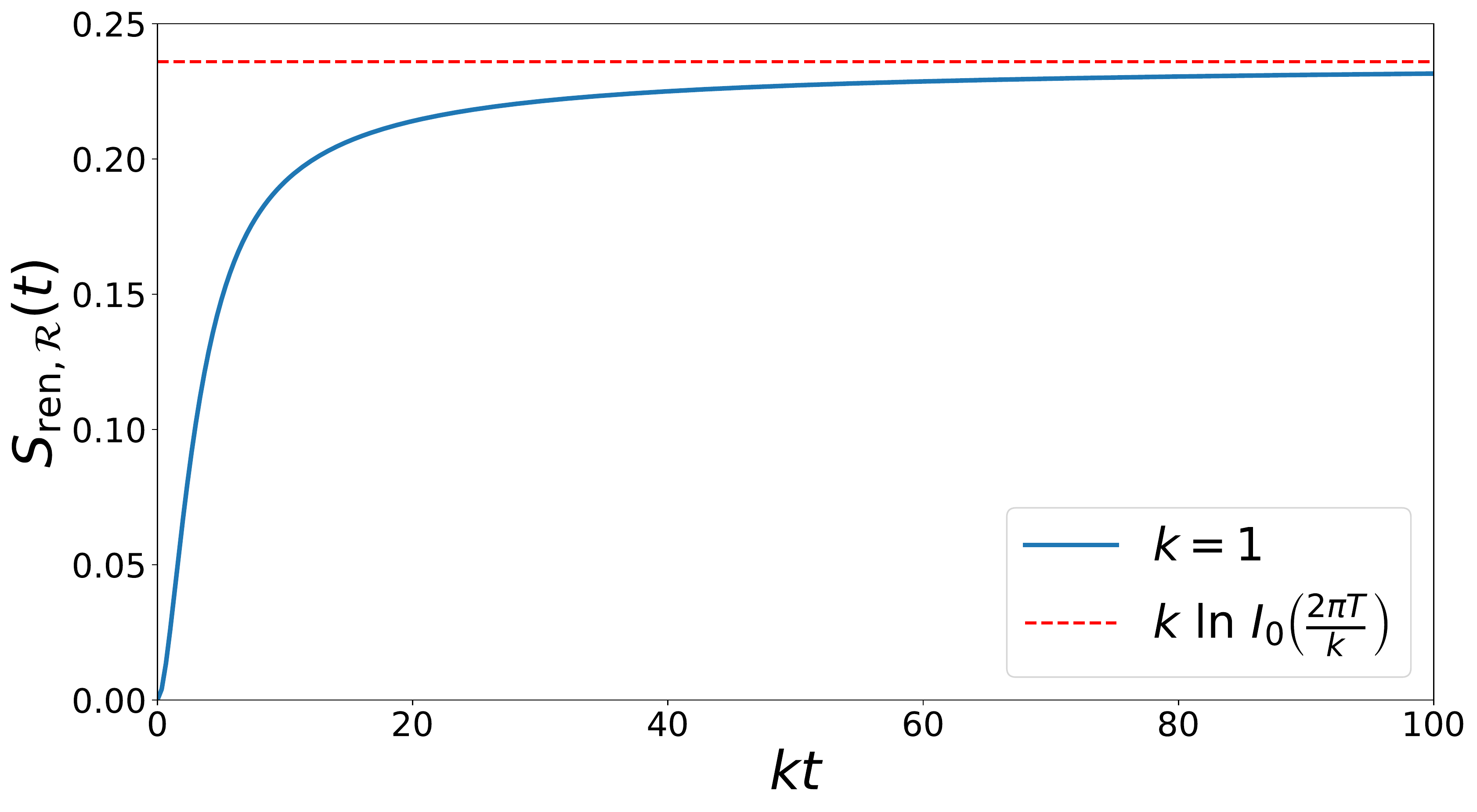}
        \caption{}
        \label{sect2:figRadiation}
    \end{subfigure}
    \hfill
    \begin{subfigure}[b]{0.49\textwidth}
        \centering
        \includegraphics[width=\textwidth]{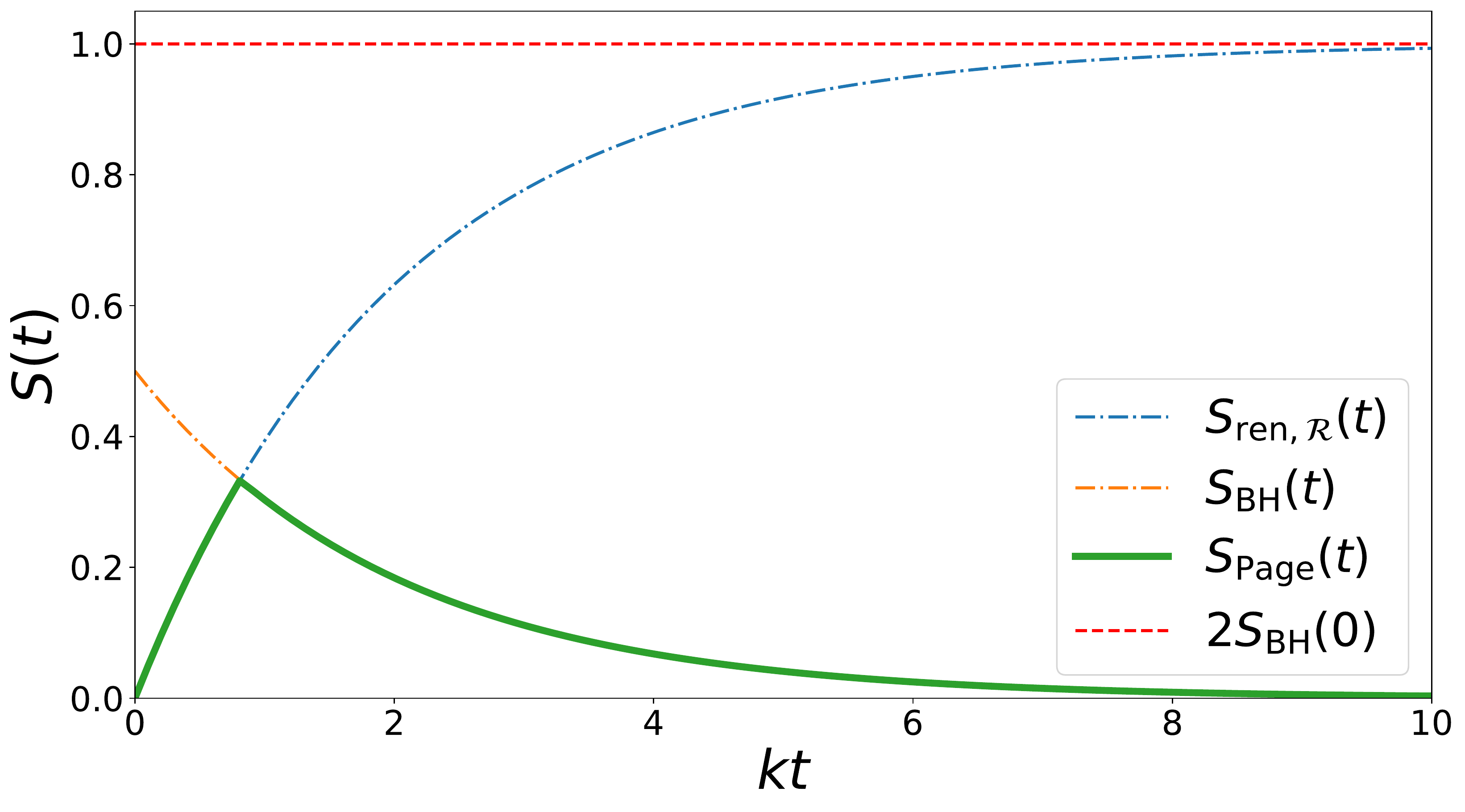}
        \caption{}
        \label{sect2:figQuasistatic}
    \end{subfigure}
    \caption{\textbf{(a)} The renormalised radiation entropy (\ref{sect5:entropyRen}) for $k = 1$ and $T = \frac{1}{2\pi}$. It starts at $0$ and asymptotes to the value given by (\ref{ch07:asympt}).
    \textbf{(b)} Following the minimum of  $S_{\text{ren},\mathcal{R}}(t)$ (\ref{sect4:Rad}) and $S_\text{BH}(t)$ (\ref{sect4:BeHa}) results in the Page curve, here drawn for $k=0.01$. Also note how the radiation asymptotes to twice the initial black hole value.}
\end{figure}

\indent We could now do a quick computation of the entropy solely attributed to this region, equal to plugging the trivial island $\mathcal{I} = \emptyset$ 
into the island formula (\ref{island}). With $\sigma \approx 0$ we have $F(\sigma) \approx 0,~F'(\sigma) \approx 1,~F''(\sigma) \approx 0$ such that the bare entropy (\ref{sect3:entropyBare}) becomes upon dismissing the cutoff terms:
\begin{equation}
    S_{\mathcal{R}}(t) = k \ln 4,
\end{equation}
which is completely time-independent. In the bulk of AdS$_2$, such an observation was made before in \cite{Callebaut2019,Callebaut:2018nlq}. The same calculation by making use of the renormalised entropy (\ref{sect3:entropyRen}) leads instead to the expression \cite{Mertens:2019bvy}:
\begin{equation}
\boxed{
    S_{\text{ren},\mathcal{R}}(t) = \frac{k}{2} \ln \frac{F(t)^2}{F'(t) t^2}}.
    \label{sect5:entropyRen}
\end{equation}
This entropy starts at $0$ and quickly reaches a fixed value, as is seen in Fig.~\ref{sect2:figRadiation}, determined by 
the limit value (\ref{sect3:horizon}) and $\lim_{t \to +\infty} F'(t) t^2$ resulting in\footnote{It was noted that $F'(t)$ indeed behaves as $t^{-2}$ for large values in \cite{Almheiri2019}, this leads to the limit 
\begin{equation*}
    \lim_{t \to +\infty } F'(t) t^2 = \left(\pi T I_1\left(\frac{2 \pi T}{k}\right)  \right)^{-2}.
\end{equation*}}
\begin{equation}
    \lim_{t \to +\infty }  S_{\text{ren},\mathcal{R}}(t) = k \ln I_0 \left(\frac{2\pi T}{k} \right).
    \label{ch07:asympt}
\end{equation}
\indent The plot \ref{sect2:figRadiation} shows $S$ as a function of $kt$. One can retrieve the two limit scenarios by only considering the ratio $k/T$:
\begin{itemize}
    \item \textbf{Static black hole limit:} this coincides with the ratio going to zero via $k \to 0$ or $T \to +\infty$. The first 
    one is directly equivalent to no evaporation. Since $k \propto c$, this can be seen as decreasing the amount of evaporation channels the black hole has access to, slowing down the evaporation.\footnote{We should be careful in this regard, we implicitly assumed a large central charge $c$ in order for the graviton contribution to be negligible.}
    The latter corresponds to shooting in a pulse with infinite energy (\ref{sect2:ADM}). It is `non-evaporating' in the sense that the energy is infinite, and hence decays sluggishly. 
    \item \textbf{Poincar\'e limit:} this corresponds to $k/T \to +\infty$. Either by instant evaporation $k\to +\infty$ or by having no pulse at all $T \to 0$. 
\end{itemize}

For a macroscopic black hole for which $k/T \ll 1$, we can make use of the approximations of the modified Bessel functions for large values to leading-order to get:
\begin{equation}
    S_{\text{ren},\mathcal{R}}(t) \approx 2 \pi T \left(1 - e^{-\frac{k}{2}t}  \right).
    \label{sect4:Rad}
\end{equation}
In this case, the Bekenstein--Hawking entropy can be quasi-statically approximated by the entropy of the static solution \eqref{sect2:staticEnergy} with the time-dependent energy \eqref{sect3:ECondition}, leading to
\begin{equation}
    S_\text{BH}(t) = \pi T e^{-\frac{k}{2}t},
    \label{sect4:BeHa}
\end{equation}
in our chosen units. By following the minimum of these two curves, shown in Fig.~\ref{sect2:figQuasistatic}, we obtain the Page curve with accompanying Page time $kt_\text{Page} = 2 \ln \frac{3}{2}$. However, without the island formula, there is no instruction telling us to take the minimum. Either way, this curve agrees qualitatively with what one would expect from unitary evolution. \\
\indent The radiation entropy goes to twice that of the initial black hole value $S_{\text{ren}, \mathcal{R}}(t \to +\infty)] = 2 S_\text{BH}(t = 0)$. This is not a coincidence, because when we look at their variation at any time, we unearth \cite{Mertens:2019bvy}\footnote{It was argued in the past \cite{Zurek1982, Fiola1994} that an evaporating black hole in empty space of $D$ spatial dimensions leads to a radiation entropy which 
is a factor $(D+1)/D$ larger than the initial value of the Bekenstein--Hawking entropy; it is an irreversible process. This factor originates from comparing the increasing entropy of a free Boson gas at $T_H$ with the decreasing entropy of a black hole 
\begin{equation*}
    \delta S = \frac{D+1}{D} \frac{E}{T_H} dt, \hspace{2cm} \delta S_\text{BH} = - \frac{E}{T_H} dt.
\end{equation*}
Note that in this case, this factor appears between the Bekenstein--Hawking entropy and a fine-grained entropy instead.}
\begin{equation}
    \delta S_{\text{ren},\mathcal{R}}(t) = - 2 \delta S_\text{BH}(t).
		\label{zurek}
\end{equation}

\section{Operational islands}
\label{sect:archipelago}

\begin{figure}[h!]
        \centering
        \includegraphics[width=0.35\textwidth]{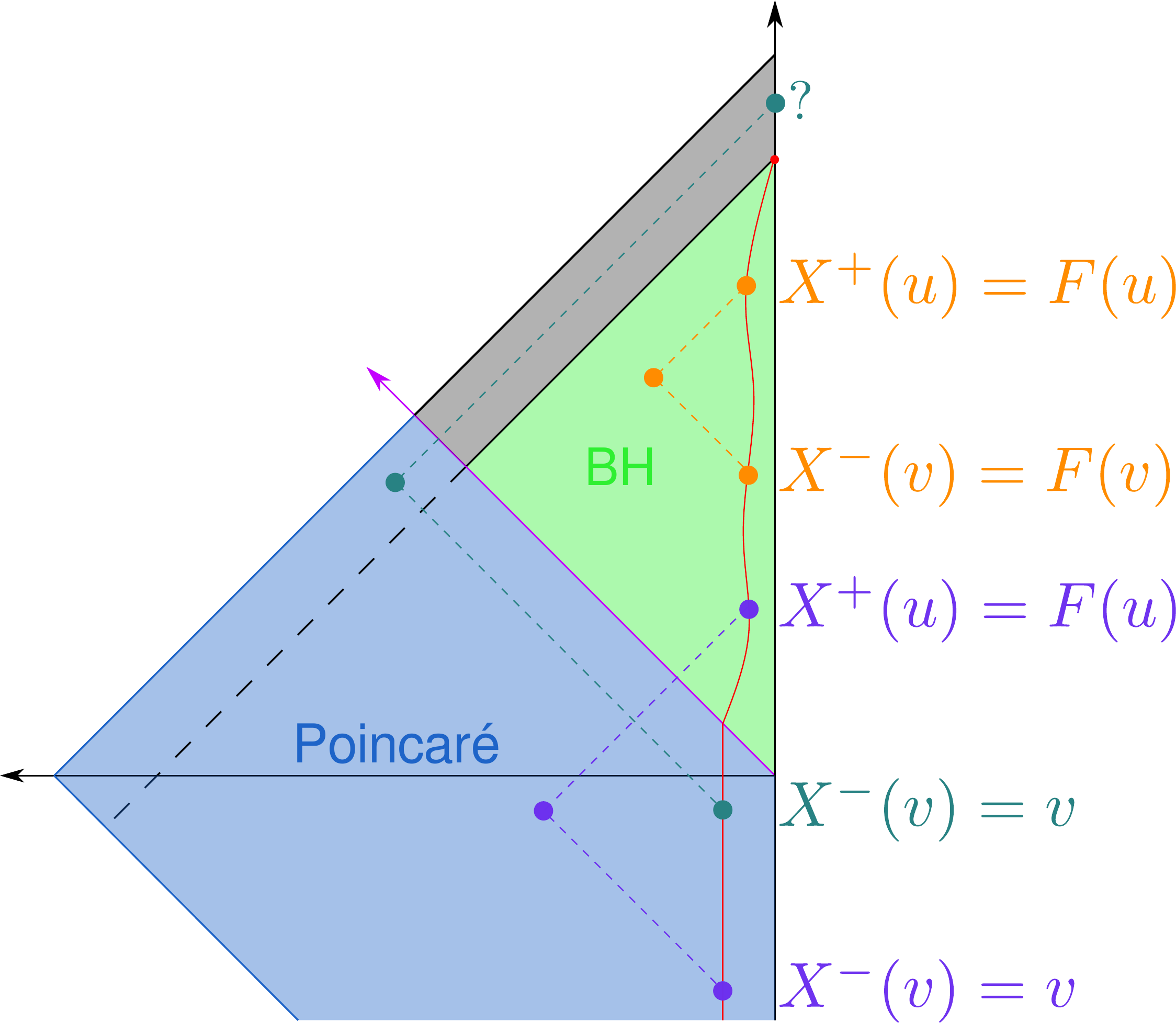}
        \caption{The radar definition of how the boundary observer would associate coordinates to a bulk point \cite{Blommaert2019}. The bulk point prior to the pulse but in the exterior corresponds to $(F(u), v)$ and for one after the pulse we get $(F(u), F(v))$. However, such a notion cannot be used for a bulk point in the interior -- lying behind $\xplus(u) = F_\infty$.}
        \label{sect5:figRadarDef}
\end{figure}

Finally, we use the QES formula to find out precisely how the above analysis is modified. \\
\indent Firstly, the expression for the dilaton can be written in terms of the time reparametrisation without use of the memory integrals found in \cite{Almheiri2015}.
One either directly integrates the general expression \cite{Goto2020} or one makes use of the EoM in the case of purely infalling matter $T_{+ +}=0$ \cite{Moitra2020} and combines this with 
the perfect absorption condition $\normord{T_{vv}}~= 0$ in our case \cite{Hollowood2020}. One ends up with the formula 
\begin{equation}
    \phi = 2 \phi_r \left[\frac{1}{2} \frac{F''(v)}{F'(v)} + \frac{F'(v)}{F(u) - F(v)} \right].
    \label{sect3:dilatonGeneral}
\end{equation}
\indent The most general expression for the entropy consisting of the dilaton contribution (\ref{sect3:dilatonGeneral}) associated 
to the island $(u, v)$, and that of the bulk quantum fields for an interval $(u_b, v_b) \to (u, v)$ (\ref{sect3:entropyRen}) is then given by:
\begin{align}
    S = \left[\frac{1}{2} \frac{F''(v)}{F'(v)} + \frac{F'(v)}{F(u) - F(v)} \right] 
    + \frac{k}{2}  \ln \frac{[F(u_b) - F(u)]^2}{F'(u_b) F'(u) (u_b - u)^2} + \frac{k}{2}  \ln \frac{[F(v_b) - F(v)]^2}{F'(v_b) F'(v)(v_b - v)^2} .
\end{align}
One end of the interval $(u_b, v_b)$ is anchored to the boundary, hence by making use of (\ref{sect2:boundary1}, \ref{sect2:boundary2}) we can set 
    \begin{align}
        u_b = v_b = t, \qquad   F(u_b) = F(v_b) = F(t), \qquad  F(u_b) - F(v_b) = 2 \eps F'(t),
    \end{align}
resulting in
\begin{align}
    S(t, u, v) = &\left[ \frac{1}{2} \frac{F''(v)}{F'(v)} + \frac{F'(v)}{F(u) - F(v)} \right] \nonumber \\
    &+ \frac{k}{2} \ln \frac{[F(t) - F(u)]^2}{F'(u) (t - u)^2} + \frac{k}{2} \ln \frac{[F(t) - F(v)]^2}{F'(v)(t - v)^2} - k \ln F'(t).
    \label{sect5:entropyGeneral}
\end{align}
\indent Next, We specify this equation to the different regions of the Penrose diagram, as in Fig. \ref{sect5:figRadarDef}. Essentially, we are looking at all points spacelike 
separated from the boundary for $t > 0$ and which lie in the exterior since the radar definition is ill-defined behind the horizon. Hence, we have two regions: a pre-pulse one 
with coordinates $(F(u), v)$ and a post-pulse one described by $(F(u), F(v))$.

\subsection{Pre-pulse island}
\label{sect:prePulseIsland}
We start with the region prior to the pulse corresponding to $\xminus \leq 0 \leq F(t) \leq \xplus$ equivalent to $v \leq 0 \leq t \leq u$ such that $F(v) = v$. The entropy (\ref{sect5:entropyGeneral}) becomes 
\begin{equation}
    S_{\text{pre}}(t, u, v) = \frac{1}{F(u) - v} + \frac{k}{2} \ln \frac{[F(t) - F(u)]^2}{F'(u) (t - u)^2} + \frac{k}{2} \ln \frac{[F(t) - v]^2}{(t - v)^2} - k \ln F'(t).
    \label{sect:entropyPre}
\end{equation}
\noindent To find islands within this frame, we extremise (\ref{sect:entropyPre}) with respect to both $u$ and $v$ independently
\begin{subequations}
    \begin{align}
        \label{sect5:preuCond}
        \partial_u S_\text{pre} = 0 &= \frac{-F'(u)}{(F(u) - v)^2} + k \left[\frac{-F'(u)}{F(t) - F(u)} + \frac{1}{t-u} - \frac{1}{2} \frac{F''(u)}{F'(u)}  \right], \\
        \label{sect5:prevCond}
        \partial_v S_\text{pre} = 0 &= \frac{1}{(F(u) - v)^2} + k \left[\frac{-1}{F(t) - v} + \frac{1}{t - v}  \right].
    \end{align}
\end{subequations}

\noindent By numerical inspection of these equations, it is easy to see that the island solution lies on the past horizon $\mathcal{I}^-$: $v_\mathcal{X} \to -\infty$, immediately giving an infinite scrambling time. It is easy to verify that this satisfies the $v$-condition (\ref{sect5:prevCond}) for finite $t$ and $u$.
The $u$-condition (\ref{sect5:preuCond}) consequently reduces to 
\begin{equation}
    0 = \frac{-F'(u)}{F(t) - F(u)} + \frac{1}{t-u} - \frac{1}{2} \frac{F''(u)}{F'(u)}, 
    \label{sect5:preIslandPastCond}
\end{equation}
and is effectively the purely matter condition $\partial_u S_\text{pre, bulk} = 0$ since the dilaton term vanishes. The solution is $u=t$. 

\begin{figure}[h!]
    \centering
    \begin{subfigure}[b]{0.49\textwidth}
        \centering
				        \includegraphics[width=\textwidth]{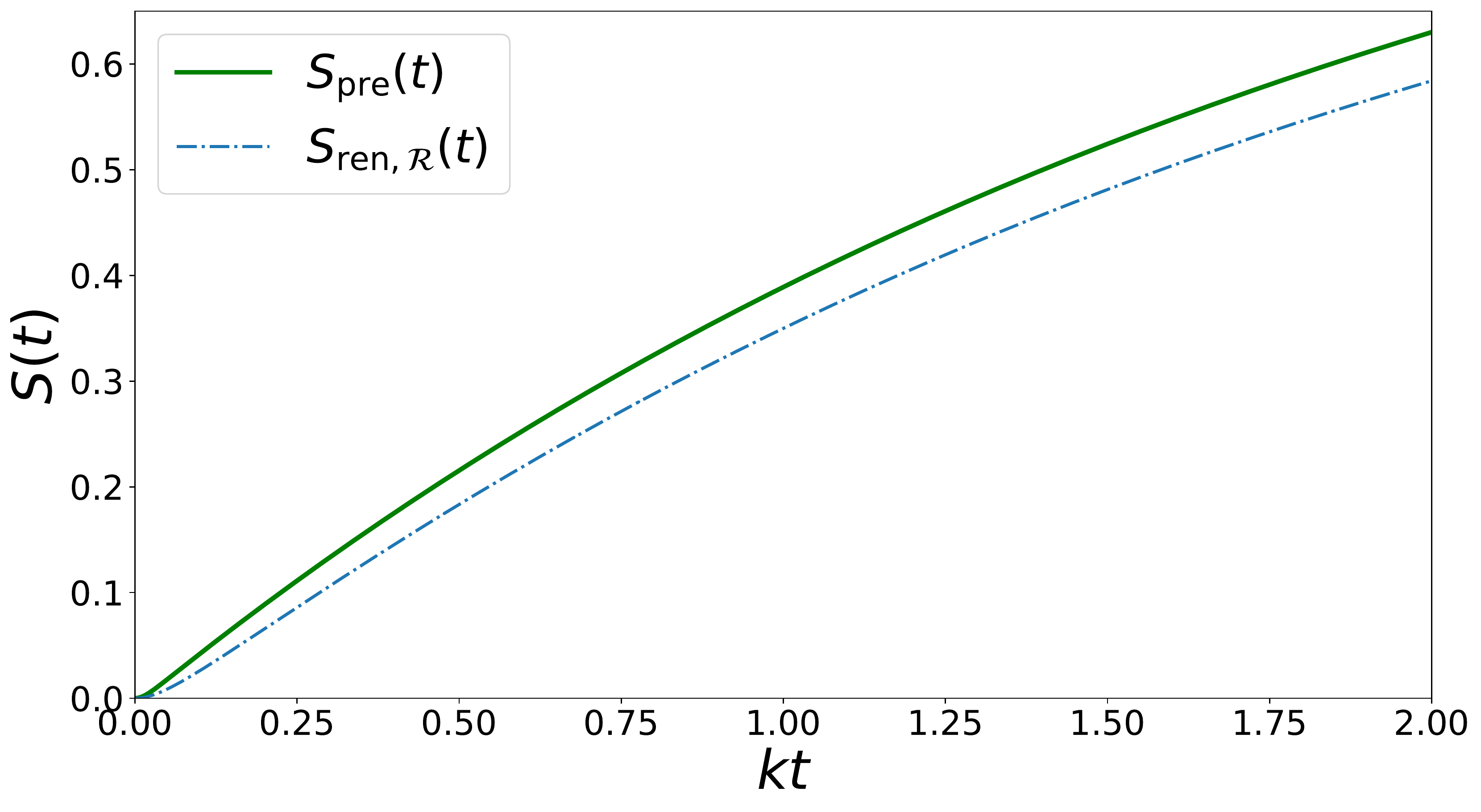}
    \caption{}
		\label{sect5:figPreIslandEntropy}
    \end{subfigure}
    \hfill
    \begin{subfigure}[b]{0.49\textwidth}
        \centering
        \includegraphics[width=\textwidth]{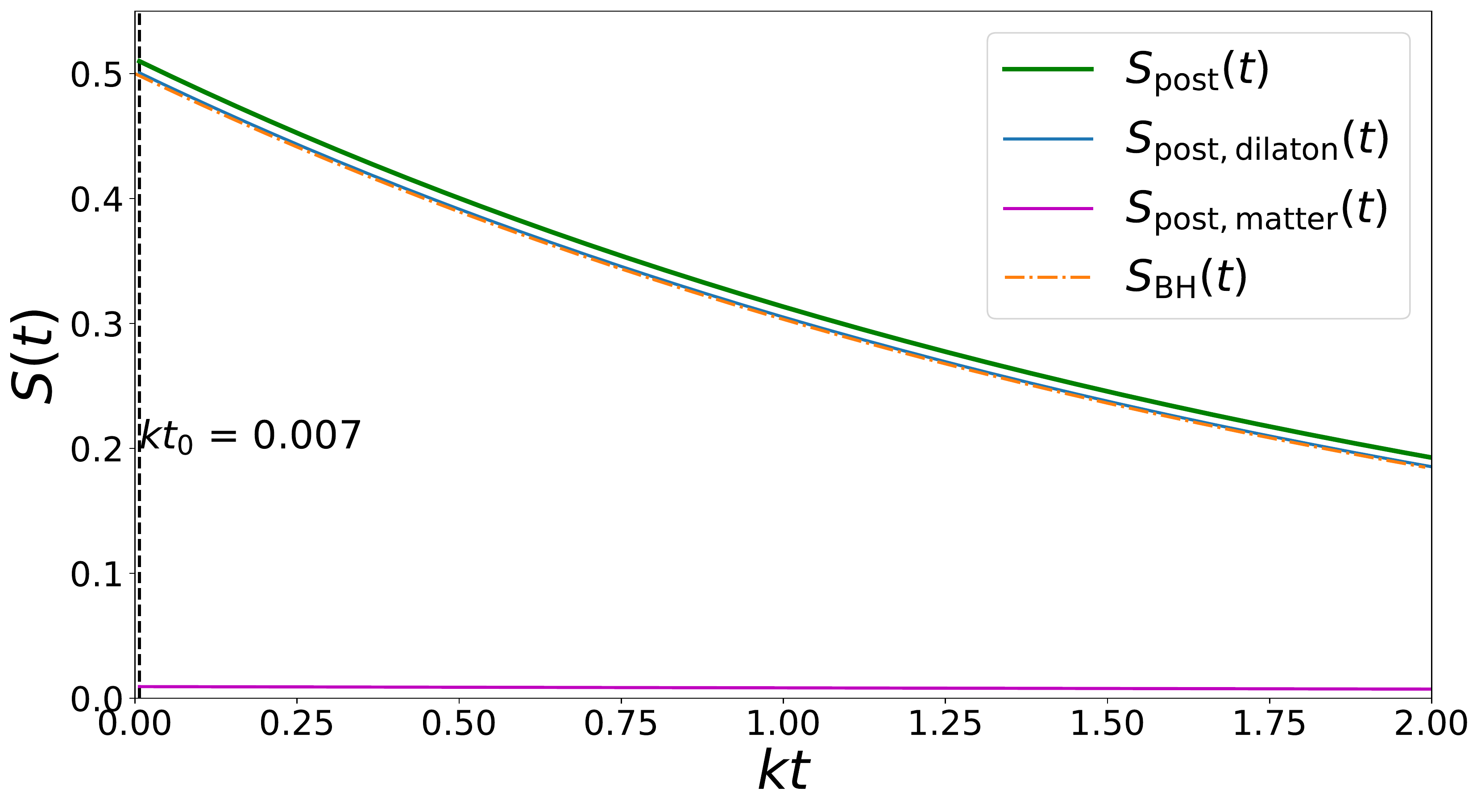}
        \caption{}
        \label{sect5:figEntropyPost}
    \end{subfigure}
    \caption{\textbf{(a)} Evolution of the entropy associated to the pre-pulse island for $k=0.01$ and $T = \frac{1}{2\pi}$. The entropy of the island on the past horizon looks similar to the radiation entropy (\ref{sect5:entropyRen}).
    \textbf{(b)} The post-island entropy for $k = 0.01,~T = \frac{1}{2\pi}$. We show how both contributions, dilaton and bulk, contribute to the generalised entropy. 
    The semiclassical Bekenstein--Hawking entropy is also plotted and almost exactly coincides with the dilaton. The initial time is depicted as the dashed, vertical line.}
\end{figure}

\indent This island exists for all $t$, starts with an entropy equal to $0$, and increases indefinitely. Eventually, it ends at $(F_\infty, -\infty)$. The entropy becomes:
\begin{equation}
    \boxed{S_\text{pre}(t) =  -\frac{k}{2} \ln F'(t)}.
    \label{sect5:entropyPast}
\end{equation}
The time evolution of this entropy is shown in Fig.~\ref{sect5:figPreIslandEntropy} for $k = 0.01$. As a comparison, the renormalised radiation entropy (\ref{sect5:entropyRen}) computed in the previous section is plotted as well.

\indent One can interpret this island as being trivial $\mathcal{I} = \emptyset$ since the classical piece vanishes, and we are only left with the entropy due to the matter fields present. 
A similar situation happens in the Schwarzschild case: the trivial island extends all the way down to $r = 0$ where it contributes no area piece since $A \propto r^2$ in such a geometry \cite{Almheiri2020c}.

For a macroscopic black hole $k/T \ll 1$, where one has the approximate form of $F(t)$ given in equation \eqref{sect3:reparamApprox}, we find for $kt$ of order 1:
\begin{align}
   S_\text{pre}(t)  = - \frac{k}{2} \ln F'(t) \, \approx \, 2 \pi T \left(1 - e^{-\frac{k}{2}t} \right)
    = S_{\text{ren},\mathcal{R}}(t),
    \label{sect5:approxPre}
\end{align}
which matches precisely onto the outgoing renormalised radiation entropy written in \eqref{sect4:Rad}.\footnote{Intriguingly, the formula \eqref{sect5:entropyPast} matches precisely with that in the CGHS/RST model \cite{Almheiri:2013wka}.}

\subsection{Aside: the island at $t<0$}
For consistency, we should find zero entropy when $t<0$ since the black hole has not even formed then. For $t<0$, there are two possible regions to find an island: the region where $u>0$ and the region where $u<0$. \\
\indent For the region $u>0$, it is clear from equation \eqref{sect:entropyPre} upon setting $F(t)=t$, that the value of the entropy decreases monotonically as $v_\mathcal{X} \to -\infty$ as before. On this past horizon, the value of the entropy functional also decreases when moving towards $u\to 0$. Hence no saddle exists in this region. \\
\indent For the region $u<0$ where additionally $F(u) = u$, the matter piece vanishes identically, and the equations \eqref{sect5:preuCond} and \eqref{sect5:prevCond}  are solved by  $v_\mathcal{X} \to -\infty$, without a specification on $u$. For this set of possible island endpoints, one has the entropy $S=0$, which is simultaneously also the global minimum of the entropy functional in these two regions ($u>0$ and $u<0$) of the Penrose diagram.

\subsection{Post-pulse island}
We finally arrive at the islands which are the hardest to discover, namely those after the pulse $0 \leq v \leq t \leq u$ with the most general entropy functional (\ref{sect5:entropyGeneral}). This leads to the island conditions 
\begin{subequations}
    \begin{align}
        \label{sect5:postuCond}
        \partial_u S_\text{post} = 0 =~&-\frac{F'(u) F'(v)}{[F(u) - F(v)]^2} + k \left[\frac{-F'(u)}{F(t) - F(u)} + \frac{1}{t-u} - \frac{1}{2} \frac{F''(u)}{F'(u)}  \right], \\
        \label{sect5:postvCond}
        \partial_v S_\text{post} = 0 =~&\frac{1}{2} \partial_v \frac{F''(v)}{F'(v)} + \frac{F''(v)}{F(u) - F(v)} + \left(\frac{F'(v)}{F(u) - F(v)} \right)^2 \nonumber \\ 
        &+ k \left[\frac{-F'(v)}{F(t) - F(v)} + \frac{1}{t - v} - \frac{1}{2} \frac{F''(v)}{F'(v)}  \right].
    \end{align}
\end{subequations}
\indent Because of the high nonlinearity, a nontrivial island needs to be found numerically. 
The result for the post-island entropy with $k=0.01$ is depicted in Fig.~\ref{sect5:figEntropyPost}. \\
\indent Notice that as indeed expected, the matter contribution to $S_\text{post}(t)$ is tiny compared to the dilaton contribution. Moreover, this decreasing part of the Page curve can be tracked quite well by the classical Bekenstein--Hawking entropy \eqref{sect4:BeHa}. \\

\noindent In a bit more detail, the resulting island is characterised by the following properties:
\begin{itemize}
\item The island is at the minimum of the entropy functional \eqref{sect5:entropyGeneral}.
\item The island ends its life on the boundary at $\xplus = F_\infty = \xminus$ where the observer ends as well, requiring $u_\mathcal{X} = v_\mathcal{X} \to + \infty$. Indeed, one can easily verify that every term vanishes.\footnote{In the limit of large times, the arguments of the modified Bessel functions go to zero and only those 
of the second kind will remain. Combining this with $F'(t) \to 0$ leads to 
\begin{equation*}
    \lim_{t \to +\infty} \frac{F''(t)}{F'(t)} = 0.
\end{equation*}
Similar manipulations are required for the other terms.} 
Furthermore, the entropy correctly vanishes in this limit. When equally dividing
the last term in expression (\ref{sect5:entropyGeneral}) among the two prior, what remains is the following limit 
\begin{align}
    \lim_{u \to t} \frac{[F(t) - F(u)]^2}{F'(t) F'(u) (t - u)^2} = \frac{F'(t)^2}{F'(t)^2} = 1.
\end{align}
At this point, the scrambling time vanishes as well since $t = v_\mathcal{X}$.
\item 
As $v_\mathcal{X}$ generally increases when $t$ does, the earliest possible island is one at the pulse itself $v_\mathcal{X} = 0$. By using Eqs.~(\ref{sect5:postuCond}, \ref{sect5:postvCond}), this is when\footnote{It is easy to 
show that by using the Schwarzian (\ref{sect3:EBHSch}) we have 
\begin{equation*}
    \partial_t  \frac{F''(t)}{F'(t)} = \frac{1}{2} \left(\frac{F''(t)}{F'(t)} \right)^2 + 2 (\pi T)^2 e^{-kt}.
\end{equation*}} 
\begin{subequations}
    \begin{align}
        -\frac{F'(u_0)}{F(u_0)^2} + k \left[\frac{-F'(u_0)}{F(t_0) - F(u_0)} + \frac{1}{t_0-u_0} - \frac{1}{2} \frac{F''(u_0)}{F'(u_0)}  \right]  &= 0,\\
        - (\pi T)^2 + \frac{1}{F(u_0)^2} + k \left[\frac{1}{t_0} - \frac{1}{F(t_0)} \right] &= 0. 
    \end{align}
\end{subequations}
The second condition can be directly solved to 
\begin{equation}
    k F(u_0) = \frac{k}{\pi T} \frac{1}{ \sqrt{1 + \left(\frac{k}{\pi T}\right)^2 \left[ \frac{1}{kF(t_0)} - \frac{1}{kt_0} \right] }}.
    \label{sect5:initialIsland}
\end{equation}
Because of $t - F(t) \geq 0$, it immediately follows that $\frac{1}{F(t)} - \frac{1}{t} \geq 0$. Irrespective of the ratio $k/T$ and $t_0$, the denominator in the expression above will be larger than unity. 
Hence the island always starts outside the would-be horizon. In the limiting case when $k/T \to 0$, the island starts on the would-be horizon.\\
\indent The initial time $kt_0$ when the island emerges on the pulse, as a function of $k/T$ can also be considered. Somewhat surprising is the appearance of a threshold value beyond which this initial time formally becomes negative. This threshold is $k/T \approx 0.547$ and the island then starts at the would-be horizon $k F(u_0) = \frac{k}{\pi T}$.
One would first conclude that for values beyond this threshold the island does not start on the pulse but rather somewhere else; the initial island conditions break down. However, numerics show that the only candidate islands lie within the lightcone of the boundary observer: $v < t$ is violated, and as such these are not suitable. However, when $k/T$ is sufficiently large, we are creating a rapidly evaporating small black hole. For such microscopic black holes, our calculation ought not to be trusted anymore in any case.

\item 
The apparent horizon is the locus where $\partial_v \phi =0$, or when the first line of \eqref{sect5:postvCond} vanishes. An intersection between the apparent horizon and the island endpoint cannot occur. Indeed, such an incidence would require the second line of \eqref{sect5:postvCond} to separately vanish as well. It is readily checked that this second line only vanishes when $v=t$. However, the numerical solution of the island endpoint $(u_\mathcal{X},v_\mathcal{X})$ is readily checked never to lead to a location where $v_\mathcal{X} = t$ unless at the very endpoint of evaporation where all quantities $\to +\infty$.
\item
For a macroscopic black hole where $k/T \ll 1$, we checked numerically that the resulting entropy limits precisely to the quasi-static Bekenstein--Hawking entropy \eqref{sect4:BeHa}:
\begin{equation}
S_\text{post}(t) \, \approx \, S_\text{BH}(t).
\label{sect5:postApprox}
\end{equation}
\end{itemize}

\subsection{The Page curve}
\label{sect:conclusion}

\begin{figure}[h!]
    \centering
    \begin{subfigure}[b]{0.35\textwidth}
        \centering
			   \includegraphics[width=0.85\linewidth]{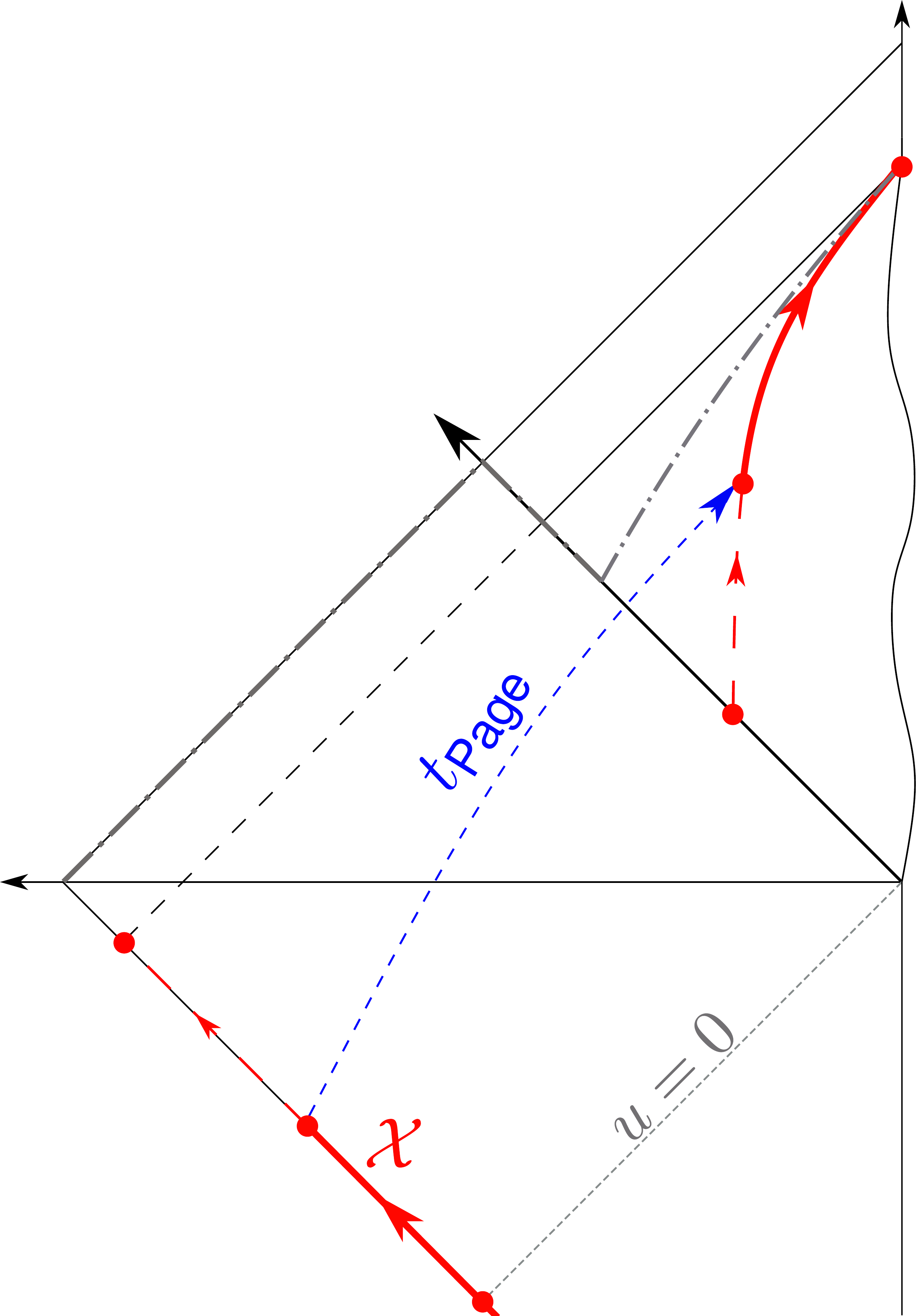}
               \caption{}
        \label{sect5:figislandTrajectory}
    \end{subfigure}
    \hfill
    \begin{subfigure}[b]{0.63\textwidth}
        \centering
				 \includegraphics[width=\textwidth]{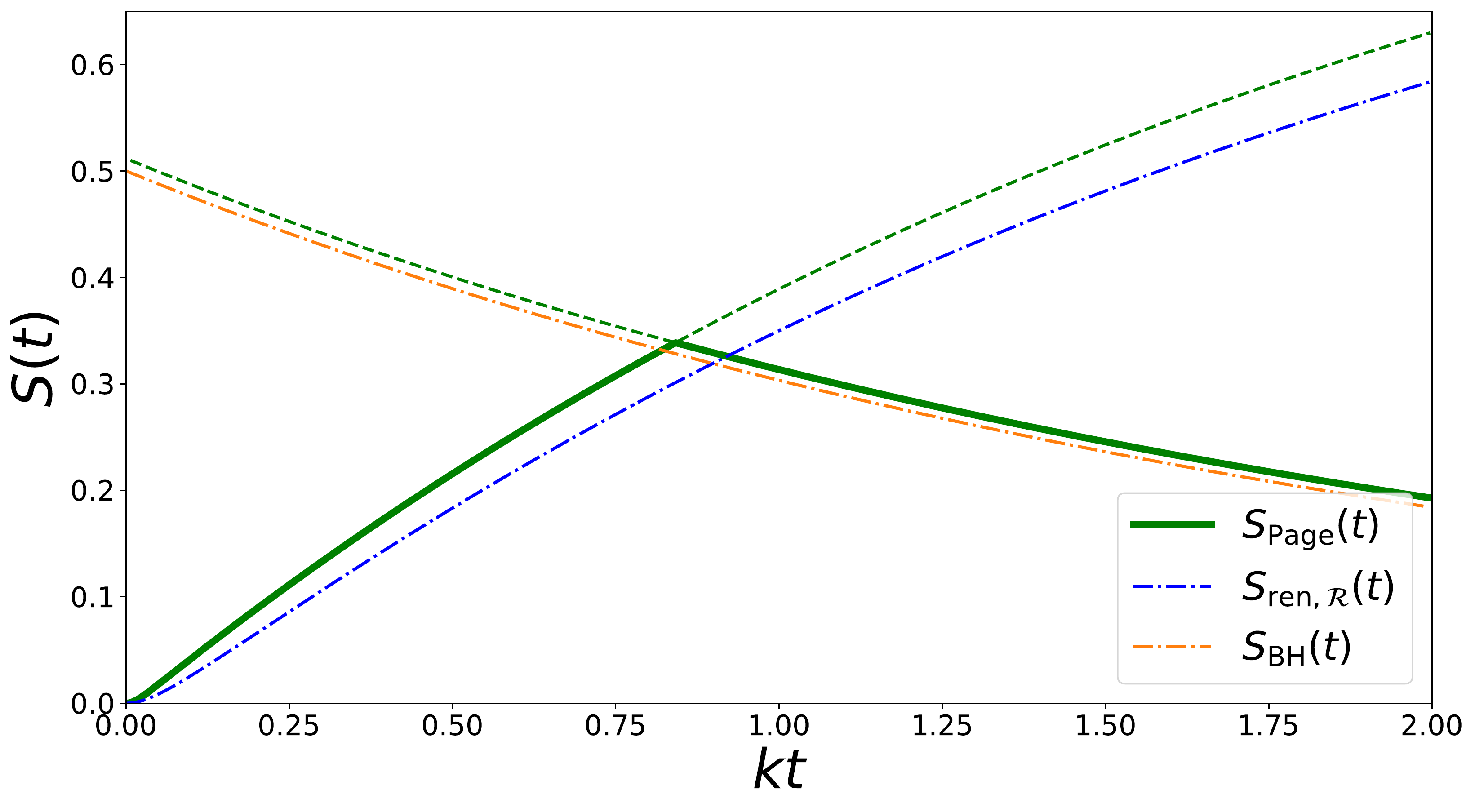}
        \caption{}
        \label{sect5:figPageCurve}
    \end{subfigure}
    \caption{\textbf{(a)} The island trajectories (in red). At the Page time, the dominance of the two islands changes. The initial island is on the past horizon. The final island is outside of the event horizon and the apparent horizon (gray dot-dashed line) for its entire lifetime. \textbf{(b)} The Page curve for $k = 0.01,~T = \frac{1}{2\pi}$ is given by the solid line in green. There is a small shift to a smaller Page time when compared with 
    the intersection of the fine-grained radiation entropy (\ref{sect5:entropyRen}) and the classical Bekenstein--Hawking entropy when deviating from the strict macroscopic $k/T \ll 1$ limit.}
\end{figure}

\noindent Combining the results of both islands, we obtain the island structure of Fig. \ref{sect5:figislandTrajectory} and the Page curve graph in Fig.~\ref{sect5:figPageCurve}.

As was established earlier, the semiclassical Bekenstein--Hawking entropy is a good fit for the entropy associated to the post-island which gives the decreasing contribution to the Page curve. 
The increasing part due to the pre-pulse island can be relatively well approximated by the radiation entropy. \\
\indent These observations become precise in the macroscopic limit where $k/T \ll 1$, and hence the Page curve for an evaporating macroscopic black hole in JT gravity is identical to that obtained by first following the Hawking radiation entropy (\ref{sect5:approxPre}), and subsequently following the semiclassical Bekenstein--Hawking entropy (\ref{sect5:postApprox}).

Because of this equality in the macroscopic limit, the downward piece of the Page curve decreases with half the slope as the upward piece, as found in \eqref{zurek}. 
This can also be found to be approximately correct from a numerical standpoint only. \\
\indent Moving away from the strict macroscopic limit, the Page time happens slightly earlier in the black hole life cycle.

\indent In general, the increasing part of the Page curve is associated to a trivial island with an entropy associated to it being purely due to the matter fields; e.g.~for Schwarzschild this is one all the way down to $r = 0$ \cite{Almheiri2020c}. The decreasing part is attributed to a nontrivial island behind the horizon. In our setup, we identified the trivial island with the island lying on the past horizon prior to the pulse where the dilaton vanishes. Moreover, the nontrivial island after the pulse inevitably lies in the exterior in our setup, due to how we defined the matter entropy in an operational way. \\

\noindent For comparison, like in many works on this topic, it is instructive to perform the same computations for the non-evaporating black hole. We present these results in Appendix \ref{app:eternal} where the resulting Page curve is also shown. Our pre- and post-pulse islands are located at similar locations in the Penrose diagrams.

\section{Charged black holes in JT gravity}
\label{sect:charged}
In this and the next section, we generalise the evaporating black hole model studied above to include additional conserved quantities. We focus on the electrically charged case first. Concretely, we will solve the JT versions of the Einstein--Maxwell system:
\begin{equation}
    G_{\mu\nu} = 8 \pi G \left\langle T_{\mu\nu}\right\rangle, \qquad \nabla^\mu F_{\mu\nu} = \left\langle J_\nu\right\rangle,
\end{equation}
where gravity and EM are treated classically, but they are sourced by charged quantum matter. 

Later on, we will embed this system within a larger supersymmetric system to illustrate that our physical results are also natural from a structural perspective.

\subsection{Setup}
The uncharged black hole system is governed by a Schwarzian boundary action, which is in turn found in the low-energy regime of the SYK model. In order to motivate the corresponding charged generalisation, we can take a look at the charged version of the SYK model instead.
The low-energy dynamics of the complex SYK model at finite temperature is governed by the following system; coupling the Euclidean thermal Schwarzian model, describing the purely gravitational sector, to a U(1) BF model describing the gauge field sector of the low-energy bulk dual \cite{Davison:2016ngz,Gu:2019jub}:
\begin{equation}
\label{lecsykprev}
S[f,\sigma] = -C \int_0^\beta d\tau \left\{\tan \frac{\pi}{\beta}f(\tau), \tau \right\} +\frac{K}{2} \int_0^\beta d\tau \left(\sigma'(\tau)- i \mu f'(\tau)\right)^2.
\end{equation}
The new field $\sigma$ parametrizes charge fluctuations, and is only defined up to a constant shift: $\sigma \sim \sigma + \text{(constant)}$, in direct analogy to the M\"obius group invariance of the Schwarzian reparameterisation $F(\tau)$.

A second way of obtaining the same model \eqref{lecsykprev} is by looking at the near-extremal near-horizon dynamics of the 4d Reissner--Nordstr\"om (RN) black hole. The resulting $s$-wave dynamics is governed by the JT model coupled to 2d Maxwell theory, which can in turn be described in this particular context by the 2d U(1) BF model. One quick way of appreciating this relation between the 2d Maxwell model (which is quasi-topological and depends solely on the area of surfaces) and the 2d U(1) BF model (which is topological up to length-dependent boundary contributions) is by realising that in AdS$_2$ areas and lengths scale in the same way (see e.g. Appendix C of \cite{Kapec:2019ecr} for an argument along these lines). Up to a rescaling of coupling coefficients, this then allows an immediate identification of amplitudes. The leading deviations from extremality are then described by the coupled bulk 2d JT + U(1) BF model:
\begin{equation}
\label{BFact}
S_{\text{bulk}} = \frac{1}{16\pi G} \int_\mathcal{M}\sqrt{-g} \Phi(R+2) + \frac{K}{2} \int_\mathcal{M} B F + \text{(boundary)},
\end{equation}
which reduces to pure boundary dynamics given by \eqref{lecsykprev} indeed upon choosing suitable boundary conditions.\footnote{To fully match with \eqref{lecsykprev} one has to introduce a suitable holonomy boundary condition on the gauge field to match with the twisting of $\sigma$ induced by having nonzero $\mu$. We make some comments below.}

This model \eqref{BFact} yields a grand canonical partition function (on the disk topology) given by
\begin{equation}
\label{gcpf}
\mathcal{Z}(\beta,\mu) = \sum_{Q \in e\mathbb{Z}}\int_{\frac{Q^2}{2K}}^{+\infty} dE e^{-\beta (E-\mu Q)} \sinh 2\pi \sqrt{2C} \sqrt{E-Q^2/2K}.
\end{equation}
The $Q$-integral is a sum due to charge quantisation. In the semiclassical large $Q$ regime, one can approximate it by an integral. The saddle equations hold in the regime $E \gg Q^2/2K$, where we can approximate sinh as exp, and are given by 
\begin{subequations}
\begin{alignat}{2}
\label{hawktemp}
\frac{\partial S}{\partial E} &= \beta ,\qquad &&\Rightarrow \quad \beta = \frac{\pi \sqrt{2C}}{\sqrt{E-Q^2/2K}}, \\
\frac{\partial S}{\partial Q} &= \mu\beta ,\qquad &&\Rightarrow \quad \mu = Q/K,
\end{alignat}
\end{subequations}
relating the temperature $\beta^{-1}$ and chemical potential $\mu$ to the mass $E$ and charge $Q$ of the black hole. This saddle approximation is rendered invalid for small values of $E-Q^2/2K$. This model hence describes the spectrum of charged black holes also in the regime where quantum gravitational corrections are important. \\
\indent The parameter $K$ can be matched to the original 4d RN black hole parameters as follows. A reversible change in the state of the 4d RN black hole follows the first law:
\begin{equation}
\label{revS}
\delta E = \frac{Q}{4\pi r_+}\delta Q,
\end{equation}
where $V(r) = \frac{Q}{4\pi r}$ is the Coulomb potential of the black hole, and $r_+$ is the radial location of the outer horizon. This corresponds to an electric field $E(r) =  \frac{Q}{4\pi r^2}$, which is approximated as constant in the near-horizon region where $r\approx r_+$. The dynamics in this near-horizon regime are described by the 2d Maxwell theory. Integrating \eqref{revS} leads to the finite relation:
\begin{equation}
E = \frac{Q^2}{8\pi r_+} + \text{constant},
\end{equation}
describing the energy of the black hole above extremality in terms of the charge added, while preserving its entropy. Note that this relation deviates from the actual 4d RN relation for sufficiently large choices of $E$ (and $Q$).
From this relation, by comparing with \eqref{gcpf} we identify $K \equiv 4\pi r_+$, see also \cite{Iliesiu:2020qvm} for a more thorough treatment of this identification. \\

\noindent We will make some simple adjustments to modify the action \eqref{lecsykprev} into a form that will be of use to our real-time dynamical evolution. Firstly, the explicit chemical potential $\mu$-term couples both sectors $f$ and $\sigma$ explicitly. It is however simple to decouple the sectors by redefining $\sigma \to \sigma - i \mu f(\tau)$. In case of the thermal partition function and correlators, this redefinition leads to a twisted periodicity of the new field $\sigma$ \cite{Chaturvedi:2018uov,Mertens:2019tcm}.\footnote{This procedure can in turn be interpreted in terms of a group-theoretic character insertion in the bulk dual BF TQFT.} Secondly, since we are interested in real-time evolution in a non-stationary (i.e. non-thermal) state, we use an arbitrary reparametrisation of the Poincar\'e frame $F(\tau)$ instead of the time reparametrisation of the black hole time coordinate related as $F(\tau) = \tan \frac{\pi}{\beta}f(\tau)$. Therefore, the Euclidean action that will be our starting point is:
\begin{equation}
\label{lecsyk}
S[F,\sigma] = -C \int d\tau \left\{F(\tau), \tau \right\} +\frac{K}{2} \int d\tau \sigma'(\tau)^2.
\end{equation}

\subsection{Charged matter fields}
Next we introduce charged matter fields in the bulk. For both simplicity and concreteness, let us consider a charged massless scalar field $\phi(\mathbf{x})$ of charge $+q$, with Lorentzian action:
\begin{equation}
S_m = -\int_\mathcal{M} d^2\mathbf{x}\, \sqrt{-g} D_\mu \phi (D^\mu \phi)^*,
\end{equation}
where $D_\mu = \partial_\mu - iq A_\mu$. In case of no background gauge field $A_\mu=0$, the field equation leads to decoupled holomorphic and antiholomorphic sectors as $\phi(u,v) = \phi(u) + \phi(v)$. A chiral component of this field transforms under chiral coordinate and gauge transformations as 
\begin{align}
\partial_U \phi(U) \, \to \, \frac{d U}{d u} \partial_u \phi(U(u)), \qquad 
D_u \phi(u) \, \to \, e^{iq \sigma(u)} D_u \phi(u).
\end{align}

After integrating out the Lagrange multipliers $\Phi$ and $B$ in \eqref{BFact}, one is reduced to a patch of AdS$_2$ with a flat connection $F=0$, or $A_\mu = \partial_\mu \chi$ is pure gauge.\footnote{We assume our manifold $\mathcal{M}$ has trivial topology.} Picking a particular bulk coordinate frame on AdS$_2$ and bulk gauge choice for $A_\mu$, is largely arbitrary. The only physically relevant information is the boundary values of these gauge transformations, since the large ``gauge'' transformations are not gauged at all. As before, we choose to work with bulk frames on AdS$_2$ that preserve holomorphicity by setting
\begin{equation}
X^+ = F(u), \qquad X^- = F(v),
\end{equation}
with the same function $F(.)$. Analogously, we restrict to pure gauge fields $A_\mu = \partial_\mu \chi$, with
\begin{equation}
\label{gaugechoice}
\chi(u,v) = \sigma(u),
\end{equation}
which reduces at the boundary $\partial \mathcal{M}$ $u=v=t$ to the physical $\sigma(t)$. 

Starting out with the reference case $A_\mu =0$, for which $\sigma = 0$,\footnote{Up to a constant shift that is meaningless for the $\sigma$-field as mentioned before.} we gauge transform to the specific gauge choice \eqref{gaugechoice} by setting  
\begin{align}
\label{matgaugefix}
D_u \phi_{\text{new}} =  e^{i q \sigma(u)} \partial_u \phi, \qquad
D_v \phi_{\text{new}} =  e^{i q \sigma(v)} \partial_v \phi,
\end{align}
and will insert these operators into correlation functions below. We will drop the ``new'' subscript from here on. Notice that for ingoing ($v$) or outgoing ($u$) modes a different gauge choice is made. For our purposes, this is natural since the boundary is chosen to be transparent, so there is no relation between them. If one would consider a reflecting boundary on the other hand, the modes would become naturally identified through this reflection. \\
\indent An alternative way of phrasing this issue is that to define a gauge-invariant observable, one needs to suitably dress the bulk operator with a Wilson line emanating from the boundary \cite{Donnelly:2015hta, Donnelly:2016rvo, Giddings:2018umg, Giddings:2019wmj,Harlow:2021dfp}. There is a natural choice on how to do this, choosing a line along the same null direction in which the matter is moving. 
The boundary gauge field is related to this as
\begin{equation}
\left.A_t\right|_{\partial \mathcal{M}} = \partial_t \sigma(t).
\end{equation}

\subsection{Matter energy density}

Now we can consider the (holomorphic) matter two-point function in the Poincar\'e vacuum state  $\left|0_+\right\rangle$:
\begin{equation}
\left\langle 0_+\right| D_{u_1}\phi(u_1) D_{u_2}\phi^*(u_2) \left|0_+\right\rangle.
\end{equation}
We can evaluate this correlator explicitly, by using the bulk gauge choice \eqref{matgaugefix} and the free boson correlator, as:
\begin{equation}
\label{twochir}
\left\langle 0_+\right| D_{u_1}\phi(u_1) D_{u_2}\phi^*(u_2) \left|0_+\right\rangle = -\frac{1}{4\pi} \frac{F'(u_1)F'(u_2)}{(F(u_1)-F(u_2))^2}e^{iq(\sigma(u_1)-\sigma(u_2))},
\end{equation}
in terms of the Poincar\'e frame $F(t)$, and the gauge transformation $\sigma(t)$. \\
\indent The chiral stress tensor components are given by:
\begin{equation}
\label{stresscompo}
T_{uu} = D_u \phi^* D_u \phi + D_u \phi D_u \phi^*, \qquad T_{vv} = D_v \phi^* D_v \phi + D_v \phi D_v \phi^*.
\end{equation}
As composite operators, these require regularisation and renormalisation. We implement the point-splitting procedure, where we subtract the vacuum contribution as:
\begin{subequations}
\begin{align}
\normord{T_{uu}(u_2)} &=  \lim_{u_1\to u_2} \left[D_u \phi(u_1) D_u \phi(u_2)^* + \frac{1}{4\pi} \frac{1}{(u_1-u_2)^2} + (cc)\right], \\
\normord{T_{vv}(v_2)} &= \lim_{v_1\to v_2} \left[ D_v\phi(v_1) D_v\phi(v_2)^* + \frac{1}{4\pi} \frac{1}{(v_1-v_2)^2} + (cc)\right].
\end{align}
\end{subequations}
Evaluated in the Poincar\'e vacuum $\left|0_+\right\rangle$, we series-expand \eqref{twochir} as $u_1 \to u_2$:
\begin{equation}
\frac{F'(u_1)F'(u_2)}{(F(u_1)-F(u_2))^2}e^{iq(\sigma(u_1) - \sigma(u_2))} = \frac{1}{u_{12}^2} + \frac{iq \sigma'(u_2)}{u_{12}} + \frac{1}{6}\left\{F,u_2\right\} + \frac{1}{2}q (-q\sigma'(u_2)^2 + i\sigma''(u_2)) + \hdots
\end{equation}
We then obtain
\begin{equation}
\label{bulkstresscharg}
\left\langle 0_+\right|  \normord{T_{uu}(u)} \left|0_+\right\rangle = -\frac{1}{12\pi}\left\{F,u\right\} + \frac{1}{4\pi}q^2 \sigma'^2(u),
\end{equation}
which has a positive contribution both from the gravitational piece, and from the gauge sector. Note that this is the left-moving energy momentum flux of both particles and antiparticles. 

\indent If we use the thermal saddle configurations, where $F(t) = \tanh \frac{\pi}{\beta} t$ and $\sigma'(t) = \mu$, we are describing the semiclassical 2d charged black hole with temperature $\beta^{-1}$ and chemical potential $\mu$, where we obtain
\begin{equation}
\label{ench}
E_{\text{bath}}(u,v) \equiv ~ \left\langle \normord{T_{uu}(u)} \right\rangle  + \left\langle \normord{T_{vv}(v)} \right\rangle ~=  \frac{\pi}{3\beta^2} + \frac{1}{2\pi} q^2 \mu^2 .
\end{equation}
This is the energy density contained in the Unruh heat bath of a charged black hole. This expression matches with that in older work by Iso, Umetsu and Wilczek \cite{Iso:2006wa}, where different techniques were used to obtain the same expression \eqref{ench} describing the Unruh heat bath energy flows of the spherically symmetric sector of 3+1d RN black holes. \\

\noindent Between different chiral frames $(U(u), V(v))$, including possibly different background gauge choices \eqref{matgaugefix}, these renormalised stress tensor components transform anomalously as
\begin{subequations}
\begin{align}
\label{ano1}
    \normord{T_{uu}^\sigma} &~= \left(\frac{d U}{d u}  \right)^2 \normord{T_{UU}^\chi} - \frac{c}{24 \pi}  \{U, u\} + \frac{q^2}{4\pi} (\sigma'^2(u) - \chi'^2(u)), \\
    \label{ano2}
    \normord{T_{vv}^\sigma} &~= \left(\frac{d V}{dv}  \right)^2 \normord{T_{VV}^\chi} - \frac{c}{24 \pi} \{V, v\} + \frac{q^2}{4\pi} (\sigma'^2(v) - \chi'^2(v)),
\end{align}
\end{subequations}
generalising \eqref{sect2:stress}. For clarity, we denoted the background gauge with the superscript. Picking one of the frames to be the Poincar\'e patch, by setting $U=X^+$, $V=X^-$, and $\chi=0$, we get:
\begin{subequations}
\begin{align}
    \normord{T_{uu}^\sigma} &~= \left(\frac{d X^+}{d u}  \right)^2 \normord{T_{++}} - \frac{c}{24 \pi}  \{X^+, u\} + \frac{q^2}{4\pi} \sigma'^2(u), \\ \normord{T_{vv}^\sigma} &~= \left(\frac{d X^-}{dv}  \right)^2 \normord{T_{--}} - \frac{c}{24 \pi} \{X^-, v\} + \frac{q^2}{4\pi} \sigma'^2(v).
\end{align}
\end{subequations}
Computing the VeV, we then immediately reproduce \eqref{bulkstresscharg}.

\indent The formulas \eqref{sect2:cov} now need to be adjusted to compensate for the contribution from background gauge transformations as:
\begin{equation}
\label{stresstensor}
    T_{uu}(u) = \frac{c}{24 \pi} \{ F(u), u \} ~- \frac{q^2}{4\pi} \sigma'^2(u) ~+ \normord{T_{uu}^\sigma(u)}, \hspace{1cm} T_{vv}(v) = \frac{c}{24 \pi} \{ F(v), v \} ~- \frac{q^2}{4\pi} \sigma'^2(v) ~+ \normord{T_{vv}^\sigma(v)}.
\end{equation}
The quantities on the left hand side transform as rank 2 symmetric tensor fields under coordinate transformations, and are invariant under gauge transformations. They serve as suitable insertions as sources in Einstein's equations. A different way of motivating such an expression is by embedding it into the superspace version of \eqref{sect2:cov}. We will come back to the supersymmetric embedding later on in Section \ref{s:susy}. 
Note that evaluating \eqref{stresstensor} in the Poincar\'e vacuum, using \eqref{bulkstresscharg} for $c=2$, we get precisely zero.

As before, the inhomogeneous terms cancel out for the net influx at the boundary $u=v=t$: $\left\langle T_{vv}(t) \right\rangle - \left\langle T_{uu}(t) \right\rangle =\,  \left\langle \normord{T_{vv}(t)} \right\rangle- \left\langle \normord{T_{uu}(t)} \right\rangle$ as sourcing equation \eqref{sect2:cov}. \\

\noindent For later reference, we also summarise the spectral occupation of the charged Unruh heat bath in 2d, see \cite{Gibbons:1975kk} for the original 4d calculations. The total occupation number of mode $\omega$ with charge $+q$ can be obtained by essentially Fourier transforming the two-point function as:
\begin{align}
\label{occ}
N_{\omega,q} = -\frac{1}{4\pi^2\omega}\int_{-\infty}^{+\infty} du_1 \int_{-\infty}^{+\infty} du_2 e^{-i \omega (u_1-u_2)} &\left[\frac{F'(u_1)F'(u_2)}{(F_1-F_2)^2}e^{iq(\sigma_1-\sigma_2)} - \left(\frac{1}{u_{12}}\right)^2\right].
\end{align}
Setting $F(t)= \tanh \frac{\pi}{\beta}t$ and $\sigma'(t) = \mu$ once more to pick the semiclassical charged black hole, the integrals can be performed and result in the spectral occupation:\footnote{Up to a volume prefactor arising from the $u_1+u_2$ integral of \eqref{occ}, that we omit.}
\begin{equation}
\label{bathfinite}
N_{\omega,q} \,=\, \frac{\omega-q\mu}{\omega} \frac{e^{-\frac{\beta}{2}(\omega-q\mu)}}{e^{\frac{\beta}{2}(\omega-q\mu)}-e^{-\frac{\beta}{2}(\omega-q\mu)}},
\end{equation}
where particles with charge $+q$ are preferably emitted for a black hole with chemical potential $+\mu$. \\
\indent The total energy in the bath can then be determined by computing:
\begin{equation}
\label{Ebath}
E_{\text{bath}} = \int_0^{+\infty}d\omega \omega \left(N_{\omega,q} + N_{\omega,-q} \right),
\end{equation}
and matches the above expression \eqref{ench} when integrating that one over space. \\
\indent We want to draw the attention to the prefactor in the expression \eqref{bathfinite}, which is the absorption coefficient of the superradiant modes, where we denote these low-frequency modes by the same name as for rotating black holes. Note that it flips sign when $\omega < q\mu$ to ensure an overall positive occupation number. At zero temperature, this expression reduces to:
\begin{equation}
\label{bathzero}
N_{\omega,q}^{\beta \to \infty} \,=\, \frac{q\mu-\omega}{\omega} \Theta(q\mu - \omega),
\end{equation}
which is the occupation of particle-antiparticle pairs due to a constant electric field in vacuum -- the Schwinger effect. Only superradiant modes with $\omega < q\mu$ are still occupied.

\subsection{Matter current density}
The matter sector also has a conserved local current density. The (normalised) U(1) conserved current density is given by the expression:
\begin{equation}
J_\mu = \frac{q}{i} \left(\phi D_\mu \phi^* - \phi^* D_\mu \phi\right). 
\end{equation}
Within quantum expressions, one again needs to point-split and renormalise the expression as:
\begin{subequations}
\begin{align}
\normord{J_u(u_2)} &= \lim_{u_1\to u_2} \frac{q}{i} \left[ \phi(u_1) D_u \phi^*(u_2) - \frac{1}{4\pi} \frac{1}{(u_1-u_2)} - (cc)\right], \\
\normord{J_v(v_2)} &= \lim_{v_1\to v_2}\frac{q}{i} \left[\phi(v_1) D_v \phi^*(v_2) - \frac{1}{4\pi} \frac{1}{(v_1-v_2)} - (cc)\right]. 
\end{align}
\end{subequations}
Using the bulk correlation function:
\begin{equation}
\label{twochir2}
\left\langle 0_+\right| \phi(u_1) D_{u_2}\phi^*(u_2) \left|0_+\right\rangle = \frac{1}{4\pi} \frac{F'(u_2)}{F(u_1)-F(u_2)}e^{iq(\sigma(u_1)-\sigma(u_2))},
\end{equation}
and the series expansion
\begin{equation}
 \frac{F'(u_2)}{(F(u_1)-F(u_2))}e^{iq(\sigma_1-\sigma_2)} = \frac{1}{u_{12}} +iq \sigma'(u_2) + \hdots,
\end{equation}
we readily find its expectation value in the Poincar\'e vacuum state as:
\begin{equation}
\label{currentnorm}
\left\langle 0_+\right|  \normord{J_{u}(u)} \left|0_+\right\rangle = \frac{1}{2\pi}q^2 \sigma'(u).
\end{equation}
\indent E.g. for the semiclassical charged black hole solution where $\sigma'(t) = \mu$, we obtain the total charge density:
\begin{equation}
\label{Qdens}
J_0(u,v) \equiv \left\langle  \normord{J_{u}(u)}  \right\rangle + \left\langle  \normord{J_{v}(v)}  \right\rangle = \frac{1}{\pi}q^2 \mu.
\end{equation}
This again matches with the expressions in \cite{Iso:2006wa} found in a different way for the RN black hole.

There is a second way to appreciate this result. Just as the total energy density \eqref{Ebath} is obtained by adding particle and antiparticle contributions to the spectral occupation \eqref{occ}, the charge density can be found by subtracting the relevant contributions from particles and antiparticles. One can then compute the total charge in the bath by integrating over $\omega$ and match this with the volume integral of the charge density \eqref{Qdens}. \\

\noindent The above normal-ordered current densities are operationally defined, but are not suitable to insert as sources into equations of motion since they are not gauge-invariant: the quantity \eqref{twochir2} is \emph{not} gauge-invariant. The reason for this apparent discrepancy is that the $\phi \phi^*$ correlator is singular, and this singularity can be compensated by a zero in the numerator. Indeed, the above calculation is easily used to prove the inhomogeneous transformation:
\begin{subequations}
\begin{align}
\label{tfnocharge}
    \normord{J_{u}^\sigma(u)} &= \left(\frac{d U}{d u}\right) \normord{J_{U}^{\chi}(U)} + \frac{1}{2\pi}q^2 (\sigma'(u)-\chi'(u)), \\
    \label{tf2nocharge}
    \normord{J_{v}^\sigma(v)} &= \left(\frac{d V}{d v}\right)\normord{J_{V}^{\chi}(V)} + \frac{1}{2\pi}q^2 (\sigma'(v)-\chi'(v)),
\end{align}
\end{subequations}
where in the notation we have explicitly kept track of both the coordinate frame in the subscript \emph{and} the background gauge choice in the superscript. 

In analogy with gravity, the resolution is to define a gauge-invariant current density by adding an inhomogeneous term that precisely compensates this anomalous transformation, as
\begin{align}
\label{gaugeinvJ}
    J_u(u) \equiv -\frac{1}{2\pi}q^2 \sigma'(u) + \normord{J_{u}^\sigma(u)}, \qquad
    J_v(v) \equiv -\frac{1}{2\pi}q^2 \sigma'(v) + \normord{J_{v}^\sigma(v)},
\end{align}
such that a change of background gauge $\sigma \to \chi$, where we use (\ref{tfnocharge},\ref{tf2nocharge}), now leaves $J_u$ and $J_v$ invariant, in the sense that e.g.:
\begin{equation}
    J_u(u) = -\frac{1}{2\pi}q^2 \sigma'(u) + \normord{J_{u}^\sigma(u)}~= -\frac{1}{2\pi}q^2 \chi'(u) + \normord{J_{u}^\chi(u)}.
\end{equation}

Note that all terms transform as rank 1 tensors under coordinate transformations, since this transformation is not anomalous for the current densities $\normord{J_{u}(u)}$ and $\normord{J_{v}(v)}$ of (\ref{tfnocharge}, \ref{tf2nocharge}).

Vacua that are unitarily equivalent in the charged case are found by restricting to transformations for which the inhomogeneous terms in both (\ref{tfnocharge}, \ref{tf2nocharge}) and (\ref{ano1}, \ref{ano2}) vanish. This requires for both chiral sectors $U= \frac{au+b}{cu+d}$ as before, and $\sigma = \chi + \text{(constant)}$. This is the PSL$(2,\mathbb{R}) \times {}$U(1) global piece of the underlying gauge groups, generalising the purely M\"obius group of the uncharged case. This is also the bosonic subalgebra of the OSp$(2|2)$ superalgebra, when we embed the charged system into the $\mathcal{N}=2$ supersymmetric system, as we will do further on.

\subsection{Charge equation of motion}

Here we will derive the correct equation of motion for the field $\sigma$, when it is sourced by quantum matter in the bulk. We start with the Lorentzian gravitational + gauge field action:
\begin{equation}
\label{lecsyk2}
S_{\text{L}}[F,\sigma] = -C \int dt \left\{F(t), t \right\} + \frac{K}{2} \int dt \sigma'(t)^2.
\end{equation}
Varying w.r.t. $\sigma$ leads to 
\begin{equation}
\label{Sbulk}
\frac{\delta S_{\text{L}}[F,\sigma]}{\delta \sigma(t)} = -K\sigma'',
\end{equation}
where $K\sigma'$ is to be interpreted as the total charge in the system. The equation of motion is hence just charge conservation.

Now we couple a bulk matter action to this boundary action $S[F,\sigma]$, minimally coupled to the gauge sector as:
\begin{equation}
S_m = \int_\mathcal{M} d^2\mathbf{x}\, \mathcal{L}_m(\phi, D_\mu \phi), \qquad D_\mu = \partial_\mu - i g \partial_\mu \sigma.
\end{equation}
In order to complete the description, we have to choose a particular gauge for the field $\sigma(t,z)$ throughout the bulk as it is a priori only defined on the boundary curve $z=0$. In this section only, we choose to extend the gauge transformation $\sigma(t)$ into the bulk in the simplest way possible by setting:\footnote{This argument is in analogy with a similar derivation for the sourced Schwarzian (gravitational) equation of motion in Appendix B of \cite{Mertens:2018fds}.}
\begin{equation}
\sigma(t,z) \equiv \sigma(t).
\end{equation}
The equation of motion for $\sigma(t)$ is invariant under small gauge transformations, i.e. those that do not influence the boundary, and the answer hence does not depend on this particular choice of bulk gauge, as long as we limit to $\sigma(t)$ as we approach the boundary  $z\to0$. Then we directly find
\begin{equation}
\frac{\delta S_m}{\delta \sigma(t)} = \int dz \int dt' \frac{\partial \mathcal{L}_m}{\partial A_t}\partial_{t'} \delta(t'-t) = -\frac{d}{dt}\int dz \frac{\partial \mathcal{L}_m}{\partial A_t} = \frac{d Q}{dt},
\end{equation}
where $Q(t) = \int dz \sqrt{-g} J^t(t,z)$, and we have defined the currents as 
\begin{equation}
J^\mu = -\frac{1}{\sqrt{-g}} \frac{\delta S_m}{\delta A_\mu}.
\end{equation} 
Using that the matter sector describes a conserved current as: 
\begin{equation}
\frac{dQ}{dt} = - \left.\sqrt{-g} J^z\right|_{\partial \mathcal{M}} = - \sqrt{-g} g^{zz} \left.J_z \right|_{\partial \mathcal{M}} = \left.J_v - J_u \right|_{\partial \mathcal{M}},
\end{equation}
and combining this with \eqref{Sbulk}, we finally obtain the sourced equation of motion describing how the total charge in the system evolves in time, as a response to charge injections or extractions from the holographic boundary:
\begin{equation}
\label{eomcharge}
K\sigma'' = \left.J_v - J_u \right|_{\partial \mathcal{M}},
\end{equation}
where the right hand side is the net charge influx from the holographic boundary ${\partial \mathcal{M}}$ at $z=0$.

\indent In our specific case of a charged massless scalar, we hence need to insert the boundary values of:
\begin{align}
J_u = \frac{q}{i} \left(\phi D_v \phi^* - \phi^* D_v \phi\right), \qquad 
J_v = \frac{q}{i} \left(\phi D_u \phi^* - \phi^* D_u \phi\right),
\end{align}
but the derivation of \eqref{eomcharge} was done more generally.

\section{Evaporating charged black holes}
\label{sect:chargedEvap}
In this section, we solve the previously derived equations of motion in the situation of absorbing boundary conditions after an initial injection of energy $E_0$ and charge $Q_0$. We will track how both energy and charge dissipate from the total system.

\subsection{Charged absorption}
\label{s:chabs}
Sourcing the charge equation of motion by the quantum expectation values of the matter current density fluxes, we write:
\begin{equation}
K \sigma''(t) = \left\langle J_v(t)\right\rangle - \left\langle J_u(t)\right\rangle = \left\langle \normord{J_v(t)}\right\rangle - \left\langle\normord{J_u(t)}\right\rangle,
\end{equation}
where we have used that at the boundary, the inhomogeneous terms in the expressions \eqref{gaugeinvJ} $J_u$ and $J_v$ are equal and cancel out, just as in \eqref{sect2:flux}.

If our boundary at $z=0$ is reflecting, then the r.h.s. of this evolution equation is zero, and the total charge in the system does not change in time. \\
\indent Imposing instead absorbing boundary conditions, we require $\left\langle \normord{J_v} \right\rangle ~= 0$, interpretable as the boundary observer removing all charge they detect on their local detector. Plugging in \eqref{currentnorm}, this gives us the equation of motion:
\begin{equation}
\label{cheom}
K \sigma'' = -\frac{1}{2\pi} q^2 \sigma' + Q_0\delta(t),
\end{equation}
where we injected a pulse with charge $Q_0$ at time $t=0$. This leads to exponential dissipation of charge from the system after the initial charge injection:
\begin{equation}
\label{disch}
\boxed{ Q(t) = K \sigma' = Q_0 e^{-\frac{q^2}{2\pi K} t}}.
\end{equation}
\indent Note that the charge depletes because the created black hole preferably emits particles that have the same sign of charge as the injection itself. This is in agreement with intuition from the static Unruh heat bath \eqref{bathfinite}.

\noindent The total energy in the system \eqref{lecsyk2} has two contributions:
\begin{equation}
    E(t) = - C \left\{F,t\right\} + K\frac{1}{2}\sigma'^2,
\end{equation}
which can be identified as a (positive) contribution coming from the black hole itself, and a (positive) contribution coming from the energy stored in its exterior electric field.\footnote{The BF model has zero electric field; here we mean the electric field of the higher-dimensional black hole, captured by the boundary charge profile $K\sigma'(t)$ in the BF description.} Next we plug in the matter source expression \eqref{bulkstresscharg} we derived in a generic frame $(F,\sigma)$ to obtain the energy equation of motion:
\begin{equation}
\label{eneom}
\frac{d}{dt} \left( - C \left\{F,t\right\} + K\frac{1}{2}\sigma'^2\right) = \frac{1}{12\pi}\left\{F,t\right\} - \frac{1}{4\pi }q^2 \sigma'^2 + E_0 \delta(t).
\end{equation}
Multiplying \eqref{cheom} by $\sigma'$ and subtracting from \eqref{eneom}, one obtains an equation describing purely the time evolution of the black hole contribution $E_{\text{BH}}(t) \equiv -C \left\{F,t\right\}$ to the energy:
\begin{equation}
\label{frameeom}
\frac{d}{dt} E_{\text{BH}}(t) = \frac{1}{12\pi}\left\{F,t\right\} + \frac{1}{4\pi }q^2 \sigma'^2 + (E_0 - Q_0 \sigma'(0)) \delta(t),
\end{equation}
where we interpret $\sigma'(0) =  \frac{Q_0}{2K}$. From \eqref{eneom} and \eqref{frameeom}, one immediately finds the jump conditions at $t=0$:
\begin{equation}
\label{jump}
\Delta E(t=0) = E_0, \qquad \Delta E_{\text{BH}}(t=0) = E_0 -  \frac{Q_0^2}{2K}.
\end{equation}
Plugging in the charge profile \eqref{disch}, the resulting black hole energy is described by a sum of two decaying exponentials:
\begin{equation}
\left\{F,t\right\} = a e^{-\frac{t}{12\pi C}} + b e^{-\frac{q^2}{\pi K}t},
\end{equation}
where\footnote{If $12q^2C=K$, both exponentials decay at the same rate, and one has the second solution $\sim t e^{-\#t}$. Also, the second constant can be written as $a = -2(\pi T_0)^2 - b$. In the uncharged case $b \to 0$, and the differential equation gets cast into 
a familiar form.}
\begin{equation}
b = \frac{3q^2Q_0^2}{K(12q^2C-K)}, \qquad a =-\frac{1}{C}\left(E_0 - \frac{Q_0^2}{2K}\right) - b
\label{enCoeff}
\end{equation}
This leads to the total energy in the system:
\begin{equation}
\label{disen}
\boxed{
E(t) = \left(\frac{Q_0^2}{2K}- bC\right) e^{-\frac{q^2}{\pi K}t} - aC e^{-\frac{t}{12\pi C}}}.
\end{equation}

For the particular case where $E_0 = \frac{Q_0^2}{2K}$, the black hole energy $E_{\text{BH}}(t)$ does not exhibit a discontinuity at $t=0$ by \eqref{jump}, but it does become nonzero during the evaporation process. We will provide the physical interpretation further on.

There is a particular choice of parameters for which the first exponential function isn't there. This happens e.g. when this system is the bosonic subsector of an $\mathcal{N}=2$ super JT gravity model as we discuss next.

\subsection{Embedding in $\mathcal{N}=2$ supersymmetric system}
\label{s:susy}
In order to check that our analysis of the charged dissipative system was correct, we here show that it can be  naturally embedded within the $\mathcal{N}=2$ supersymmetric JT gravity model if one makes a particular choice of the coupling coefficients of the charged system. \\

\noindent The observation we make is the following. The bosonic dissipative system \eqref{sect3:ODE} has a particularly nice structure. This system then has an immediate generalisation to higher supersymmetry, by writing the analogous equation in superspace:
\begin{equation}
\label{superdissip}
\frac{d}{dt} \text{S}(t,\theta,\bar{\theta}) = - \frac{c}{24\pi C} \text{S}(t,\theta,\bar{\theta}), \qquad t>0,
\end{equation}
motivated by having a net outgoing flux that is nonzero in the evaporating superframe.

\indent It is instructive to work this out in components. For $\mathcal{N}=2$ superspace, the anomalous transformation law of the holomorphic stress tensor in superspace is given by \cite{Cohn:1986wn}:
\begin{equation}
T(\mathbf{z}) = T(\mathbf{z}') (D \theta')(\bar{D} \bar{\theta'}) + \frac{c}{24\pi} \text{S}(\mathbf{z},\mathbf{z}'), \qquad \mathbf{z} \equiv (z,\theta,\bar{\theta}).
\end{equation}
Denoting the bosonic coordinate as $t$ instead of $z$, the $\mathcal{N}=2$ super-Schwarzian derivative is given by the expression:
\begin{equation}
\text{S}(\mathbf{z},\mathbf{z}') \equiv \text{S}(t,\theta,\bar{\theta}) = \frac{\partial_t \bar{D} \bar{\theta'}}{\bar{D}\bar{\theta'}} - \frac{\partial_t D \theta'}{D\theta'} - 2 \frac{\partial_t \theta' \partial_t \bar{\theta'}}{(\bar{D} \bar{\theta')} (D \theta')},
\end{equation}
and $D = \partial_\theta + \bar{\theta} \partial_t, \quad \bar{D} = \partial_{\bar{\theta}} + \theta \partial_t$. With a suitable parametrization of the transformed coordinates $\theta'$ and $\bar{\theta'}$ as in \cite{Fu:2016vas}, the top component of S is
\begin{equation}
\text{S}_{\text{top}} = \frac{F'''}{F'} - \frac{3}{2} \left(\frac{F''}{F'}\right)^2 - 2 \sigma'^2 + \text{fermion bilinears},
\end{equation}
in terms of two bosonic functions $F(t)$ and $\sigma(t)$, and fermionic fields that we do not write explicitly. The bottom component is then likewise determined to be
\begin{equation}
\text{S}_{\text{bot}} 
= -2i \sigma' + \text{fermion bilinears}.
\end{equation}

\noindent The superspace dissipation equation \eqref{superdissip} then reduces to the coupled component equations
\begin{align}
\nonumber
C&\frac{d}{dt} \left( \left\{F,t\right\} - 2 \sigma'^2 + (\text{fermion bil.}) \right) =  - \frac{c}{24\pi}\left( \left\{F,t\right\} - 2 \sigma'^2 + (\text{fermion bil.}) \right) + E_0 \delta(t), \\
\label{susyeq2}
C&\frac{d}{dt}\left( \sigma' + (\text{fermion bil.}) \right)= - \frac{c}{24\pi} \left(\sigma' + (\text{fermion bil.}) \right) + Q_0 \delta(t), \\
\nonumber
C&\frac{d}{dt} \text{fermion} = -  \frac{c}{24\pi} \text{fermion} + \mathcal{Q}_0\delta(t),
\end{align}
where we allowed an injection of energy $E_0$, supercharge $\mathcal{Q}_0$, and U(1) R-charge $Q_0$ into our system. Our notation has also been slightly schematic in the sense that we have not written out the fermions and the fermion bilinear combinations (``fermion bil.") explicitly. 

Let us make some comments on the structure of this system of equations:
\begin{itemize}
    \item In the case where no supercharge is sent in  $\mathcal{Q}_0=0$, the fermionic fields are never turned on. 
    \item Stronger, even if $\mathcal{Q}_0 \neq 0$, it is easy to show that all fermion bilinears cancel out in the bosonic equations in \eqref{susyeq2}. The way to show this is to multiply these equations \eqref{susyeq2} by the fermion bilinears themselves, and using that they square to zero. Then either the fermion bilinear itself is zero, mapping the initial problem to the purely bosonic (`body') subsector; or their prefactor is zero, which is again the same body piece of these differential equations.\footnote{
    Intuitively, the `soul' parts in the equations should be thought of as infinitesimal compared to the body piece, allowing us to drop them compared to the body pieces.}
    \item For the simpler case of $\mathcal{N}=1$ supersymmetry, we explicitly illustrate the above arguments in Appendix \ref{app:nis1} and solve for the dissipation of both the black hole energy and the supercharge, after an initial injection of both $E_0$ and $\mathcal{Q}_0$.
\end{itemize}

Decoupling the fermions as stated above, we then obtain the purely bosonic coupled ODEs:
\begin{subequations}
\begin{align}
C&\frac{d}{dt} \left( -\left\{F,t\right\} + 2 \sigma'^2 \right) =  - \frac{c}{24\pi}\left( -\left\{F,t\right\} + 2 \sigma'^2 \right) + E_0 \delta(t), \\
C&\sigma''= - \frac{c}{24\pi}\sigma' +  Q_0 \delta(t).
\end{align}
\end{subequations}
For the particular choice of coupling coefficients in the charged black hole system of Section \ref{s:chabs}
\begin{equation}
\label{susyparam}
    K=4C, \qquad q^2=2/3,
\end{equation} 
the bosonic charged dissipative system solved in Section \ref{s:chabs} matches with the above supersymmetric dissipation of a $c=2$ system. This provides indirect evidence that our analysis of the charged dissipative system was indeed consistent, by viewing it as embedded in the $\mathcal{N}=2$ system, where the EM U(1) gauge group gets identified with the $R$-symmetry group. \\

\noindent It is actually not that hard to write down the generic evaporating equations of motion for an arbitrary amount of supersymmetry. In $\mathcal{N}=n$ extended superspace, the dissipation equation is still
\begin{equation}
\frac{d}{dt} \text{S}(t,\theta_1,\theta_2,\hdots) = - \frac{c}{24\pi C} \text{S}(t,\theta_1,\theta_2,\hdots),
\end{equation}
where we have $n$ distinct $\theta_i$'s, $i=1\hdots n$.  Schematically, this leads to the component equations:
\begin{align}
\nonumber
C&\frac{d}{dt} \left( \left\{F,t\right\} - C_R(t) + (\text{fermion bil.}) \right) =  - \frac{c}{24\pi}\left( \left\{F,t\right\} - C_R(t) + (\text{fermion bil.}) \right) + E_0 \delta(t), \\
\label{susyeq}
C&\frac{d}{dt}\left( J_a + (\text{fermion bil.}) \right)= - \frac{c}{24\pi} \left(J_a + (\text{fermion bil.}) \right) + Q_0^a \delta(t), \quad a = 1 \hdots \text{dim } G_R, \\
\nonumber
C&\frac{d}{dt} \text{fermion} = -  \frac{c}{24\pi} \text{fermion} + \mathcal{Q}_0\delta(t),
\end{align}
where the index $a$ runs across all generators $X^a$ of the R-symmetry group $G_R$. The non-Abelian R-charges and their Casimir are given by:
\begin{equation}
J^a = \text{Tr} g^{-1} \partial_t g X^a, \qquad C_R(t) = \text{Tr} (g^{-1}\partial_t g)^2 = \sum_a J^a J^a,
\end{equation}
in terms of the group variable $g(t)$ describing the bosonic degrees of freedom that describe the R-symmetry group in the super-Schwarzian.
The fermions decouple again, and the conserved charges dissipate according to
\begin{equation}
J^a(t) = Q_0^a e^{- \frac{c}{24\pi} t},  \quad a = 1 \hdots \text{dim } G_R.
\end{equation}
Multiplying the second equation of \eqref{susyeq} by $2 J^a$ and summing over $a$, we plug this in the first equation and find the black hole energy $E_{\text{BH}}(t)$ dissipating according to a very similar equation as earlier in \eqref{frameeom}:
\begin{equation}
C\frac{d}{dt} \left\{F,t\right\} =  - \frac{c}{24\pi}\left( \left\{F,t\right\} + C_R(t) \right) + \left(E_0 - \frac{\sum_a Q_0^a Q_0^a}{2}\right) \delta(t).
\end{equation}

We can summarise this by stating that the solution $E_{\text{BH}}(t)$ for \emph{all} such supersymmetric, dissipative equations is of the familiar form:
\begin{equation}
\left\{F,t\right\} = a e^{- \frac{c}{24\pi C} t} + b e^{- \frac{c}{12\pi C} t},
\label{diffeq} 
\end{equation}
for some coefficients $a$ and $b$, that are easily determined. Notice that both exponentials differ by a factor of 2 in their decay rate, which is universal for the supersymmetric systems. For the bosonic charged system of Section \ref{s:chabs} this last property is not necessarily true. \\

\begin{figure}[h]
  \centering
  \includegraphics[width=0.6\textwidth]{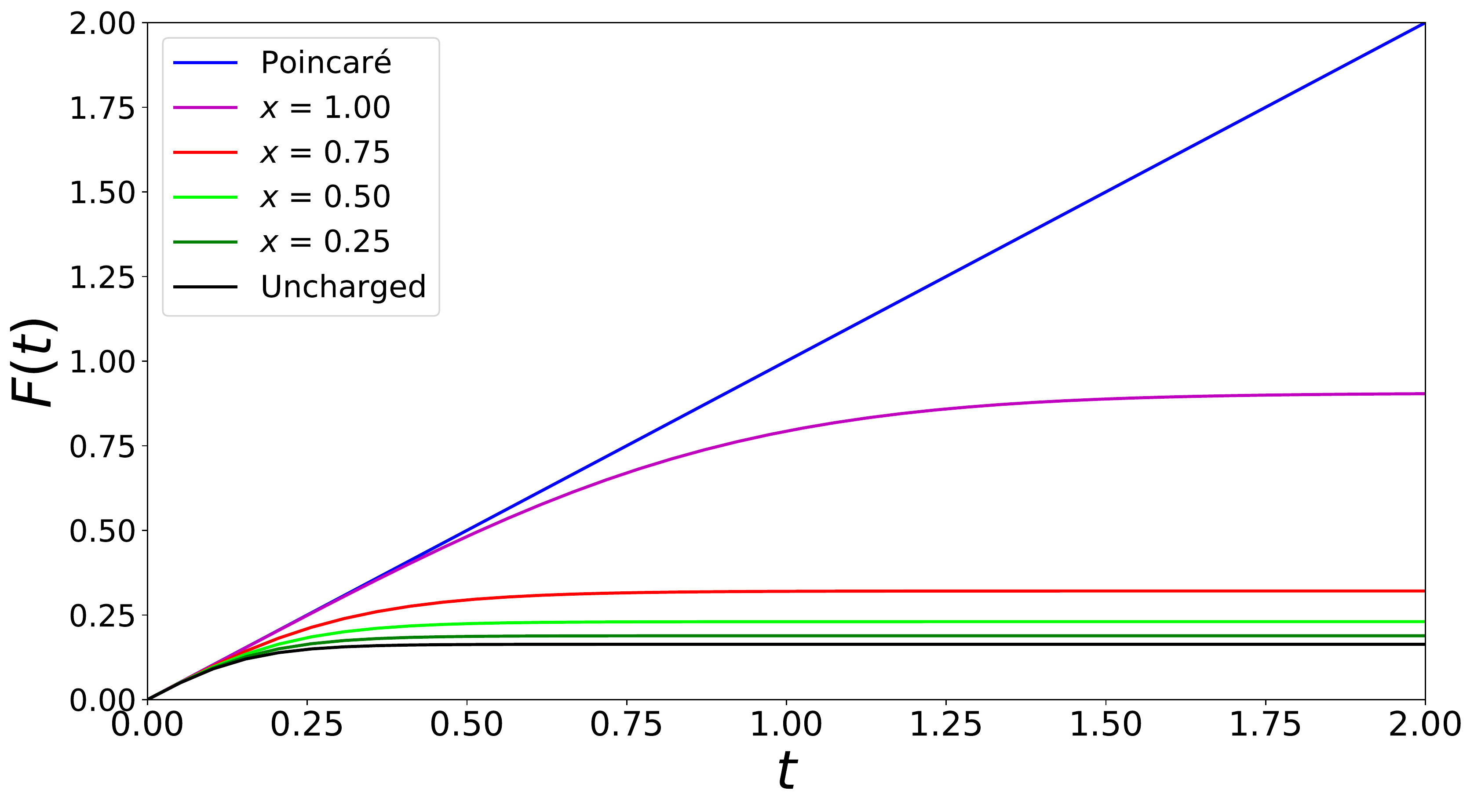}
  \caption{The reparametrisation $F(t)$ (\ref{reparam}) plotted for various $x$ with fixed energy $E_0 = 20$, fixed $k = 0.1$ and $K=4C$. As the charge increases, the profile $F(t)$ gets closer to the Poincar\'e solution $F(t) = t$ than the uncharged solution.}
  \label{fig:reparam}
\end{figure}

\noindent One can now go further and solve \eqref{diffeq} explicitly for the time reparametrisation $F(t)$, with $a$ and $b$ as in \eqref{enCoeff}. Restricting to the supersymmetric case where $6q^2C = K$ and $K = 4C$, the  solution to this equation with the correct gluing conditions at $t=0$ is 
\begin{equation}
  F(t) = \frac{\sqrt{KC}}{i Q_0} \left[\frac{-W_{i \kappa,0}\left(4 i \kappa x \right) M_{i \kappa, 0}\left(4 i \kappa x e^{-kt}\right) + M_{i \kappa, 0}\left(4 i \kappa x\right) W_{i \kappa, 0}\left(4 i \kappa xe^{-kt}\right)}{\mathcal{C} M_{i \kappa, 0}\left(4 i \kappa x e^{-kt}\right) + \mathcal{D} W_{i \kappa, 0}\left(4 i \kappa x e^{-kt}\right)}\right],
  \label{reparam}
\end{equation}
in terms of combinations of Whittaker-$W$ and -$M$ functions, with evaporation rate $k \equiv \frac{c G_\text{N}}{3 \phi_r} = \frac{c}{24\pi C}$, and where we introduced the dimensionless constants 
\begin{equation}
  \kappa \equiv 12 \pi \frac{E_0 \sqrt{KC}}{c Q_0}, \quad x \equiv \frac{Q_0^2}{2 E_0 K},
  \label{dimParam}
\end{equation}
and (\ref{Ceq}, \ref{Deq}). The future horizon is located at:
\begin{equation}
  F_\infty \equiv F(t\to+\infty) = \frac{\sqrt{KC}}{i Q_0} \frac{ M_{i \kappa, 0}\left(4 i \kappa x\right) }{ M'_{i \kappa, 0}\left(4 i \kappa x \right) - \frac{1}{8 i \kappa x} M_{i \kappa, 0}\left(4 i \kappa x \right)}.
\end{equation}
\indent Further details and checks are provided in Appendix \ref{app:DE}, and 
the reparametrisation for various $x$ at fixed $E_0$ is plotted in Fig.~\ref{fig:reparam}.

\subsection{Black hole entropy $S_{\text{BH}}$ and superradiance}
For the eternal, charged black hole in JT gravity,  the Hawking temperature is given by the relation \eqref{hawktemp}:
\begin{equation}
T = \frac{1}{\pi \sqrt{2C}} \sqrt{E - \frac{Q^2}{2K}}.
\label{tempBH}
\end{equation}
This sets a bound on the value of the parameter $x$ introduced in (\ref{dimParam}). This same bound is encountered when solving \eqref{frameeom}. Demanding $T_0$ to be positive we obtain $0 \leq x \leq 1$ where $x = 0$ is the uncharged case and $x = 1$ the charged, extremal black hole case. \\
\indent We can define an ``instantaneous'' temperature as
\begin{equation}
T(t) \equiv \frac{1}{\pi \sqrt{2C}} \sqrt{E(t) - \frac{Q(t)^2}{2K}},
\end{equation}
upon plugging in the expressions for the dissipating energy \eqref{disen} and charge \eqref{disch}. Using the black hole first law for a charged system in a quasi-static approximation
\begin{equation}
dE = T(t) dS + \frac{Q(t)}{K} dQ,
\end{equation}
the instantaneous Bekenstein--Hawking entropy $S_{\text{BH}}(t)$ is given by
\begin{equation}
\label{SBH}
S_{\text{BH}}(t) = 2\pi \sqrt{2C} \sqrt{E(t) - \frac{Q(t)^2}{2K}} = 2\pi \sqrt{2C} \sqrt{E_{\text{BH}}(t)}.
\end{equation}

\begin{figure}[h!]
  \centering
   \includegraphics[width=0.55\textwidth]{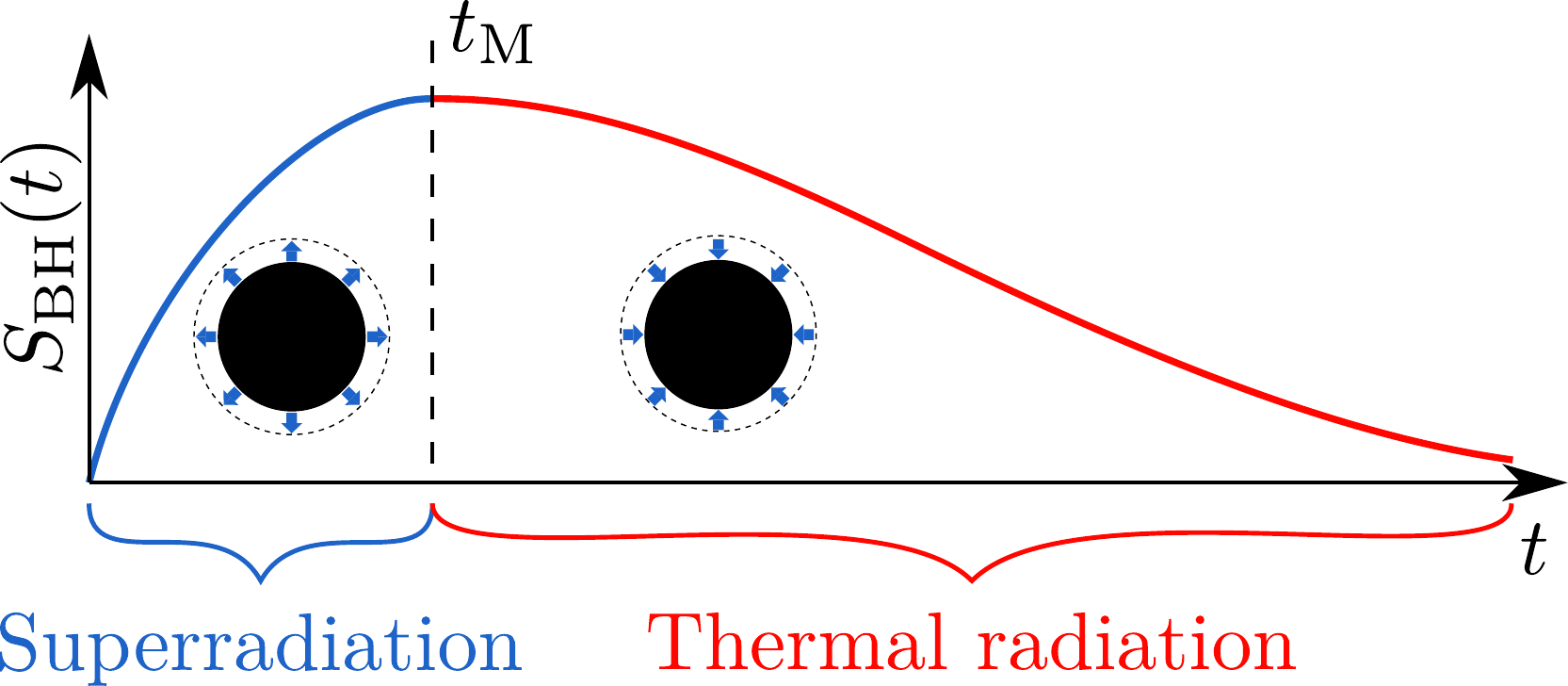}
  \caption{At times before $t_M$, the black hole emits mainly superradiant modes causing it to heat up. After $t_M$, thermal Hawking emission dominates, the black hole loses energy, and the system dissipates completely. The curve in this figure is specific for $E_0 = \frac{Q_0^2}{2K},~x=1$ for which there is no jump at $t=0$.}
    \label{Fig:superradiance}
  \end{figure}

Let's focus on the choice of parameters \eqref{susyparam} relevant for the supersymmetric system again. In this case, the charge, energy and black hole energy profiles are
\begin{subequations}
\begin{align}
Q(t) &= Q_0 e^{-\frac{ct}{24\pi C}}, \\
E(t) &= E_0 e^{-\frac{ct}{24\pi C}}, \\
E_{\text{BH}}(t) &\equiv E(t) - \frac{Q(t)^2}{2K} = E_0 e^{-\frac{ct}{24\pi C}} - \frac{Q_0^2}{2K}e^{-\frac{ct}{12\pi C}}.
\end{align}
\end{subequations}
\indent Somewhat surprisingly, whereas the total energy in the system $E(t)$ always decreases monotonically, the black hole energy $E_{\text{BH}}(t)$ and its entropy $S_{\text{BH}}(t)$ are \emph{not} necessarily monotonically decreasing, and instead can have a local maximum at the time:
\begin{equation}
\label{tm}
t_{\text{M}} = \frac{24 \pi C}{c} \ln \frac{Q_0^2}{E_0 K} = \frac{1}{k} \ln 2x.
\end{equation}
This time $t_{\text{M}}$ is positive (and hence physical) iff $Q_0^2 > E_0 K$ or $x > 0.5$. We can distinguish two qualitative cases: 
\begin{itemize}
  \item $0.5 < x \leq 1$: the resulting black hole has a temperature such that during $0 \leq t \leq t_{\text{M}}$ the spontaneous emission of superradiant modes dominates, causing the black hole to heat up and increase in size. This is reflected in its Bekenstein--Hawking entropy. For $t \geq t_{\text{M}}$, the thermal Hawking modes will be more densely populated such that the black hole now starts to dissipate. 
  \item $0 \leq x \leq 0.5$: superradiant modes are suppressed at this temperature such that the thermal Hawking radiation dominates. The black hole immediately starts shrinking.
\end{itemize}
In both cases the charge decays monotonically. This prevents the black hole from going through a cyclic life of growth and shrinking in the first case $x > 0.5$. 
After $t_{\text{M}}$, the temperature decreases, but it will not result in repeated domination of the superradiant modes due to the decaying charge. \\
\indent To appreciate these statements, it is instructive to think about an initial extremal pulse with $E_0 = \frac{Q_0^2}{2K}$, or $x=1$. For such a pulse, we have no jump in temperature and entropy at $t=0^+$:
\begin{equation}
T(0^+)=0, \qquad S_{\text{BH}}(0^+) =0.
\end{equation}
The instantaneous bath surrounding the black hole is then given by equation \eqref{bathzero}, which only contains superradiant modes up to $\omega = \frac{Q_0}{K} q$. Superradiant mode emission causes the starting extremal black hole to heat up as $t > 0$, this unlocks ordinary thermal Hawking radiation. Such physics was also found in a higher-dimensional holographic context in  \cite{Dias:2007nj}. As time progresses and the black hole heats up more and more, normal Hawking radiation starts taking over, depleting the black hole again. All the while the electric charge of the black hole monotonically decreases. We sketch the evolution of the black hole and its size in Fig. \ref{Fig:superradiance}.
   
\subsection{Early-late matter entanglement entropy}

Finally, we will calculate the early-late entanglement entropy profile for our charged or supersymmetric system. 

It can be shown that the 2d CFT entanglement entropy formula for an interval $(u_1,v_1)$, $(u_2,v_2)$:
\begin{equation}
S = \frac{c}{12} \ln \frac{(F(u_1)-F(u_2))^2}{\delta_1\delta_2 F'(u_1)F'(u_2)} + \,\, (u \leftrightarrow v), 
\end{equation}
remains the same when considering a system with an enlarged local symmetry group that goes beyond conformal coordinate transformations, i.e. in our case including U(1) gauge transformations, or superconformal transformations. We show this statement from several perspectives in Appendix \ref{sect:SCFTEE}.\footnote{This is the flat space formula, but we have seen in \eqref{renisflat} that the renormalised versions in flat space and in AdS$_2$ space match.}

\begin{figure}[h!]
  \centering
  \includegraphics[width=0.8\textwidth]{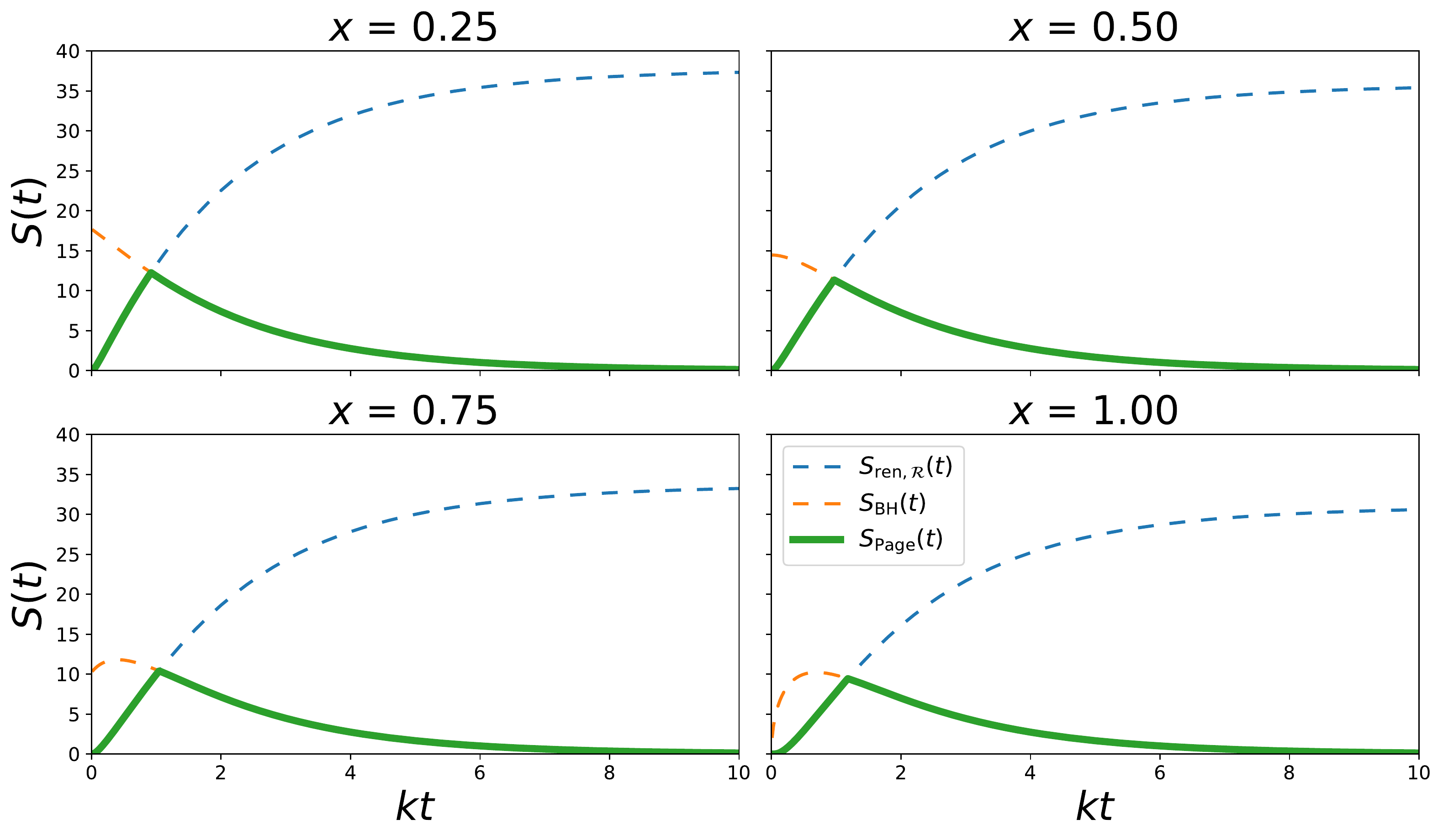}
  \caption{The Bekenstein--Hawking (\ref{SBH}) and matter radiation entropy $S_{\text{ren},\mathcal{R}}(t)$ \eqref{sect5:entropyRen} plotted for different values of $x=\frac{Q_0^2}{2 E_0 K}$ with fixed $E_0 = 20$, $k = 0.1$ and $K=4C$. 
  For $x = 0.75,~1.00$ we see the appearance of a maximum at $t_\text{M}$ as was elaborated upon in the main text.}
  \label{fig:pageCurves}
\end{figure}

As such, we can still use the formula \eqref{sect5:entropyRen} for the renormalised radiation entropy $S_{\text{ren},\mathcal{R}}(t)$ by plugging in the form of $F(t)$ we derived above in \eqref{reparam}. Both this renormalised entanglement entropy, and the above Bekenstein--Hawking entropy \eqref{SBH} are plotted in Fig.~\ref{fig:pageCurves} for different values of $x=\frac{Q_0^2}{2 E_0 K}$. Following the minimum of both curves would lead to the Page curve for the evaporating, charged black hole. This is also the result of following the quantum extremal surface formula using our operational islands prescription from Section \ref{sect:archipelago} for a macroscopic, evaporating, charged black hole. 
Notice that for larger values of the charge (or $x$), depending on the parameters, one might worry about a situation where the Page time occurs \emph{before} $t_M$. However, numerical investigations have shown us that this scenario does not seem to occur, at least not for the supersymmetric scenario \eqref{susyparam}.

Let us end with some comments on these expressions concerning their late-time and macroscopic limiting behaviours. For an arbitrary matter CFT with central charge $c$, the late-time renormalised entanglement entropy is given by
\begin{equation}
  \lim_{t\to +\infty}\frac{c}{12} \ln \frac{F(t)^2}{t^2F'(t)} = \frac{c}{12} \ln \left( \frac{1}{4 i \kappa x} M^2_{i \kappa ,0}\left( 4i \kappa x \right)\right),
  \label{entropyHawk}
\end{equation}
since the late-time behaviour of $F'(t)$ goes as $t^{-2}$, just as in the uncharged case.

\indent To investigate the macroscopic limit, we go to the regime where $E_0 C \approx E_0K \gg 1,~Q_0 \gg 1$ in such a way that the initial temperature $T_0$ (\ref{tempBH}) remains of order $\mathcal{O}(1)$. Hence, $E_0\sim \mathcal{O}(Q_0^2)$ resulting 
in $\kappa \to +\infty$ and $x$ remaining finite. Subsequently, we can make use of \eqref{Mlim}:
\begin{align}
  \lim_{t\to +\infty}\frac{c}{12} \ln \frac{F(t)^2}{t^2F'(t)} \quad \underset{\kappa \to + \infty}{\to} \quad \frac{c}{12} \ln \left( \frac{1}{24 \pi^2 Q_0}\sqrt{\frac{K}{C}}\left(\frac{x}{1-x}\right)^{1/2 }e^{8 \kappa \sqrt{\rho}} \right) , 
\end{align}
where
\begin{equation}
2\sqrt{\rho} = \sqrt{x-x^2} - \text{arccos} \sqrt{x} + \frac{\pi}{2}.
\end{equation}
From this, we obtain the final radiation entropy:
\begin{equation}
\label{finalrad}
S_{\text{ren},\mathcal{R}}(t\to+\infty) = \frac{2c}{3} \kappa \sqrt{\rho} = \frac{4\pi E_0 \sqrt{KC}}{Q_0} \left(  \frac{Q_0}{\sqrt{2K E_0}} \sqrt{E_0 - \frac{Q_0^2}{2K}}- \text{arccos}\frac{Q_0}{\sqrt{2K E_0}} + \frac{\pi}{2} \right).
\end{equation}
Taking the first term alone, we can simplify the formula into
\begin{equation}
\label{1st}
S_{\text{ren},\mathcal{R}}^{\text{$1^{st}$ term}}(t\to+\infty) = 2\pi \sqrt{2C} \sqrt{E_0 - \frac{Q_0^2}{2K}},
\end{equation}
which matches the initial Bekenstein--Hawking entropy \eqref{SBH} of the created black hole. 
The corrections in \eqref{finalrad} adjust this, and show that entropy is created in this evaporation process. Moreover, there is no simple relation between the final radiation entropy and the initial black hole entropy, unlike in the uncharged case as discussed around \eqref{zurek}. 

In the uncharged limit, where $\rho \approx x = \frac{Q_0^2}{2 K E_0}$, we can simplify \eqref{finalrad} to
\begin{equation}
S_{\text{ren},\mathcal{R}}(t\to+\infty) = 4 \pi \sqrt{ 2 C E_0 },
\end{equation}
agreeing with \cite{Mertens:2019bvy}.\footnote{This can also be found in the macroscopic limit of the expression \eqref{ch07:asympt}.
}

\indent In the opposite maximally charged regime where $E_0 = Q_0^2/2K$ or $x=1$, the final matter entropy becomes
\begin{equation}
\label{Sfin}
S_{\text{ren},\mathcal{R}}(t\to+\infty) = \frac{2\pi^2 E_0 \sqrt{KC}}{Q_0}.
\end{equation}
It is useful to compare this to the maximum black hole entropy during the evaporation. Plugging the time \eqref{tm} in \eqref{SBH}, we get
\begin{equation}
S_{\text{BH}}(t_{\text{M}}) = \frac{2\pi E_0 \sqrt{KC}}{Q_0},
\end{equation}
which is a factor of $\pi$ smaller than \eqref{Sfin}.

\section{Concluding remarks}
In this work, we have investigated black hole evaporation in the 1+1 dimensional JT gravity model. Our main endeavour has been twofold: 
\begin{enumerate}
\item[(1)]  We attempted to understand how the older concept of renormalised matter entanglement entropy fits into the recent island paradigm. We found that plugging in this renormalised entanglement entropy into the island or quantum extremal surface formula yields the expected physics, with a second saddle dominating after the Page time. In this case, both contributions to the entropy are physically meaningful. It remains to be seen precisely how this alternative proposal fits into the bigger story.
\item[(2)] We generalised the purely energetic considerations of black hole evaporation to include more general conserved quantites: charge and supercharge in particular. In both cases, we solved the dissipative equation of motion and considered the entanglement entropy profiles during evaporation.
\end{enumerate}

\noindent We will end by providing a couple of open problems, for which we will suggest concrete routes forward. 

\subsection*{Brane-world models and massive graviton}
There has been a considerable amount of confusion about the set-up of coupling a flat space bath to the gravitational region. This set-up has an interpretation in terms of brane-world Karch-Randall-Sundrum configurations \cite{Randall:1999ee,Karch:2000ct}. It is known however that for such models, the graviton in the bulk becomes massive due to quantum effects \cite{Porrati:2003sa}. Within the current context, works addressing this are e.g. \cite{Geng:2020fxl,Geng:2020qvw,Geng:2021hlu,Bhattacharya:2021jrn}.
In 1+1d, there is no bulk graviton to begin with, so the application of the above observation directly to 2d is more subtle. However, our absorbing set-up did not use this explicit coupling to a flat space bath to begin with, and it therefore seems to open a path towards understanding the massless graviton story directly.

\subsection*{Quantum effects of gauge fields}
Throughout this work, we have considered gravity and the bulk gauge sector to be classical. 
Let's include the quantum dynamics of the gauge sector here for the non-evaporating case. To isolate its effect, we work at zero temperature but include the quantum effects of the gauge sector ($K$ finite). We can write for the spectral occupation number in the Poincar\'e vacuum:\footnote{One can also allow arbitrary $F(t)$ in this expression and path integrate over the Schwarzian degrees of freedom to obtain the Unruh heat bath in full quantum gravity \cite{Mertens:2019bvy,Blommaert:2020yeo}.}
\begin{align}
\label{occnum}
N^{\beta \to \infty}_\omega[\sigma] = -\frac{1}{4\pi^2\omega}\int du_1 \int du_2 e^{-i \omega (u_1-u_2)} &\left[\frac{1}{u_{12}^2}e^{iq(\sigma(u_1)-\sigma(u_2))} - \left(\frac{1}{u_{12}}\right)^2\right],
\end{align}
where an (off-shell) choice of gauge frame $\sigma(.)$ is still present. Within the path integral over $\sigma$, the operator insertion $e^{iq(\sigma(u_1)-\sigma(u_2))}$ is Gaussian and can be readily done into \cite{Mertens:2018fds}:\footnote{We are assuming the gauge group $\mathbb{R}$ here for simplicity, otherwise the charge integral would be a sum.}
\begin{equation}
\int_{-\infty}^{+\infty} dq ~e^{-\frac{q^2}{2K}\tau}e^{-\frac{(q-Q)^2}{2K}(\beta- \tau)}e^{\mu \beta (Q-q)}.
\end{equation}
Performing then the Fourier transform in \eqref{occnum},\footnote{And using
\begin{equation}
\label{PCref}
-\frac{1}{2\pi\omega}\int_{-\infty}^{+\infty} dt \frac{1}{(t \mp i \epsilon)^2}e^{-i\omega t} =  \mp\Theta(\mp\omega)
\end{equation}
}
we obtain after the $\sigma$ path integral:
\begin{equation}
\label{backrocc}
\left\langle N_\omega^{\beta \to \infty}\right\rangle \,=\, \frac{q\mu-\left(\omega-\frac{q^2}{2K}\right)}{\omega} \Theta \left[q\mu - \left(\omega-\frac{q^2}{2K}\right)\right].
\end{equation}
\indent This formula has the following interpretation. The total energy $\omega$ of an emitted quantum consists of the kinetic and rest energy of the particle itself, and the energy of the electric field it sources. The quantity $\omega-\frac{q^2}{2K}$ is then the energy of the quantum \emph{without} its electric self-energy. 

\begin{figure}[h]
\centering
\includegraphics[width=0.3\textwidth]{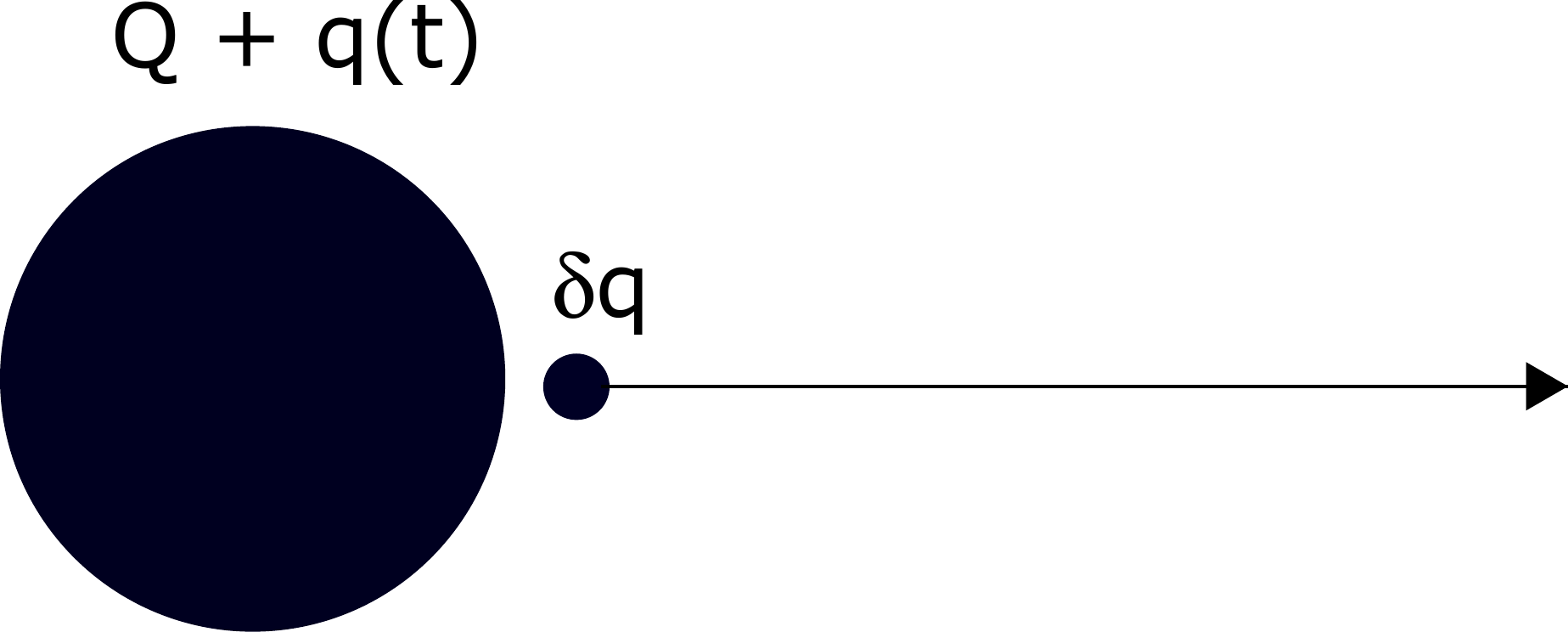}
\caption{Continuous extraction of charge $\delta q$ from a charged system of charge $Q+q(t)$ where $q(t)$ starts off at $q$ and then shrinks as charge is extracted until it reaches $0$.}
\label{contq}
\end{figure}

This can be appreciated as follows. Consider the removal of a total charge $q$ from a (extremal, or zero-temperature) black hole horizon with remaining charge $Q$ after the particle is extracted. In line with the Hawking evaporation process, this procedure is done continuously and leads to the following differential energy:
\begin{equation}
dE = \frac{Q + q(t)}{K} \delta q(t),
\end{equation}
where $q(t)$ ranges from $q$ to $0$ as the charge is extracted (Fig. \ref{contq}).
This leads to a total energy gained in the process:
\begin{equation}
\Delta E = \mu q + \frac{q^2}{2K}, \qquad \mu = \frac{Q}{K}.
\end{equation}
This backreaction effect then corrects the Schwinger energetic cutoff in the occupation number in \eqref{backrocc} by including the electric self-energy $q^2/2K$ of the particle as it is emitted.

\subsection*{Dissipative action}
One of the main points of this work has been implementing a purely dissipative interaction at the level of the classical equations of motion. One route towards a quantum implementation of this would be to first write an action principle for this equation of motion.
It is a famous result that dissipative equations of motion cannot be deduced from a local Lagrangian but can be ``localised'' by integrating in bath degrees of freedom that soak up the emitted energy, which was essentially the approach of the flat space bath coupled to JT gravity. So instead of doing this, can we directly write down the non-local action whose equations of motion directly yield
\begin{equation}
   \frac{d}{dt} \left\{F,t \right\} = - k \left\{F,t \right\}\, ?
\end{equation}
One direct route is to use the mapping $F' = e^\varphi$ in terms of which the Schwarzian action reduces essentially to that of a 1d free boson $\varphi$. The required dissipative model then also reduces to the textbook dissipative action, see e.g. \cite{Weiss1999}, where one introduces the Caldeira-Leggett non-local term:
\begin{equation}
    S_{\text{CL}} \sim \int dt \int dt' \frac{(\varphi(t)-\varphi(t'))^2}{(t-t')^2}.
\end{equation}
It remains to be seen how useful this observation is for the development of the quantum dissipative models at hand.

\section*{Acknowledgements}
We thank A. Blommaert, N. Callebaut, J. Engels\"oy, Y. Fan, J. Maldacena, M. Mezei and J. Turiaci for discussions and comments on various topics related to this work. TM acknowledges financial support from Research Foundation Flanders (FWO Vlaanderen) and the European Research Council (grant BHHQG-101040024). Funded by the European Union. Views and opinions expressed are however those of the author(s) only and do not necessarily reflect those of the European Union or the European Research Council. Neither the European Union nor the granting authority can be held responsible for them. JDV is grateful for the financial support received through a Research Assistantship granted by the Okinawa Institute of Science and Technology Graduate University (OIST). 
This work was partly written as a result from JDV's thesis to obtain the degree of MSc in Physics \& Astronomy at Ghent University 2020-2021. As such, JDV 
appreciates the supervision given by TM during that year.

\appendix

\section{Non-evaporating black hole}
\label{app:eternal}
\begin{figure}[h!]
\centering
        \includegraphics[width=0.7\textwidth]{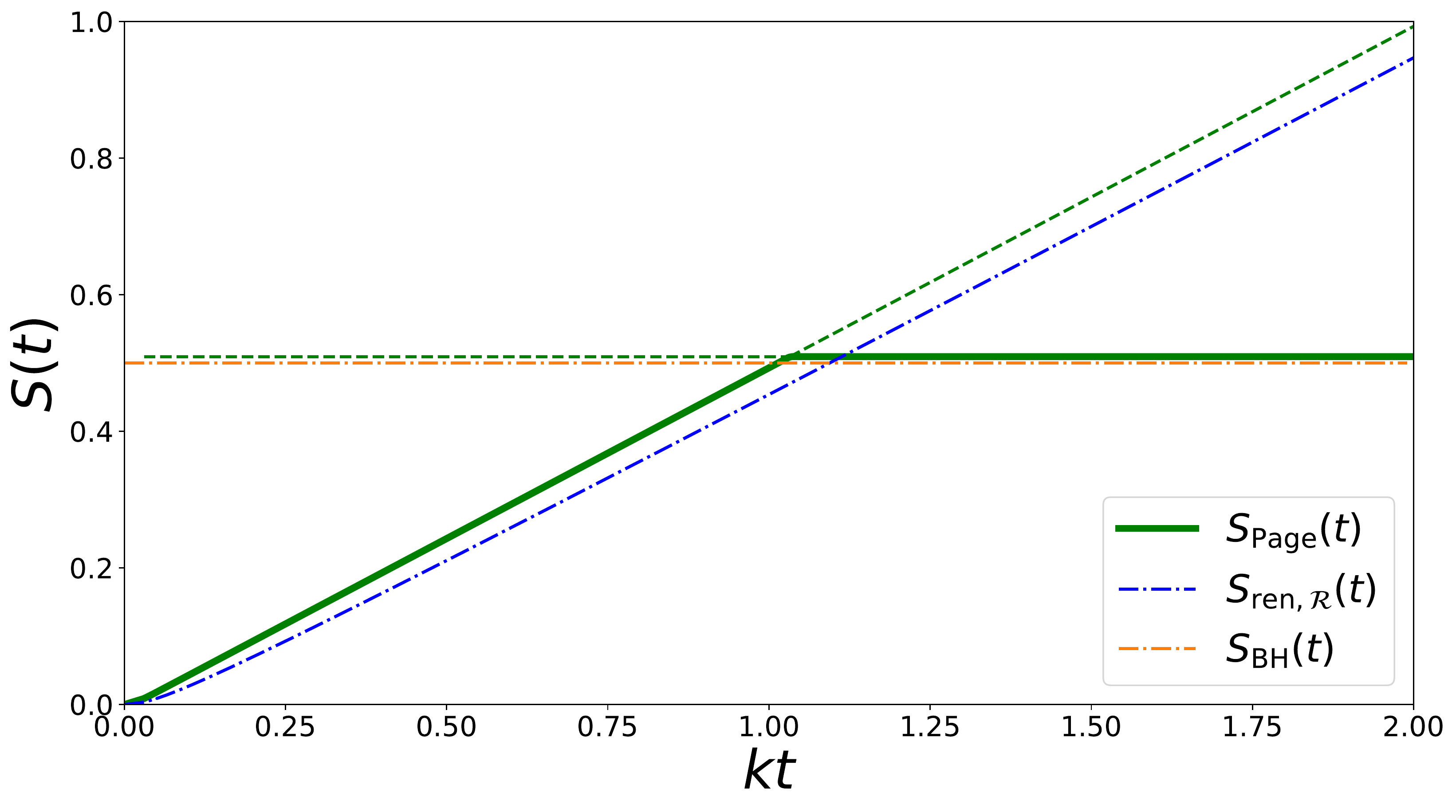}
    \caption{The Page curve for $k = 0.01,~T = \frac{1}{2\pi}$ for a non-evaporating black hole is given by the solid line in green. For $t_\text{scr} = 3.057$ one finds the same constant part of the Page curve via \eqref{app:Spost}. There is a small shift to a smaller Page time when compared with 
    the intersection of the fine-grained radiation entropy and the classical Bekenstein--Hawking entropy when deviating from the strict macroscopic $k/T \ll 1$ limit.}
        \label{app:figPageEter}
\end{figure}

The case for a non-evaporating black hole is readily made. We merely replace the evaporating time reparametrisation by the eternal one of \eqref{sect2:thermal}. One can interpret the resulting equations both as associated to the physical situation of an infalling pulse creating a non-evaporating black hole, or as an eternal black hole that was always present. The radiation and Bekenstein--Hawking entropies become now:
\begin{equation}
S_{\text{ren},\mathcal{R}}(t) =  \frac{k}{2} \ln \frac{F(t)^2}{t^2F'(t)} = k \ln \frac{\sinh \pi T t}{\pi T t},
\end{equation}
and
\begin{equation}
S_{\text{BH}}(t) =  \pi T.
\end{equation}
For a macroscopic black hole, the radiation entropy rises linearly as $k \pi T t$. \\
\indent From the QES prescription, we find a pre-pulse island at the same location as earlier, with generalised entropy:
\begin{equation}
S_\text{pre}(t) =  -\frac{k}{2} \ln F'(t) = k \ln \cosh \pi T t,
\end{equation}
and for the post-pulse island, the expression for the entropy \eqref{sect5:entropyGeneral} with \eqref{sect2:thermal} results in\footnote{Note that these equations are time translation invariant, as it should for the static black hole metric and dilaton background \eqref{staticmetric}.} 
\begin{equation}
    S_\text{post}(t, u, v) = \pi T \coth[\pi T (u-v)] + S_{\text{ren},\mathcal{R}}(u-t) + S_{\text{ren},\mathcal{R}}(t-v),
\end{equation}
numerically leading to an island starting just outside the event horizon that asymptotes to it as time goes on, with a nearly constant generalised entropy. Moreover, numerical analysis shows that 
the island stays at a fixed radial position $z$ in the black hole coordinates of \eqref{staticmetric}, namely at $u = t + t_\text{scr},~v = t - t_\text{scr}$. Taking the ansatz $u = t + a,~v = t - b$ it can be analytically checked that the island 
equations (\ref{sect5:postuCond}, \ref{sect5:postvCond}) indeed lead to the solution $a = b \equiv t_\text{scr}$. We will call this radial location, $z_{\text{s.h.}} \equiv t_\text{scr}$, the stretched horizon. In turn, this leads to a fixed entropy for this island 
\begin{equation}
		    S_\text{post} = \pi T \coth(2 \pi T z_{\text{s.h.}}) + 2 S_{\text{ren},\mathcal{R}}(z_{\text{s.h.}}).
    \label{app:Spost}
\end{equation} 
The constant scrambling time $t_\text{scr}$ (or stretched horizon location) can be found numerically by solving \eqref{sect5:initialIsland} for which $u_0 = 2 t_\text{scr},~t_0 = t_\text{scr}$, and generally decreases 
for an increasing value of $k/T$.\footnote{Eq.~\eqref{sect5:initialIsland} reduces to the extremisation of this expression.}\\
\indent For a macroscopic black hole where $k/T \ll 1$, both generalised entropies match again with the Hawking radiation and quasi-static Bekenstein--Hawking entropies. In this case, the scrambling time can be computed assuming $z_{\text{s.h.}} \gg 1/T$, by extremising \eqref{app:Spost}, leading to the result
\begin{equation}
t_\text{scr} = z_{\text{s.h.}} = \frac{1}{4\pi T} \ln \frac{4 \pi T}{k}  = \frac{1}{4\pi T} \ln \frac{24}{c} S,
\end{equation}
which indeed satisfies the assumption $z_{\text{s.h.}} \gg 1/T$ for a macroscopic black hole $k/T \ll 1$. This location corresponds in the metric \eqref{staticmetric} to the proper distance $\ell \approx \sqrt{\frac{k}{\pi T}} = \sqrt{\frac{c}{6S}}$ from the black hole horizon. \\
\indent We plot all entropies and the resulting Page curve in Fig. \ref{app:figPageEter}.

\section{Black hole dissipation in $\mathcal{N}=1$ JT supergravity}
\label{app:nis1}

The JT supergravity action can be written in superspace as \cite{Chamseddine:1991fg}:
\begin{equation}
\label{SJT}
S = -\frac{1}{16\pi G}\left[i \int d^2z d^2\theta E \Phi (R_{+-} + 2) + 2 \int dtd\theta \Phi_b K\right].
\end{equation}
It was shown in \cite{Forste:2017kwy} that the dynamics reduces to the boundary term in \eqref{SJT}, which can be written as a super-Schwarzian action:
\begin{equation}
\label{SSch}
S = 2C \int dt d\theta \text{Sch}(\tau,\theta).
\end{equation}
The super-Schwarzian is defined in $\mathcal{N}=1$ ($\tau,\theta$) superspace by:
\begin{equation}
\text{Sch} \equiv \text{Sch}_f + \theta \text{Sch}_b = \frac{D^4\theta'}{D\theta} - 2 \frac{D^3\theta'D^2\theta'}{(D\theta')^2},
\end{equation}
with $D= \partial_\theta + \theta \partial_\tau$ the superderivative and where the transformed coordinates are parametrised as 
\begin{subequations}
\begin{align}
    \tau' &= F(\tau + \theta \eta(\tau)), \\
    \theta'&= \sqrt{\partial_\tau F}\left(\theta + \eta + \frac{1}{2}\theta\eta\partial_\tau \eta\right),
\end{align} 
\end{subequations}
in terms of a bosonic function $F(\tau)$ and a fermionic function $\eta(\tau)$. This action describes the dynamics of the superframe $(F,\eta)$ of a boundary super-clock.

Written in component fields (the reparametrisation $F(\tau)$ and its superpartner $\eta(\tau)$), one writes
\begin{subequations}
\begin{align}
\label{schb}
\text{Sch}_b &= \frac{1}{2}\left[\left\{F,\tau\right\} + \eta \eta''' + 3 \eta'\eta'' - \left\{F,\tau\right\}\eta\eta'\right], \\
\label{schf}
\text{Sch}_f &= \eta'' + \frac{1}{2}\eta\eta'\eta'' + \frac{1}{2}\eta \left\{F,\tau\right\}.
\end{align}
\end{subequations}
The action \eqref{SSch} is then written in bosonic space as:
\begin{equation}
S = C\int d\tau \text{Sch}_b.
\end{equation}

\noindent There is a set of symmetry transformations of the action \eqref{SSch}: the super-M\"obius group OSp$(1|2)$ acting as \cite{Arvis:1982tq,Fu:2016vas}\footnote{When writing the transformations in this way they do not compose as a group. For our purposes here this is sufficient, but we refer to \cite{Fan:2021wsb} for more details in a related context where this does matter.}
\begin{equation}
\label{superMob}
\tau \to \frac{a\tau+b -(a\delta - b\beta) \theta}{c\tau+d - (c\delta - d\beta) \theta}, \qquad \theta \to \frac{\beta \tau + \delta +(1+\frac{3}{2}\delta\beta) \theta}{c \tau + d - (c\delta - d\beta) \theta },
\end{equation}
with $ad-bc=1$, in terms of three bosonic parameters $a,b,c$ and two fermionic parameters  $\beta,\delta$. This is a superconformal mapping with reparametrisation functions:
\begin{equation}
\label{superMobfunc}
F(\tau) = \frac{a\tau + b}{c\tau+d}, \qquad \eta(\tau) =  \delta + \beta \tau.
\end{equation}
The $\mathcal{N}=1$ super-Schwarzian has a fermionic ($f$) and a bosonic ($b$) component in its decomposition as $\text{Sch}(\tau,\theta) \equiv \text{Sch}_f(\tau) + \theta \text{Sch}_b(\tau)$. Absorbing boundary conditions are then implemented by the following dissipation of energy and supercharge, after an initial pulse $E_0, \mathcal{Q}_0$:
\begin{subequations}
\begin{align}
\frac{d}{dt} 2C\text{Sch}_b = - \frac{c}{12 \pi} \text{Sch}_b - E_0 \delta(t), \\
\label{superchdis}
\frac{d}{dt} 2C\text{Sch}_f = - \frac{c}{12 \pi} \text{Sch}_f - \mathcal{Q}_0 \delta(t).
\end{align}
\end{subequations}
Explicitly, when plugging in the expressions for the super-Schwarzian components \eqref{schb} and \eqref{schf}, 
these equations of motion look like:
\begin{align}
\nonumber
C \left( 4 \eta' \eta''' \!+ \eta \eta''''  \!+ (1\! - \! \eta \eta') \left\{F,t\right\}' -\! \eta \eta'' \left\{F,t\right\} \right) &= - \frac{c}{24 \pi} \left( \left\{F,t\right\} + \!\eta \eta''' \!+ 3 \eta'\eta'' - \!\left\{F,t\right\}\eta\eta' \right) - E_0 \delta(t), \\
\label{eq2}
C \left(\eta'''+  \frac{1}{2} \eta \eta' \eta''' + \frac{1}{2} \eta' \left\{F,t\right\} + \frac{1}{2} \eta \left\{F,t\right\}'\right) &= - \frac{c}{24 \pi} \left(\eta'' +\! \frac{1}{2}\eta\eta'\eta'' + \!\frac{1}{2}\eta \left\{F,t\right\}\right) + \frac{\mathcal{Q}_0}{2} \delta(t),
\end{align}
and represent two coupled differential equations, where one function $\eta(t)$ is Grassmann-valued. Multiplying the first equation in \eqref{eq2} by $\eta\eta'$, we find:
\begin{equation}
\eta \eta'\left( C \left\{F,t\right\}' + \frac{c}{24 \pi} \left\{F,t\right\} + E_0  \delta(t)\right)=0.
\end{equation}
Either $\eta \eta' \neq 0$, implying the bosonic prefactor between brackets vanishes, or $\eta \eta'=0$ in which case $\eta \sim \eta' \sim \eta''$, and hence all products of $\eta$'s vanish and the first equation in \eqref{eq2} directly boils down to again the same equation. Hence, w.l.o.g. we find the same dissipating equation of motion as in the bosonic case:
 \begin{equation}
\label{eqfra}
\boxed{C \left\{F,t\right\}' = - \frac{c}{24 \pi}\left\{F,t\right\} - E_0 \delta(t)}.
\end{equation}
solved by the bosonic dissipating solution:
\begin{equation}
-C\left\{F,t\right\} = E_0 e^{- \frac{c}{24 \pi C}t}.
\end{equation}

Multiplying the second equation in \eqref{eq2} by $\eta\eta'$, we obtain:
\begin{equation}
C \eta \eta' \eta''' = - \frac{c}{24 \pi} \eta \eta' \eta'' + \frac{\eta \eta'}{2} \mathcal{Q}_0 \delta(t).
\end{equation}
The second equation in \eqref{eq2} then becomes:
\begin{align}
\label{eqet}
\boxed{C \left(\eta''' + \frac{1}{2} \eta' \left\{F,t\right\} \right) = - \frac{c}{24 \pi} \eta''  + \left(1-\frac{\eta\eta'}{2}\right) \frac{\mathcal{Q}_0}{2} \delta(t) - E_0 \eta \delta(t)}.
\end{align}
Integrating this equation from $-\epsilon$ to $+\epsilon$, one finds no jump in $\eta$ or its first derivative, but a nontrivial jump in the second derivative:
\begin{align}
    \Delta \eta =0 ,\qquad 
    \Delta \eta' = 0, \qquad 
    \Delta \eta'' = \left(1-\frac{\eta(0)\eta'(0)}{2}\right) \frac{\mathcal{Q}_0}{2C} - \eta(0) E_0.
\end{align}

As an example, let's solve this differential equation for a pure supercharge injection $\mathcal{Q}_0$, where $E_0=0$. In that case, by \eqref{eqfra}, $\left\{F,t\right\} = 0$ for all times $t$. We need to solve only \eqref{eqet} which boils down to:
\begin{align}
C \eta''' = - \frac{c}{24 \pi} \eta''  + (1-\eta\eta') \mathcal{Q}_0 \delta(t).
\end{align}
One finds the solution:
\begin{align}
\eta(t) = A + Bt + D e^{- \frac{c}{24\pi C} t} \,\, = \,\, \frac{12 \pi \mathcal{Q}_0}{c} \left( t  -\frac{24 \pi C}{c}\left( 1 - e^{- \frac{c}{24 \pi C} t}\right)\right),
\end{align}
where in the second line we have implemented the jump conditions to determine the a priori three Grassmann integration constants $A$, $B$ and $D$ in terms of the single Grassmann variable $\mathcal{Q}_0$. This means in particular that the gluing conditions have forced $\eta \sim \eta' \sim \eta''$ and all terms with more than one $\eta$ to vanish automatically. \\
\indent The supercharge is then
\begin{equation}
2C \text{Sch}_f =  2C \eta'' = \mathcal{Q}_0 e^{- \frac{c}{24 \pi C} t} \theta(t),
\end{equation}
jumping to the value $\mathcal{Q}_0$ and then dissipating exponentially from the system. 

For $E_0 \neq 0$, the solution is readily generalised. One solves \eqref{eqet} into
\begin{equation}
\eta(t) = C_1 + C_2 I_0 \left( \frac{24 \pi  \sqrt{2C E_0}}{c}e^{- \frac{c }{48 \pi C}t} \right) + C_3 K_0\left( \frac{24 \pi  \sqrt{2C E_0}}{c}e^{- \frac{c }{48 \pi C}t} \right),
\end{equation}
where
\begin{subequations}
\begin{align}
C_1 &= -\frac{\mathcal{Q}_0}{E_0}, \\
C_2 &= \frac{24 \pi \mathcal{Q}_0 \sqrt{2C}}{c \sqrt{E_0}}K_1 \left(\frac{24 \pi  \sqrt{2C E_0}}{c} \right), \\
C_3 &= \frac{24 \pi \mathcal{Q}_0 \sqrt{2C}}{c \sqrt{E_0}}I_1 \left(\frac{24 \pi  \sqrt{2C E_0}}{c} \right),
\end{align}
\end{subequations}
satisfying $\eta(0)=0$, $ \eta'(0)=0$ and $\eta''(0) = \mathcal{Q}_0/2C$. This again leads to the same exponentially decaying supercharge
\begin{equation}
2C \text{Sch}_f =  2C \eta'' + C \eta \left\{f,t\right\} = \mathcal{Q}_0 e^{- \frac{c}{24 C} t} \theta(t),
\end{equation}
as it should by directly solving \eqref{superchdis}.
Notice the appearance of the same type of special functions as those appearing in the bosonic reparameterisation \eqref{sect3:reparam}.

\section{Solving the differential equation}
\label{app:DE}
A general analytic solution for the differential equation (\ref{diffeq}):
\begin{equation}
  \left\{F,t\right\} = a e^{-k t} + b e^{- 2k t},
\end{equation}
can be found when indeed restricting to the case where both exponentials decay at rates that differ by a factor of 2, and where $F(t)$ satisfying the gluing conditions 
\begin{equation}
    F(0)=0, \quad F'(0) = 1, \quad F''(0)= 0,
\end{equation}
at the infalling pulse. It consists can be written down in terms of the Whittaker-$W$ and -$M$ functions 
\begin{align}
  F(t) &= \frac{1}{\Gamma\left(\frac{1}{2} + \frac{ia}{\sqrt{8b}k}\right)^2}\int^t_0 dt \frac{e^{-kt}}{\left(\mathcal{C} M_{-\frac{ia}{\sqrt{8b}k},0}\left(i\frac{\sqrt{2b}}{k}e^{-kt}\right) + \mathcal{D} W_{-\frac{ia}{\sqrt{8b}k},0}\left(i\frac{\sqrt{2b}}{k}e^{-kt}\right)\right)^2} \nonumber \\
  &= \frac{1}{i\sqrt{2b}}\left[\frac{-W_{-\frac{ia}{\sqrt{8b}k},0}\left(i\frac{\sqrt{2b}}{k}\right) M_{-\frac{ia}{\sqrt{8b}k},0}\left(i\frac{\sqrt{2b}}{k}e^{-kt}\right) + M_{-\frac{ia}{\sqrt{8b}k},0}\left(i\frac{\sqrt{2b}}{k}\right) W_{-\frac{ia}{\sqrt{8b}k},0}\left(i\frac{\sqrt{2b}}{k}e^{-kt}\right)}{\mathcal{C} M_{-\frac{ia}{\sqrt{8b}k},0}\left(i\frac{\sqrt{2b}}{k}e^{-kt}\right) + \mathcal{D} W_{-\frac{ia}{\sqrt{8b}k},0}\left(i\frac{\sqrt{2b}}{k}e^{-kt}\right)}\right].
\end{align}
In this expression, the constants $\mathcal{C},~\mathcal{D}$ take the following form 
\begin{subequations}
\begin{align}
\label{Ceq}
  \mathcal{C} &= -W'_{-\frac{ia}{\sqrt{8b}k},0}\left(i\frac{\sqrt{2b}}{k}\right) + \frac{k}{\sqrt{8b} i}W_{-\frac{ia}{\sqrt{8b}k},0}\left(i\frac{\sqrt{2b}}{k}\right), \\
  \label{Deq}
  \mathcal{D} &= M'_{-\frac{ia}{\sqrt{8b}k},0}\left(i\frac{\sqrt{2b}}{k}\right) - \frac{k}{\sqrt{8b}i}M_{-\frac{ia}{\sqrt{8b}k},0}\left(i\frac{\sqrt{2b}}{k}\right),
\end{align}
\end{subequations}
where we have denoted the derivatives of the Whittaker function w.r.t.~their arguments by a prime.\\
\indent As a check, when setting either $a$ or $b$ to $0$ the frame should end up in a more familiar form consisting of the modified Bessel functions of the first and second kind. 
The easiest case is $a \to 0$ by making use of \cite{NIST:DLMF}
\begin{subequations}
\begin{alignat}{2}
  M_{0,0}(x) &= \sqrt{x} I_0\left(\frac{x}{2} \right), \qquad &&M'_{0,0}(x) =\frac{1}{2\sqrt{x}}I_0\left(\frac{x}{2} \right) + \frac{1}{2}\sqrt{x} I'_0\left(\frac{x}{2} \right), \\
  W_{0,0}(x) &=\frac{1}{\sqrt{\pi}}\sqrt{x} K_0\left(\frac{x}{2} \right), \qquad &&W'_{0,0}(x) =\frac{1}{2\sqrt{\pi}\sqrt{x}}K_0\left(\frac{x}{2} \right) + \frac{1}{2\sqrt{\pi}}\sqrt{x}. K'_0\left(\frac{x}{2} \right). 
\end{alignat}
\end{subequations}
\indent Just as in the uncharged case, the reparametrisation eventually reaches an asymptotic, fixed value when $t \to +\infty$
\begin{equation}
  F(t \to +\infty) = \frac{1}{i\sqrt{2b}} \frac{M_{-\frac{ia}{\sqrt{8b}k},0}\left(i\frac{\sqrt{2b}}{k}\right)}{M'_{-\frac{ia}{\sqrt{8b}k},0}\left(i\frac{\sqrt{2b}}{k}\right) - \frac{k}{\sqrt{8b}i}M_{-\frac{ia}{\sqrt{8b}k},0}\left(i\frac{\sqrt{2b}}{k}\right)},
  \label{app:horizonValue}
\end{equation}
stemming from the behaviour of the Whittaker functions for small arguments
\begin{equation}
  M_{\mu, 0}(z) \to \sqrt{z} + \mathcal{O}(z), \quad W_{\mu, 0}(z) \to - \frac{\sqrt{z} \ln z}{\Gamma(\frac{1}{2} - \mu)} + \mathcal{O}(z).
\end{equation}

\subsection{Limiting forms}
In certain cases where we take a limit of one or more of the parameters, the following limiting forms will be useful
\begin{align}
M_{i/x,0}(i a x) &\,\, \underset{x\to 0}{\to} \,\, \sqrt{i a x}\,  I_0 (2 \sqrt{a}), \\
W_{i/x,0}(i a x) &\,\, \underset{x\to 0}{\to} \,\, \frac{2}{\Gamma\left(\frac{1}{2} - \frac{i}{x}\right)}\sqrt{i a x} \, K_0 (2 \sqrt{a}).
\end{align}
Here, the first equality is derived using the Buchholz expansion of $M_{\kappa,\mu}(z)$ whereas the second one can be guessed from matching onto the correct $F(t)$ in the uncharged limit where $Q_0 \to 0$.\\
\indent Another useful limit is
\begin{align}
\nonumber
M_{i\kappa,0}(4i\kappa x) \,\, &\underset{\kappa \to +\infty}{\to} \,\, 2 \sqrt{i \kappa} \left(\frac{x \rho}{1-x} \right)^{1/4} I_0(4\kappa \sqrt{\rho}) \\
\label{Mlim}
\,\, &\underset{\kappa \to +\infty}{\to} \,\, \left(\frac{x}{1-x}\right)^{1/4} \frac{e^{4 \kappa \sqrt{\rho}+i\frac{\pi}{4}}}{\sqrt{2\pi}},
\end{align}
where 
\begin{equation}
\label{Mwhit}
2\sqrt{\rho} = \sqrt{x-x^2} - \text{arccos} \sqrt{x} + \frac{\pi}{2}.
\end{equation}
Especially when also $x \to 0$, we obtain $\rho \approx x$.
The limit for the Whittaker-$W$ function has a similar expression
\begin{equation}
W_{\kappa,0}(4\kappa x) \,\, \underset{\kappa \to +\infty}{\to} \,\, \frac{2 \Gamma\left(\frac{1}{2}+\kappa\right)}{\sqrt{2\pi}} \left(\frac{x}{1-x}\right)^{1/4} \cos \left( 4 \kappa \sqrt{\rho} - \frac{\pi}{4}\right),
\end{equation}
where $2\sqrt{\rho}$ differs from \eqref{Mwhit} by an additive term $\pi /2$
\begin{equation}
2\sqrt{\rho} = \sqrt{x-x^2} - \text{arccos} \sqrt{x}.
\end{equation}

\section{(S)CFT entanglement entropy}
\label{sect:SCFTEE}
In this section, we derive a general formula for the entanglement entropy in a 2d (S)CFT, where we include the (super)frame dependence of the result. We first present a general argument in subsection \ref{app:general} and then provide a direct proof for the specific case of $\mathcal{N}=1$ supersymmetry in subsection \ref{app:nis1e}.

\subsection{General argument}
\label{app:general}
Our main goal is to prove the following statement: 
\begin{quote}
\emph{The entanglement entropy of an interval in any given state does not depend on any compact internal symmetries characterising said state.}
\end{quote}
This is a generalisation of a statement by Cardy \cite{Cardy:talk} that the entanglement entropy in 2d CFT in the grand canonical ensemble does not depend on any chemical potentials. \\

\noindent Of course, since any 2d SCFT has a bosonic Virasoro subalgebra, the entanglement entropy of an interval with endpoints $(u_1,v_1)$ and $(u_2,v_2)$ in the vacuum state $\left|0_u\right\rangle$ defined by these coordinates, is given by the expression
\begin{equation}
S = \frac{c}{12} \ln \frac{(u_1-u_2)^2}{\delta_1\delta_2} + \,\, (u \leftrightarrow v). 
\end{equation}
Under a conformal transformation $X^+=X^+(u)$ and $X^-=X^-(v)$, one can derive the entanglement entropy in the state $\left|0_U\right\rangle$ as
\begin{equation}
S = \frac{c}{12} \ln \frac{(X^+(u_1)-X^+(u_2))^2}{\partial_u X^{+}(u_1)\partial_u X^{+}(u_2)\delta_1\delta_2} + \,\, (u \leftrightarrow v),
\end{equation}
where we crucially described the cutoff $\epsilon$ in the $(u,v)$ coordinates. However, when we have an extended superconformal algebra, we have at our disposal more general superconformal transformation to map to a larger class of states. It is this more general extension of this formula that we are after. \\
\indent An alternative but equivalent way of phrasing this is the following. The entanglement entropy is readily determined by the replica trick and twist operator insertions. For larger superconformal algebras, these twist operators are characterised by more than just their conformal weight. For e.g. $\mathcal{N}=2$, we would need to know the charge of the twist operators, or their R-symmetry representation for $\mathcal{N} >2$ superconformal algebras. We can then perform a gauge transformation and find out how the entanglement entropy depends on a change of gauge as well as a change of (bosonic) frame. Both of these are combined in a superconformal algebra in terms of a specification of superframe. \\ 

\begin{figure}[h]
\centering
\includegraphics[width=0.3\textwidth]{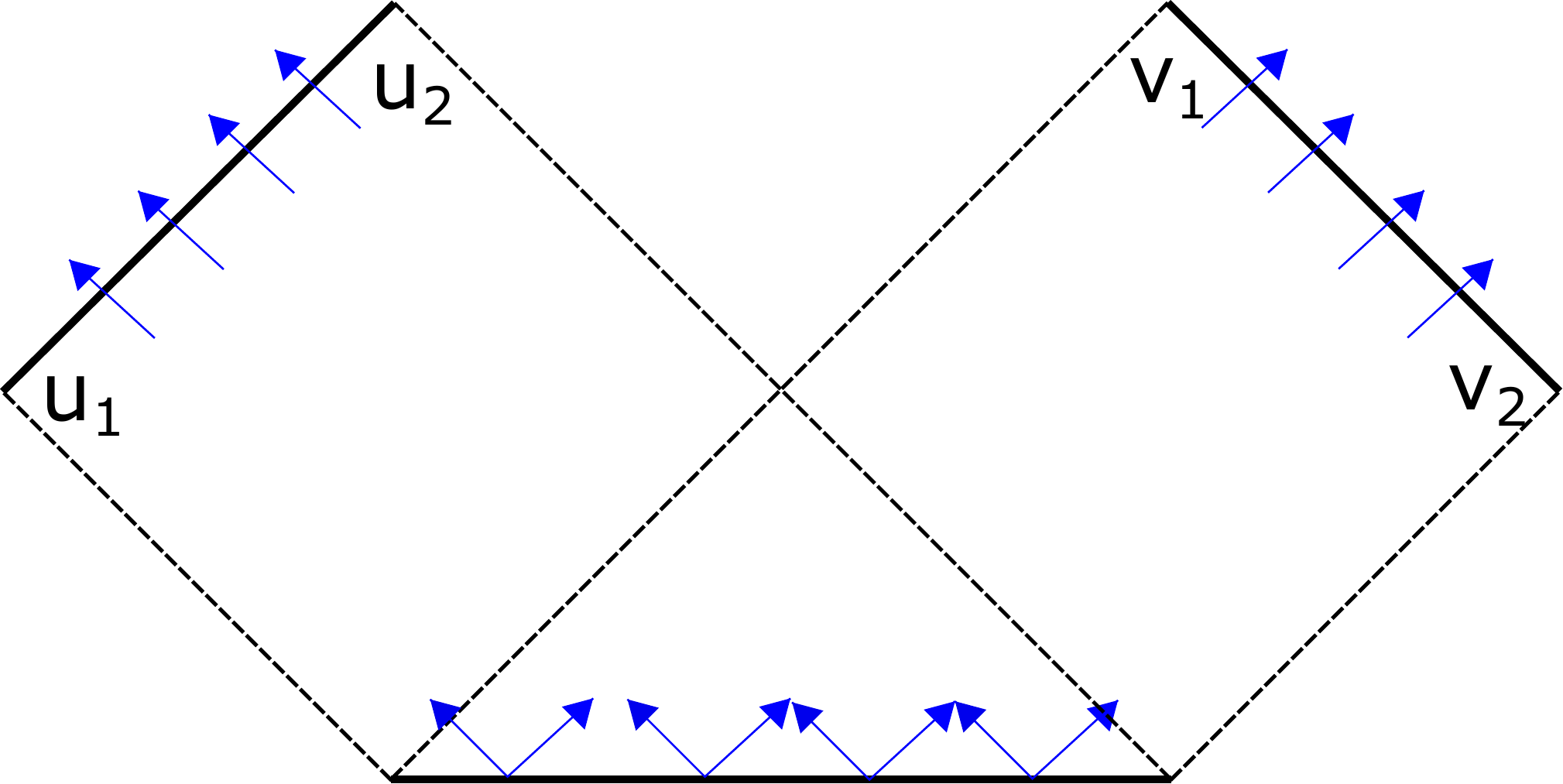}
\caption{Decomposition of entanglement entropy in 2d CFT in decoupled left- and right-moving sectors.}
\label{2dCFTS}
\end{figure}

\noindent In a bosonic 2d CFT, there are several arguments to obtain the entanglement entropy of an interval. The argument that will be directly suitable to generalisation is the following where
we focus on the left-moving sector solely in what follows, the right-moving sector being treated identically (Fig. \ref{2dCFTS}).

\indent Consider the lightcone interval $(u_1,u_2)$ in a 2d CFT. Then performing the M\"obius conformal mapping
\begin{equation}
Z = \frac{u - u_1}{u - u_2},
\end{equation}
maps the lightcone interval $(u_1,u_2)$ to the real half-line $\mathbb{R}^-$ with $u_1$ mapped to the origin and $u_2$ mapped to $-\infty$. This can in a second step be mapped to the entire axis by an exponential transform:
\begin{equation}
e^{\frac{2\pi}{\beta}w} = -Z,
\end{equation}
mapping $u_1$ to $-\infty$ and $u_2$ to $+\infty$. The final coordinate $w$ has periodicity $\beta$ for its imaginary part, implying that the $w$-frame is just a thermal system on the entire real axis. \\
\indent Regularising the left end at $z=u_1-\delta_1$, one finds the relation
\begin{equation}
\label{IRrel}
e^{-\frac{2\pi}{\beta} L_{1}} = -\frac{\delta_1}{u_1-u_2},
\end{equation}
in terms of an IR regulator $L_{1}$ in the $w$-coordinates as $w\to-\infty$. Since the $w$-system is thermal, the thermal energy and entropy in the $w$-frame are then readily computed as
\begin{equation}
E = \frac{c}{12\pi }\frac{2\pi^2}{\beta^2}(L_{1}+L_{2}), \qquad S = \frac{ c \pi}{3\beta}(L_{1}+L_{2}),
\end{equation}
where the IR cut-off $L_{1}+L_{2}$ is the total range of the $w$ coordinate. Finally, using \eqref{IRrel}, we get
\begin{equation}
\label{bosen}
S = \frac{ c \pi}{3\beta}(L_{1}+L_{2}) = \frac{c}{12} \ln \left( \frac{(F(u_1)-F(u_2))^2}{\delta_1\delta_2 F'(u_1)F'(u_2)} \right).
\end{equation}
\indent One can view this entanglement entropy between points $u_1$ and $u_2$ as that in the vacuum state annihilated by positive frequency modes w.r.t. the time coordinate $F \equiv F(t)$. The right-moving piece is simply added to it due to the complete decoupling between left- and right-moving modes. The resulting expression is SL$(2,\mathbb{R})$ invariant, reflecting the unitary equivalence of vacua related by M\"obius transformations. \\

\noindent To generalise to both charged and supersymmetric systems, one can immediately write down the supersymmetric extension as
\begin{equation}
\label{genent}
S = -\frac{c}{12} \text{Re}\left.\ln \left( \delta_1\delta_2 \left\langle O_1(u_1) O_2(u_2)\right\rangle \right) \right|_\text{bottom} + \,\, (u \leftrightarrow v), 
\end{equation}
where $O$ are chiral superconformal primaries of weight $1$, and we take the bottom component of this superspace expression. This formula is uniquely determined by the following two properties:
\begin{itemize}
\item It is invariant under the suitable global superconformal group, reflecting super-M\"obius transformations preserving the vacuum state.
\item Restricting to the bosonic subgroup of conformal transformations, we have
\begin{equation}
\left.\left\langle O_1(u_1,\theta_{u1}, \hdots) O_2(u_2,\theta_{u2}, \hdots )\right\rangle \right|_\text{body, Virasoro} =  \frac{F'(u_1)F'(u_2)}{(F(u_1) - F(u_2))^2}.
\end{equation}
\end{itemize}
\indent The presence of the real part and the bottom component arise from the relation \eqref{IRrel} relating the IR cutoff $L_{i}$  to the UV cutoff $\delta_i$. After superreparametrisation, the r.h.s. of that relation becomes generically a complex supernumber. Finding the physical IR cutoff then requires restricting to the real part of the bottom component. 

In case of nontrivial R-symmetry group, as appears starting with $\mathcal{N}= 2$ supersymmetry, the correct `body' superconformal two-point function is given by:
\begin{equation}
\left.\left\langle O_{M}{}^\alpha(u_1,\theta_{u1}, \hdots) O^\alpha{}_M(u_2,\theta_{u2}, \hdots )\right\rangle \right|_\text{body} =  \frac{F'(u_1)F'(u_2)}{(F(u_1)-F(u_2))^2}\left[g^{-1}(u_1)g(u_2)\right]_{MM},
\end{equation}
where one considers the operators transforming in a suitable unitary representation of the (compact) R-symmetry group, and we take a fixed diagonal element $MM$.\footnote{We do not need to describe this quantity in more detail since the result will follow for any choice of representation and index $M$.} This form is prescribed by the first property above. \\
\indent Taking the log of this quantity, the $g$-dependent piece is purely imaginary since it is writable as $\exp(i X_{MM})$ where $X$ is a Hermitian matrix, with hence real diagonal elements. Therefore, the contribution from the compact R-symmetry group cancels out. For the case of $\mathcal{N}=2$, the contribution from the U(1)$_R$ charge $\sigma'$ cancels out explicitly since
\begin{equation}
\label{chargecancel}
\frac{c}{6}\left[i q (\sigma_1-\sigma_2) - i q (\sigma_1-\sigma_2)\right] = 0,
\end{equation}
for any charge $q$. One can view this as a cancellation between particle and antiparticle contributions, or mathematically between complex representations and their conjugates. 

Indeed, Cardy proved in \cite{Cardy:talk} that the entanglement entropy in 2d CFT in the grand canonical ensemble does not depend on the chemical potential for any (possibly non-Abelian) charge, in particular holding for an R-symmetry group within an extended superalgebra.\footnote{Cardy also presented a quick insightful argument as follows. Take $\mathcal{Z} = \text{Tr}e^{-\beta H + \beta \mu Q}$ and perturb around $\mu=0$. The lowest correction occurs at order $\mu^2$ and leads to the thermal entropy correction:
\begin{equation}
\delta S = \mu^2 (\beta \partial_\beta -1 ) \beta^2\left\langle Q^2\right\rangle_{\mu=0}.
\end{equation}
Since in a CFT we have $\left\langle Q^2\right\rangle_{\mu=0} \sim \frac{L^{d-1}}{\beta^{d-1}}$, we finally get:
\begin{equation}
\delta S = - (d-2) \mu^2\beta^2 \frac{L^{d-1}}{\beta^{d-1}},
\end{equation}
which vanishes in $d=2$. As another explicit perspective, the total energy density in the presence of a chemical potential is \eqref{ench}. Since the second term does not depend on the temperature, it doesn't contribute to the entropy and one indeed finds the same as in the uncharged case. } The above derivation provides the generalisation beyond the grand canonical ensemble to arbitrary states, at least when embedded in superconformal algebras. \\

\noindent If we restrict to cases where our ingoing pulse carries no fermionic charge, the fermions are identically zero at all times, and the final result just boils down to the same entanglement entropy formula as in the $\mathcal{N}=0$ case:
\begin{equation}
\label{entend}
S = \frac{c}{12} \ln \left( \frac{(F(u_1)-F(u_2))^2}{\delta_1 \delta_2  F'(u_1)F'(u_2)} \right) + \,\, (u \leftrightarrow v).
\end{equation}
\indent We can show that at least for the $\mathcal{N}=1$ case, this same formula is true even when injecting fermionic charge. We do this by showing that $\eta(t) \sim \mathcal{Q}_0$ for a single Grassmann-odd variable $\mathcal{Q}_0$ in Appendix \ref{app:nis1}.\footnote{One can imagine that for extended supersymmetry, multiple Grassmann-odd parameters exist, corresponding to the different supercharge injections. In that case, fermion bilinears can be nonzero.}
However, when computing the bulk matter entanglement entropy in the full quantum gravity theory \cite{Mertens:2019bvy}, one has to integrate over the bosonic and fermionic reparametrisations. In that case, one has to incorporate the bilinear fermionic corrections in the bottom component. This is of no concern for this work but is important to point out for the eventual full story.

\indent For the particular case of $\mathcal{N}=1$ supersymmetry, we provide a more direct derivation below.

\subsection{Entanglement entropy in $\mathcal{N}=1$ 2d SCFT}
\label{app:nis1e}
We illustrate the formula \eqref{genent} for $\mathcal{N}=1$ supersymmetry by explicitly deriving it along the same lines as the bosonic formula in the beginning of this section. Consider a superspace interval $((u_1,\theta_{u_1}), \, (u_2,\theta_{u_2}))$ in a 2d SCFT, where we specify holomorphic coordinates in the form $(z,\theta)$, and where we will set $\theta_{u_1} = \theta_{u_2}$ in the end to restrict to the physical (= bosonic) interval. We start by using the $\mathcal{N}=1$ super-M\"obius mapping with reparametrisations \eqref{superMobfunc}:
\begin{equation}
F(z) = \frac{z-u_1}{z-u_2}, \qquad \eta(z) = \frac{\theta_{u_1} - \theta_{u_2}}{u_2-u_1} + \frac{u_1\theta_{u_2}-u_2\theta_{u_1}}{u_2-u_1}z,
\end{equation}
to get as in \eqref{superMob}\footnote{
This simplifies in the limit $\theta_{u_1} = \theta_{u_2}$ to:
\begin{equation}
Z = \frac{z-u_1 + \theta_{u_1} \theta}{z-u_2 + \theta_{u_1} \theta}, \qquad \Theta = \frac{-1 + \theta}{z-u_2+\theta_{u_1} \theta}.
\end{equation}
}
\begin{equation}
Z = \frac{z-u_1 + \theta_{u_1} \theta}{z-u_2 + \theta_{u_2} \theta}, \qquad \Theta = \frac{\frac{\theta_{u_1}-\theta_{u_2}}{u_2-u_1}z + \frac{u_1\theta_{u_2}-u_2\theta_{u_1}}{u_2-u_1} + \left(1+ \frac{3}{2} \frac{\theta_{u_1}\theta_{u_2}}{u_2-u_1}\right)\theta}{z-u_2+\theta_{u_2} \theta}.
\end{equation}
The new $Z$-coordinate is the ratio of two superdistances. In particular, close to the left endpoint where $z=u_1+ \delta_b, \,\, \theta = \theta_{u_1} + \delta_f$ or to the right endpoint $z = u_2 - \delta_b, \,\, \theta = \theta_{u_2} - \delta_f$, one gets:
\begin{equation}
e^{-\frac{2\pi}{\beta}L_1} = -\frac{\delta}{u_1-u_2 - \theta_{u_1} \theta_{u_2}}, \qquad e^{\frac{2\pi}{\beta}L_2} = \frac{u_2 - u_1 - \theta_{u_2} \theta_{u_1}}{\delta},
\end{equation}
where $\delta = \delta_b + \theta_{u_1} \delta_f$ or $\delta = \delta_b + \theta_{u_2} \delta_f$ is the regularisation in the original superframe $(z,\theta)$.\footnote{Note that one can set $\delta_f=0$ by choice of regulator. However, it is worthwhile keeping an open mind.} Next, we reparametrise these expressions to:
\begin{equation}
\label{sconfma}
e^{-\frac{2\pi}{\beta}L_1} = -\frac{ (D_{u_1} \theta_{u_1}')\delta}{u_1'-u_2' - \theta_{u_1}' \theta_{u_2}'}, \qquad e^{\frac{2\pi}{\beta}L_2} = \frac{u_2' -u_1' - \theta_{u_2}' \theta_{u_1}'}{(D_{u_2} \theta_{u_2}') \delta}.
\end{equation}
This corresponds to starting in another frame $(z',\theta')$ related to the original one by a superconformal transformation. The regulator $\delta \to (D_z \theta') \delta$ in the process. Indeed:
\begin{equation}
\tau_1' -\tau_2' - \theta_1'\theta_2' \underset{{\tiny \left.\begin{array}{c}
\tau_1 \to \tau_2 \\
\theta_1 \to \theta_2
 \end{array}\right.}}{\approx} (\tau_1-\tau_2 - \theta_1\theta_2) (D_\theta \theta')^2,
\end{equation}
generalising the bosonic:
\begin{equation}
\tau_1' -\tau_2' \equiv F(\tau_1) - F(\tau_2) \underset{\tau_1\to\tau_2}{\approx} (\tau_1-\tau_2) F'(\tau_2).
\end{equation}
\noindent We compute the total thermal energy and entropy in the $w$-frame as:
\begin{equation}
E = \frac{c}{12\pi }\frac{2\pi^2}{\beta^2}(L_1+L_2), \qquad S = \frac{ c \pi}{3\beta}(L_1+L_2).
\end{equation}
Finally, using \eqref{sconfma} we get the anticipated:
\begin{equation}
\boxed{S =  \frac{c}{6} \ln \left.\left( \frac{(u_1'-u_2'-\theta_{u_1}'\theta_{u_2}')^2}{\delta^2 D_{u_1} \theta_{u_1}' D_{u_2} \theta_{u_2}'} \right)\right|_{\text{bottom}}},
\end{equation}
where we need to extract the bottom component of this expression when expanding in the odd variables $\theta_{u_1}$ and $\theta_{u_2}$. 
For semiclassical evaporation, we explicitly solved for the decaying frames in Appendix \ref{app:nis1}, where in particular the fermionic superpartner is proportional to a single Grassmann variable $\eta(t) \sim \mathcal{Q}_0$. All fermion bilinears vanish because of this, and the expression reduces to \eqref{entend}.
When using a matter superfield as the agent of Hawking radiation, the only difference is then the shift in central charge $c=1 \to c=3/2$ by including the emission channel from the fermion. 

\mciteSetMidEndSepPunct{}{\ifmciteBstWouldAddEndPunct.\else\fi}{\relax}
\bibliographystyle{utphys}
{\small \bibliography{ReferencesJulian}{}}

\providecommand{\href}[2]{#2}\begingroup\raggedright\begin{thebibliography}{100}

\bibitem{Hawking1971}
S.~W. Hawking, ``Gravitational radiation from colliding black holes,''
  \href{http://dx.doi.org/10.1103/PhysRevLett.26.1344}{{\em Phys. Rev. Lett.}
  {\bfseries 26} (May, 1971) 1344--1346}.
  \url{https://link.aps.org/doi/10.1103/PhysRevLett.26.1344}.

\bibitem{Hawking1975}
S.~W. {Hawking}, ``Particle creation by black holes,'' {\em Comun. math. Phys
  43} {\bfseries 199-220} (12 april, 1975) .

\bibitem{Hawking1976}
S.~W. Hawking, ``Breakdown of predictability in gravitational collapse,''
  \href{http://dx.doi.org/10.1103/PhysRevD.14.2460}{{\em Phys. Rev. D}
  {\bfseries 14} (Nov, 1976) 2460--2473}.
  \url{https://link.aps.org/doi/10.1103/PhysRevD.14.2460}.

\bibitem{Maldacena1999}
J.~M. Maldacena, ``{The Large N limit of superconformal field theories and
  supergravity},'' \href{http://dx.doi.org/10.1023/A:1026654312961}{{\em Adv.
  Theor. Math. Phys.} {\bfseries 2} (1998) 231--252},
  \href{http://arxiv.org/abs/hep-th/9711200}{{\ttfamily arXiv:hep-th/9711200}}.

\bibitem{Almheiri2020a}
A.~{Almheiri}, R.~{Mahajan}, J.~{Maldacena}, and Y.~{Zhao}, ``{The Page curve
  of Hawking radiation from semiclassical geometry},''
  \href{http://dx.doi.org/10.1007/JHEP03(2020)149}{{\em Journal of High Energy
  Physics} {\bfseries 2020} no.~3, (Mar., 2020) 149},
  \href{http://arxiv.org/abs/1908.10996}{{\ttfamily arXiv:1908.10996
  [hep-th]}}.

\bibitem{Almheiri2020c}
A.~Almheiri, T.~Hartman, J.~Maldacena, E.~Shaghoulian, and A.~Tajdini, ``{The
  entropy of Hawking radiation},''
  \href{http://dx.doi.org/10.1103/RevModPhys.93.035002}{{\em Rev. Mod. Phys.}
  {\bfseries 93} no.~3, (2021) 035002},
  \href{http://arxiv.org/abs/2006.06872}{{\ttfamily arXiv:2006.06872
  [hep-th]}}.

\bibitem{Penington2019}
G.~Penington, S.~H. Shenker, D.~Stanford, and Z.~Yang, ``{Replica wormholes and
  the black hole interior},''
  \href{http://dx.doi.org/10.1007/JHEP03(2022)205}{{\em JHEP} {\bfseries 03}
  (2022) 205}, \href{http://arxiv.org/abs/1911.11977}{{\ttfamily
  arXiv:1911.11977 [hep-th]}}.

\bibitem{Almheiri2020}
A.~Almheiri, T.~Hartman, J.~Maldacena, E.~Shaghoulian, and A.~Tajdini,
  ``{Replica Wormholes and the Entropy of Hawking Radiation},''
  \href{http://dx.doi.org/10.1007/JHEP05(2020)013}{{\em JHEP} {\bfseries 05}
  (2020) 013}, \href{http://arxiv.org/abs/1911.12333}{{\ttfamily
  arXiv:1911.12333 [hep-th]}}.

\bibitem{Engelhardt2021}
N.~Engelhardt, S.~Fischetti, and A.~Maloney, ``{Free energy from replica
  wormholes},'' \href{http://dx.doi.org/10.1103/PhysRevD.103.046021}{{\em Phys.
  Rev. D} {\bfseries 103} no.~4, (2021) 046021},
  \href{http://arxiv.org/abs/2007.07444}{{\ttfamily arXiv:2007.07444
  [hep-th]}}.

\bibitem{Goto2020}
K.~Goto, T.~Hartman, and A.~Tajdini, ``{Replica wormholes for an evaporating 2D
  black hole},'' \href{http://dx.doi.org/10.1007/JHEP04(2021)289}{{\em JHEP}
  {\bfseries 04} (2021) 289}, \href{http://arxiv.org/abs/2011.09043}{{\ttfamily
  arXiv:2011.09043 [hep-th]}}.

\bibitem{Page1993a}
D.~N. Page, ``{Information in black hole radiation},''
  \href{http://dx.doi.org/10.1103/PhysRevLett.71.3743}{{\em Phys. Rev. Lett.}
  {\bfseries 71} (1993) 3743--3746},
  \href{http://arxiv.org/abs/hep-th/9306083}{{\ttfamily arXiv:hep-th/9306083}}.

\bibitem{Page2013}
D.~N. Page, ``{Time Dependence of Hawking Radiation Entropy},''
  \href{http://dx.doi.org/10.1088/1475-7516/2013/09/028}{{\em JCAP} {\bfseries
  09} (2013) 028}, \href{http://arxiv.org/abs/1301.4995}{{\ttfamily
  arXiv:1301.4995 [hep-th]}}.

\bibitem{Almheiri2013}
A.~Almheiri, D.~Marolf, J.~Polchinski, and J.~Sully, ``{Black Holes:
  Complementarity or Firewalls?},''
  \href{http://dx.doi.org/10.1007/JHEP02(2013)062}{{\em JHEP} {\bfseries 02}
  (2013) 062}, \href{http://arxiv.org/abs/1207.3123}{{\ttfamily arXiv:1207.3123
  [hep-th]}}.

\bibitem{Czech2012}
B.~Czech, J.~L. Karczmarek, F.~Nogueira, and M.~Van~Raamsdonk, ``{The Gravity
  Dual of a Density Matrix},''
  \href{http://dx.doi.org/10.1088/0264-9381/29/15/155009}{{\em Class. Quant.
  Grav.} {\bfseries 29} (2012) 155009},
  \href{http://arxiv.org/abs/1204.1330}{{\ttfamily arXiv:1204.1330 [hep-th]}}.

\bibitem{Wall2014}
A.~C. Wall, ``{Maximin Surfaces, and the Strong Subadditivity of the Covariant
  Holographic Entanglement Entropy},''
  \href{http://dx.doi.org/10.1088/0264-9381/31/22/225007}{{\em Class. Quant.
  Grav.} {\bfseries 31} no.~22, (2014) 225007},
  \href{http://arxiv.org/abs/1211.3494}{{\ttfamily arXiv:1211.3494 [hep-th]}}.

\bibitem{Headrick2014a}
M.~Headrick, V.~E. Hubeny, A.~Lawrence, and M.~Rangamani, ``{Causality \&
  holographic entanglement entropy},''
  \href{http://dx.doi.org/10.1007/JHEP12(2014)162}{{\em JHEP} {\bfseries 12}
  (2014) 162}, \href{http://arxiv.org/abs/1408.6300}{{\ttfamily arXiv:1408.6300
  [hep-th]}}.

\bibitem{Hayden2007}
P.~Hayden and J.~Preskill, ``{Black holes as mirrors: Quantum information in
  random subsystems},''
  \href{http://dx.doi.org/10.1088/1126-6708/2007/09/120}{{\em JHEP} {\bfseries
  09} (2007) 120}, \href{http://arxiv.org/abs/0708.4025}{{\ttfamily
  arXiv:0708.4025 [hep-th]}}.

\bibitem{Sekino2008}
Y.~Sekino and L.~Susskind, ``{Fast Scramblers},''
  \href{http://dx.doi.org/10.1088/1126-6708/2008/10/065}{{\em JHEP} {\bfseries
  10} (2008) 065}, \href{http://arxiv.org/abs/0808.2096}{{\ttfamily
  arXiv:0808.2096 [hep-th]}}.

\bibitem{Maldacena2013}
J.~Maldacena and L.~Susskind, ``{Cool horizons for entangled black holes},''
  \href{http://dx.doi.org/10.1002/prop.201300020}{{\em Fortsch. Phys.}
  {\bfseries 61} (2013) 781--811},
  \href{http://arxiv.org/abs/1306.0533}{{\ttfamily arXiv:1306.0533 [hep-th]}}.

\bibitem{vanRaamsdonk2010}
M.~Van~Raamsdonk, ``{Building up spacetime with quantum entanglement},''
  \href{http://dx.doi.org/10.1142/S0218271810018529}{{\em Gen. Rel. Grav.}
  {\bfseries 42} (2010) 2323--2329},
  \href{http://arxiv.org/abs/1005.3035}{{\ttfamily arXiv:1005.3035 [hep-th]}}.

\bibitem{Engelhardt2015}
N.~Engelhardt and A.~C. Wall, ``{Quantum Extremal Surfaces: Holographic
  Entanglement Entropy beyond the Classical Regime},''
  \href{http://dx.doi.org/10.1007/JHEP01(2015)073}{{\em JHEP} {\bfseries 01}
  (2015) 073}, \href{http://arxiv.org/abs/1408.3203}{{\ttfamily arXiv:1408.3203
  [hep-th]}}.

\bibitem{Takayanagi2017}
M.~Rangamani and T.~Takayanagi,
  \href{http://dx.doi.org/10.1007/978-3-319-52573-0}{{\em {Holographic
  Entanglement Entropy}}}.
\newblock Springer International Publishing, 1~ed., 26 jan, 2017.

\bibitem{Hubeny2007}
V.~E. Hubeny, M.~Rangamani, and T.~Takayanagi, ``{A Covariant holographic
  entanglement entropy proposal},''
  \href{http://dx.doi.org/10.1088/1126-6708/2007/07/062}{{\em JHEP} {\bfseries
  07} (2007) 062}, \href{http://arxiv.org/abs/0705.0016}{{\ttfamily
  arXiv:0705.0016 [hep-th]}}.

\bibitem{Jafferis:2015del}
D.~L. Jafferis, A.~Lewkowycz, J.~Maldacena, and S.~J. Suh, ``{Relative entropy
  equals bulk relative entropy},''
  \href{http://dx.doi.org/10.1007/JHEP06(2016)004}{{\em JHEP} {\bfseries 06}
  (2016) 004}, \href{http://arxiv.org/abs/1512.06431}{{\ttfamily
  arXiv:1512.06431 [hep-th]}}.

\bibitem{Bekenstein1972}
J.~D. {Bekenstein}, ``Black holes and the second law,''
  \href{http://dx.doi.org/https://doi.org/10.1007/BF02757029}{{\em Lettere al
  Nuovo Cimento (1971-1985)} {\bfseries 4, 737-740} (1972) }.

\bibitem{Bekenstein1973}
J.~D. Bekenstein, ``{Black holes and entropy},''
  \href{http://dx.doi.org/10.1103/PhysRevD.7.2333}{{\em Phys. Rev. D}
  {\bfseries 7} (1973) 2333--2346}.

\bibitem{Holzhey1994}
C.~Holzhey, F.~Larsen, and F.~Wilczek, ``{Geometric and renormalized entropy in
  conformal field theory},''
  \href{http://dx.doi.org/10.1016/0550-3213(94)90402-2}{{\em Nucl. Phys. B}
  {\bfseries 424} (1994) 443--467},
  \href{http://arxiv.org/abs/hep-th/9403108}{{\ttfamily arXiv:hep-th/9403108}}.

\bibitem{Fiola1994}
T.~M. Fiola, J.~Preskill, A.~Strominger, and S.~P. Trivedi, ``{Black hole
  thermodynamics and information loss in two-dimensions},''
  \href{http://dx.doi.org/10.1103/PhysRevD.50.3987}{{\em Phys. Rev. D}
  {\bfseries 50} (1994) 3987--4014},
  \href{http://arxiv.org/abs/hep-th/9403137}{{\ttfamily arXiv:hep-th/9403137}}.

\bibitem{Almheiri:2013wka}
A.~Almheiri and J.~Sully, ``{An Uneventful Horizon in Two Dimensions},''
  \href{http://dx.doi.org/10.1007/JHEP02(2014)108}{{\em JHEP} {\bfseries 02}
  (2014) 108}, \href{http://arxiv.org/abs/1307.8149}{{\ttfamily arXiv:1307.8149
  [hep-th]}}.

\bibitem{Mertens:2019bvy}
T.~G. Mertens, ``{Towards Black Hole Evaporation in Jackiw-Teitelboim
  Gravity},'' \href{http://dx.doi.org/10.1007/JHEP07(2019)097}{{\em JHEP}
  {\bfseries 07} (2019) 097}, \href{http://arxiv.org/abs/1903.10485}{{\ttfamily
  arXiv:1903.10485 [hep-th]}}.

\bibitem{Almheiri2019}
A.~Almheiri, N.~Engelhardt, D.~Marolf, and H.~Maxfield, ``{The entropy of bulk
  quantum fields and the entanglement wedge of an evaporating black hole},''
  \href{http://dx.doi.org/10.1007/JHEP12(2019)063}{{\em JHEP} {\bfseries 12}
  (2019) 063}, \href{http://arxiv.org/abs/1905.08762}{{\ttfamily
  arXiv:1905.08762 [hep-th]}}.

\bibitem{Almheiri2019a}
A.~Almheiri, R.~Mahajan, and J.~Maldacena, ``{Islands outside the horizon},''
  \href{http://arxiv.org/abs/1910.11077}{{\ttfamily arXiv:1910.11077
  [hep-th]}}.

\bibitem{Geng:2020fxl}
H.~Geng, A.~Karch, C.~Perez-Pardavila, S.~Raju, L.~Randall, M.~Riojas, and
  S.~Shashi, ``{Information Transfer with a Gravitating Bath},''
  \href{http://dx.doi.org/10.21468/SciPostPhys.10.5.103}{{\em SciPost Phys.}
  {\bfseries 10} no.~5, (2021) 103},
  \href{http://arxiv.org/abs/2012.04671}{{\ttfamily arXiv:2012.04671
  [hep-th]}}.

\bibitem{Chen2020}
H.~Z. Chen, Z.~Fisher, J.~Hernandez, R.~C. Myers, and S.-M. Ruan,
  ``{Information Flow in Black Hole Evaporation},''
  \href{http://dx.doi.org/10.1007/JHEP03(2020)152}{{\em JHEP} {\bfseries 03}
  (2020) 152}, \href{http://arxiv.org/abs/1911.03402}{{\ttfamily
  arXiv:1911.03402 [hep-th]}}.

\bibitem{ZheChen2020}
H.~Z. Chen, Z.~Fisher, J.~Hernandez, R.~C. Myers, and S.-M. Ruan,
  ``{Evaporating Black Holes Coupled to a Thermal Bath},''
  \href{http://dx.doi.org/10.1007/JHEP01(2021)065}{{\em JHEP} {\bfseries 01}
  (2021) 065}, \href{http://arxiv.org/abs/2007.11658}{{\ttfamily
  arXiv:2007.11658 [hep-th]}}.

\bibitem{Hollowood2020}
T.~J. Hollowood and S.~P. Kumar, ``{Islands and Page Curves for Evaporating
  Black Holes in JT Gravity},''
  \href{http://dx.doi.org/10.1007/JHEP08(2020)094}{{\em JHEP} {\bfseries 08}
  (2020) 094}, \href{http://arxiv.org/abs/2004.14944}{{\ttfamily
  arXiv:2004.14944 [hep-th]}}.

\bibitem{Hollowood2020a}
T.~J. Hollowood, S.~Prem~Kumar, and A.~Legramandi, ``{Hawking radiation
  correlations of evaporating black holes in JT gravity},''
  \href{http://dx.doi.org/10.1088/1751-8121/abbc51}{{\em J. Phys. A} {\bfseries
  53} no.~47, (2020) 475401}, \href{http://arxiv.org/abs/2007.04877}{{\ttfamily
  arXiv:2007.04877 [hep-th]}}.

\bibitem{Geng:2021hlu}
H.~Geng, A.~Karch, C.~Perez-Pardavila, S.~Raju, L.~Randall, M.~Riojas, and
  S.~Shashi, ``{Inconsistency of islands in theories with long-range
  gravity},'' \href{http://dx.doi.org/10.1007/JHEP01(2022)182}{{\em JHEP}
  {\bfseries 01} (2022) 182}, \href{http://arxiv.org/abs/2107.03390}{{\ttfamily
  arXiv:2107.03390 [hep-th]}}.

\bibitem{Alishahiha:2020qza}
M.~Alishahiha, A.~Faraji~Astaneh, and A.~Naseh, ``{Island in the presence of
  higher derivative terms},''
  \href{http://dx.doi.org/10.1007/JHEP02(2021)035}{{\em JHEP} {\bfseries 02}
  (2021) 035}, \href{http://arxiv.org/abs/2005.08715}{{\ttfamily
  arXiv:2005.08715 [hep-th]}}.

\bibitem{Azarnia:2021uch}
S.~Azarnia, R.~Fareghbal, A.~Naseh, and H.~Zolfi, ``{Islands in flat-space
  cosmology},'' \href{http://dx.doi.org/10.1103/PhysRevD.104.126017}{{\em Phys.
  Rev. D} {\bfseries 104} no.~12, (2021) },
  \href{http://arxiv.org/abs/2109.04795}{{\ttfamily arXiv:2109.04795
  [hep-th]}}.

\bibitem{Omidi:2021opl}
F.~Omidi, ``{Entropy of Hawking radiation for two-sided hyperscaling violating
  black branes},'' \href{http://dx.doi.org/10.1007/JHEP04(2022)022}{{\em JHEP}
  {\bfseries 04} (2022) 022}, \href{http://arxiv.org/abs/2112.05890}{{\ttfamily
  arXiv:2112.05890 [hep-th]}}.

\bibitem{Anderson2021}
L.~Anderson, O.~Parrikar, and R.~M. Soni, ``{Islands with gravitating baths:
  towards ER = EPR},'' \href{http://dx.doi.org/10.1007/JHEP10(2021)226}{{\em
  JHEP} {\bfseries 10} (2021) 226},
  \href{http://arxiv.org/abs/2103.14746}{{\ttfamily arXiv:2103.14746
  [hep-th]}}.

\bibitem{Engelsoey2016}
J.~Engels\"oy, T.~G. Mertens, and H.~Verlinde, ``{An investigation of AdS$_{2}$
  backreaction and holography},''
  \href{http://dx.doi.org/10.1007/JHEP07(2016)139}{{\em JHEP} {\bfseries 07}
  (2016) 139}, \href{http://arxiv.org/abs/1606.03438}{{\ttfamily
  arXiv:1606.03438 [hep-th]}}.

\bibitem{jackiw}
R.~Jackiw, ``Lower dimensional gravity,''
  \href{http://dx.doi.org/https://doi.org/10.1016/0550-3213(85)90448-1}{{\em
  Nuclear Physics B} {\bfseries 252} (1985) 343 -- 356}.
  \url{http://www.sciencedirect.com/science/article/pii/0550321385904481}.

\bibitem{teitelboim}
C.~Teitelboim, ``Gravitation and hamiltonian structure in two space-time
  dimensions,'' {\em Physical Review Letters D \textbf{126} (1983) 41} .

\bibitem{Almheiri2015}
A.~Almheiri and J.~Polchinski, ``{Models of AdS$_{2}$ backreaction and
  holography},'' \href{http://dx.doi.org/10.1007/JHEP11(2015)014}{{\em JHEP}
  {\bfseries 11} (2015) 014}, \href{http://arxiv.org/abs/1402.6334}{{\ttfamily
  arXiv:1402.6334 [hep-th]}}.

\bibitem{Jensen2016}
K.~Jensen, ``{Chaos in AdS$_2$ Holography},''
  \href{http://dx.doi.org/10.1103/PhysRevLett.117.111601}{{\em Phys. Rev.
  Lett.} {\bfseries 117} no.~11, (2016) 111601},
  \href{http://arxiv.org/abs/1605.06098}{{\ttfamily arXiv:1605.06098
  [hep-th]}}.

\bibitem{Maldacena2016a}
J.~Maldacena, D.~Stanford, and Z.~Yang, ``{Conformal symmetry and its breaking
  in two dimensional Nearly Anti-de-Sitter space},''
  \href{http://dx.doi.org/10.1093/ptep/ptw124}{{\em PTEP} {\bfseries 2016}
  no.~12, (2016) 12C104}, \href{http://arxiv.org/abs/1606.01857}{{\ttfamily
  arXiv:1606.01857 [hep-th]}}.

\bibitem{Kitaev:2017awl}
A.~Kitaev and S.~J. Suh, ``{The soft mode in the Sachdev-Ye-Kitaev model and
  its gravity dual},'' \href{http://dx.doi.org/10.1007/JHEP05(2018)183}{{\em
  JHEP} {\bfseries 05} (2018) 183},
  \href{http://arxiv.org/abs/1711.08467}{{\ttfamily arXiv:1711.08467
  [hep-th]}}.

\bibitem{Blommaert2019}
A.~Blommaert, T.~G. Mertens, and H.~Verschelde, ``{Clocks and Rods in
  Jackiw-Teitelboim Quantum Gravity},''
  \href{http://dx.doi.org/10.1007/JHEP09(2019)060}{{\em JHEP} {\bfseries 09}
  (2019) 060}, \href{http://arxiv.org/abs/1902.11194}{{\ttfamily
  arXiv:1902.11194 [hep-th]}}.

\bibitem{Davies:1976ei}
P.~C.~W. Davies, S.~A. Fulling, and W.~G. Unruh, ``{Energy Momentum Tensor Near
  an Evaporating Black Hole},''
  \href{http://dx.doi.org/10.1103/PhysRevD.13.2720}{{\em Phys. Rev. D}
  {\bfseries 13} (1976) 2720--2723}.

\bibitem{Christensen:1977jc}
S.~M. Christensen and S.~A. Fulling, ``{Trace Anomalies and the Hawking
  Effect},'' \href{http://dx.doi.org/10.1103/PhysRevD.15.2088}{{\em Phys. Rev.
  D} {\bfseries 15} (1977) 2088--2104}.

\bibitem{Fabbri2005}
A.~Fabbri and J.~Navarro-Salas, {\em {Modeling black hole evaporation}}.
\newblock Imperial College Press, 26 jan, 2005.

\bibitem{Spradlin:1999bn}
M.~Spradlin and A.~Strominger, ``{Vacuum states for AdS(2) black holes},''
  \href{http://dx.doi.org/10.1088/1126-6708/1999/11/021}{{\em JHEP} {\bfseries
  11} (1999) 021}, \href{http://arxiv.org/abs/hep-th/9904143}{{\ttfamily
  arXiv:hep-th/9904143}}.

\bibitem{NIST:DLMF}
F.~W.~J. Olver, A.~B. Olde~Daalhuis, D.~W. Lozier, B.~I. Schneider, R.~F.
  Boisvert, C.~W. Clark, B.~R. Miller, H.~S. Saunders, B.~V. Sand~Cohl, M.~A.
  McClain, and eds., ``{\it NIST Digital Library of Mathematical Functions}.''
  Release 1.1.6 of 2022-06-30.
\newblock \url{http://dlmf.nist.gov/}.

\bibitem{Affleck1994}
I.~{Affleck} and A.~W.~W. {Ludwig}, ``{The Fermi edge singularity and boundary
  condition changing operators},''
  \href{http://dx.doi.org/10.1088/0305-4470/27/16/007}{{\em Journal of Physics
  A Mathematical General} {\bfseries 27} no.~16, (Aug., 1994) 5375--5392},
  \href{http://arxiv.org/abs/cond-mat/9405057}{{\ttfamily
  arXiv:cond-mat/9405057 [cond-mat]}}.

\bibitem{Bousso2016}
R.~Bousso, Z.~Fisher, S.~Leichenauer, and A.~C. Wall, ``{Quantum focusing
  conjecture},'' \href{http://dx.doi.org/10.1103/PhysRevD.93.064044}{{\em Phys.
  Rev. D} {\bfseries 93} no.~6, (2016) 064044},
  \href{http://arxiv.org/abs/1506.02669}{{\ttfamily arXiv:1506.02669
  [hep-th]}}.

\bibitem{Chen2020a}
H.~Z. Chen, R.~C. Myers, D.~Neuenfeld, I.~A. Reyes, and J.~Sandor, ``{Quantum
  Extremal Islands Made Easy, Part II: Black Holes on the Brane},''
  \href{http://dx.doi.org/10.1007/JHEP12(2020)025}{{\em JHEP} {\bfseries 12}
  (2020) 025}, \href{http://arxiv.org/abs/2010.00018}{{\ttfamily
  arXiv:2010.00018 [hep-th]}}.

\bibitem{Rozali2020}
M.~Rozali, J.~Sully, M.~Van~Raamsdonk, C.~Waddell, and D.~Wakeham,
  ``{Information radiation in BCFT models of black holes},''
  \href{http://dx.doi.org/10.1007/JHEP05(2020)004}{{\em JHEP} {\bfseries 05}
  (2020) 004}, \href{http://arxiv.org/abs/1910.12836}{{\ttfamily
  arXiv:1910.12836 [hep-th]}}.

\bibitem{Gautason2020}
F.~F. Gautason, L.~Schneiderbauer, W.~Sybesma, and L.~Thorlacius, ``{Page Curve
  for an Evaporating Black Hole},''
  \href{http://dx.doi.org/10.1007/JHEP05(2020)091}{{\em JHEP} {\bfseries 05}
  (2020) 091}, \href{http://arxiv.org/abs/2004.00598}{{\ttfamily
  arXiv:2004.00598 [hep-th]}}.

\bibitem{Hartman2020}
T.~Hartman, E.~Shaghoulian, and A.~Strominger, ``{Islands in Asymptotically
  Flat 2D Gravity},'' \href{http://dx.doi.org/10.1007/JHEP07(2020)022}{{\em
  JHEP} {\bfseries 07} (2020) 022},
  \href{http://arxiv.org/abs/2004.13857}{{\ttfamily arXiv:2004.13857
  [hep-th]}}.

\bibitem{Hashimoto2020}
K.~Hashimoto, N.~Iizuka, and Y.~Matsuo, ``{Islands in Schwarzschild black
  holes},'' \href{http://dx.doi.org/10.1007/JHEP06(2020)085}{{\em JHEP}
  {\bfseries 06} (2020) 085}, \href{http://arxiv.org/abs/2004.05863}{{\ttfamily
  arXiv:2004.05863 [hep-th]}}.

\bibitem{Almheiri:2019psy}
A.~Almheiri, R.~Mahajan, and J.~E. Santos, ``{Entanglement islands in higher
  dimensions},'' \href{http://dx.doi.org/10.21468/SciPostPhys.9.1.001}{{\em
  SciPost Phys.} {\bfseries 9} no.~1, (2020) 001},
  \href{http://arxiv.org/abs/1911.09666}{{\ttfamily arXiv:1911.09666
  [hep-th]}}.

\bibitem{Geng:2020qvw}
H.~Geng and A.~Karch, ``{Massive islands},''
  \href{http://dx.doi.org/10.1007/JHEP09(2020)121}{{\em JHEP} {\bfseries 09}
  (2020) 121}, \href{http://arxiv.org/abs/2006.02438}{{\ttfamily
  arXiv:2006.02438 [hep-th]}}.

\bibitem{Susskind1993}
L.~Susskind, L.~Thorlacius, and J.~Uglum, ``{The Stretched horizon and black
  hole complementarity},''
  \href{http://dx.doi.org/10.1103/PhysRevD.48.3743}{{\em Phys. Rev. D}
  {\bfseries 48} (1993) 3743--3761},
  \href{http://arxiv.org/abs/hep-th/9306069}{{\ttfamily arXiv:hep-th/9306069}}.

\bibitem{Headrick2014}
M.~Headrick, ``{General properties of holographic entanglement entropy},''
  \href{http://dx.doi.org/10.1007/JHEP03(2014)085}{{\em JHEP} {\bfseries 03}
  (2014) 085}, \href{http://arxiv.org/abs/1312.6717}{{\ttfamily arXiv:1312.6717
  [hep-th]}}.

\bibitem{Chen2020b}
H.~Z. Chen, R.~C. Myers, D.~Neuenfeld, I.~A. Reyes, and J.~Sandor, ``{Quantum
  Extremal Islands Made Easy, Part I: Entanglement on the Brane},''
  \href{http://dx.doi.org/10.1007/JHEP10(2020)166}{{\em JHEP} {\bfseries 10}
  (2020) 166}, \href{http://arxiv.org/abs/2006.04851}{{\ttfamily
  arXiv:2006.04851 [hep-th]}}.

\bibitem{Almheiri2018}
A.~Almheiri, ``{Holographic Quantum Error Correction and the Projected Black
  Hole Interior},'' \href{http://arxiv.org/abs/1810.02055}{{\ttfamily
  arXiv:1810.02055 [hep-th]}}.

\bibitem{Callebaut2019}
N.~Callebaut, ``{The gravitational dynamics of kinematic space},''
  \href{http://dx.doi.org/10.1007/JHEP02(2019)153}{{\em JHEP} {\bfseries 02}
  (2019) 153}, \href{http://arxiv.org/abs/1808.10431}{{\ttfamily
  arXiv:1808.10431 [hep-th]}}.

\bibitem{Callebaut:2018nlq}
N.~Callebaut and H.~Verlinde, ``{Entanglement Dynamics in 2D CFT with Boundary:
  Entropic origin of JT gravity and Schwarzian QM},''
  \href{http://dx.doi.org/10.1007/JHEP05(2019)045}{{\em JHEP} {\bfseries 05}
  (2019) 045}, \href{http://arxiv.org/abs/1808.05583}{{\ttfamily
  arXiv:1808.05583 [hep-th]}}.

\bibitem{Zurek1982}
W.~H. Zurek, ``Entropy evaporated by a black hole,''
  \href{http://dx.doi.org/10.1103/PhysRevLett.49.1683}{{\em Phys. Rev. Lett.}
  {\bfseries 49} (Dec, 1982) 1683--1686}.
  \url{https://link.aps.org/doi/10.1103/PhysRevLett.49.1683}.

\bibitem{Moitra2020}
U.~Moitra, S.~K. Sake, S.~P. Trivedi, and V.~Vishal, ``{Jackiw-Teitelboim Model
  Coupled to Conformal Matter in the Semi-Classical Limit},''
  \href{http://dx.doi.org/10.1007/JHEP04(2020)199}{{\em JHEP} {\bfseries 04}
  (2020) 199}, \href{http://arxiv.org/abs/1908.08523}{{\ttfamily
  arXiv:1908.08523 [hep-th]}}.

\bibitem{Davison:2016ngz}
R.~A. Davison, W.~Fu, A.~Georges, Y.~Gu, K.~Jensen, and S.~Sachdev,
  ``{Thermoelectric transport in disordered metals without quasiparticles: The
  Sachdev-Ye-Kitaev models and holography},''
  \href{http://dx.doi.org/10.1103/PhysRevB.95.155131}{{\em Phys. Rev. B}
  {\bfseries 95} no.~15, (2017) 155131},
  \href{http://arxiv.org/abs/1612.00849}{{\ttfamily arXiv:1612.00849
  [cond-mat.str-el]}}.

\bibitem{Gu:2019jub}
Y.~Gu, A.~Kitaev, S.~Sachdev, and G.~Tarnopolsky, ``{Notes on the complex
  Sachdev-Ye-Kitaev model},''
  \href{http://dx.doi.org/10.1007/JHEP02(2020)157}{{\em JHEP} {\bfseries 02}
  (2020) 157}, \href{http://arxiv.org/abs/1910.14099}{{\ttfamily
  arXiv:1910.14099 [hep-th]}}.

\bibitem{Kapec:2019ecr}
D.~Kapec, R.~Mahajan, and D.~Stanford, ``{Matrix ensembles with global
  symmetries and \textquoteright{}t Hooft anomalies from 2d gauge theory},''
  \href{http://dx.doi.org/10.1007/JHEP04(2020)186}{{\em JHEP} {\bfseries 04}
  (2020) 186}, \href{http://arxiv.org/abs/1912.12285}{{\ttfamily
  arXiv:1912.12285 [hep-th]}}.

\bibitem{Iliesiu:2020qvm}
L.~V. Iliesiu and G.~J. Turiaci, ``{The statistical mechanics of near-extremal
  black holes},'' \href{http://dx.doi.org/10.1007/JHEP05(2021)145}{{\em JHEP}
  {\bfseries 05} (2021) 145}, \href{http://arxiv.org/abs/2003.02860}{{\ttfamily
  arXiv:2003.02860 [hep-th]}}.

\bibitem{Chaturvedi:2018uov}
P.~Chaturvedi, Y.~Gu, W.~Song, and B.~Yu, ``{A note on the complex SYK model
  and warped CFTs},'' \href{http://dx.doi.org/10.1007/JHEP12(2018)101}{{\em
  JHEP} {\bfseries 12} (2018) 101},
  \href{http://arxiv.org/abs/1808.08062}{{\ttfamily arXiv:1808.08062
  [hep-th]}}.

\bibitem{Mertens:2019tcm}
T.~G. Mertens and G.~J. Turiaci, ``{Defects in Jackiw-Teitelboim Quantum
  Gravity},'' \href{http://dx.doi.org/10.1007/JHEP08(2019)127}{{\em JHEP}
  {\bfseries 08} (2019) 127}, \href{http://arxiv.org/abs/1904.05228}{{\ttfamily
  arXiv:1904.05228 [hep-th]}}.

\bibitem{Donnelly:2015hta}
W.~Donnelly and S.~B. Giddings, ``{Diffeomorphism-invariant observables and
  their nonlocal algebra},''
  \href{http://dx.doi.org/10.1103/PhysRevD.93.024030}{{\em Phys. Rev. D}
  {\bfseries 93} no.~2, (2016) 024030},
  \href{http://arxiv.org/abs/1507.07921}{{\ttfamily arXiv:1507.07921
  [hep-th]}}. [Erratum: Phys.Rev.D 94, 029903 (2016)].

\bibitem{Donnelly:2016rvo}
W.~Donnelly and S.~B. Giddings, ``{Observables, gravitational dressing, and
  obstructions to locality and subsystems},''
  \href{http://dx.doi.org/10.1103/PhysRevD.94.104038}{{\em Phys. Rev. D}
  {\bfseries 94} no.~10, (2016) 104038},
  \href{http://arxiv.org/abs/1607.01025}{{\ttfamily arXiv:1607.01025
  [hep-th]}}.

\bibitem{Giddings:2018umg}
S.~B. Giddings and A.~Kinsella, ``{Gauge-invariant observables, gravitational
  dressings, and holography in AdS},''
  \href{http://dx.doi.org/10.1007/JHEP11(2018)074}{{\em JHEP} {\bfseries 11}
  (2018) 074}, \href{http://arxiv.org/abs/1802.01602}{{\ttfamily
  arXiv:1802.01602 [hep-th]}}.

\bibitem{Giddings:2019wmj}
S.~Giddings and S.~Weinberg, ``{Gauge-invariant observables in gravity and
  electromagnetism: black hole backgrounds and null dressings},''
  \href{http://dx.doi.org/10.1103/PhysRevD.102.026010}{{\em Phys. Rev. D}
  {\bfseries 102} no.~2, (2020) 026010},
  \href{http://arxiv.org/abs/1911.09115}{{\ttfamily arXiv:1911.09115
  [hep-th]}}.

\bibitem{Harlow:2021dfp}
D.~Harlow and J.-q. Wu, ``{Algebra of diffeomorphism-invariant observables in
  Jackiw-Teitelboim gravity},''
  \href{http://dx.doi.org/10.1007/JHEP05(2022)097}{{\em JHEP} {\bfseries 05}
  (2022) 097}, \href{http://arxiv.org/abs/2108.04841}{{\ttfamily
  arXiv:2108.04841 [hep-th]}}.

\bibitem{Iso:2006wa}
S.~Iso, H.~Umetsu, and F.~Wilczek, ``{Hawking radiation from charged black
  holes via gauge and gravitational anomalies},''
  \href{http://dx.doi.org/10.1103/PhysRevLett.96.151302}{{\em Phys. Rev. Lett.}
  {\bfseries 96} (2006) 151302},
  \href{http://arxiv.org/abs/hep-th/0602146}{{\ttfamily arXiv:hep-th/0602146}}.

\bibitem{Gibbons:1975kk}
G.~W. Gibbons, ``{Vacuum Polarization and the Spontaneous Loss of Charge by
  Black Holes},'' \href{http://dx.doi.org/10.1007/BF01609829}{{\em Commun.
  Math. Phys.} {\bfseries 44} (1975) 245--264}.

\bibitem{Mertens:2018fds}
T.~G. Mertens, ``{The Schwarzian theory \textemdash{} origins},''
  \href{http://dx.doi.org/10.1007/JHEP05(2018)036}{{\em JHEP} {\bfseries 05}
  (2018) 036}, \href{http://arxiv.org/abs/1801.09605}{{\ttfamily
  arXiv:1801.09605 [hep-th]}}.

\bibitem{Cohn:1986wn}
J.~D. Cohn, ``{N=2 Superriemann surfaces},''
  \href{http://dx.doi.org/10.1016/0550-3213(87)90039-3}{{\em Nucl. Phys. B}
  {\bfseries 284} (1987) 349--364}.

\bibitem{Fu:2016vas}
W.~Fu, D.~Gaiotto, J.~Maldacena, and S.~Sachdev, ``{Supersymmetric
  Sachdev-Ye-Kitaev models},''
  \href{http://dx.doi.org/10.1103/PhysRevD.95.026009}{{\em Phys. Rev. D}
  {\bfseries 95} no.~2, (2017) 026009},
  \href{http://arxiv.org/abs/1610.08917}{{\ttfamily arXiv:1610.08917
  [hep-th]}}. [Addendum: Phys.Rev.D 95, 069904 (2017)].

\bibitem{Dias:2007nj}
O.~J.~C. Dias, R.~Emparan, and A.~Maccarrone, ``{Microscopic theory of black
  hole superradiance},''
  \href{http://dx.doi.org/10.1103/PhysRevD.77.064018}{{\em Phys. Rev. D}
  {\bfseries 77} (2008) 064018},
  \href{http://arxiv.org/abs/0712.0791}{{\ttfamily arXiv:0712.0791 [hep-th]}}.

\bibitem{Randall:1999ee}
L.~Randall and R.~Sundrum, ``{A Large mass hierarchy from a small extra
  dimension},'' \href{http://dx.doi.org/10.1103/PhysRevLett.83.3370}{{\em Phys.
  Rev. Lett.} {\bfseries 83} (1999) 3370--3373},
  \href{http://arxiv.org/abs/hep-ph/9905221}{{\ttfamily arXiv:hep-ph/9905221}}.

\bibitem{Karch:2000ct}
A.~Karch and L.~Randall, ``{Locally localized gravity},''
  \href{http://dx.doi.org/10.1088/1126-6708/2001/05/008}{{\em JHEP} {\bfseries
  05} (2001) 008}, \href{http://arxiv.org/abs/hep-th/0011156}{{\ttfamily
  arXiv:hep-th/0011156}}.

\bibitem{Porrati:2003sa}
M.~Porrati, ``{Higgs phenomenon for the graviton in ADS space},''
  \href{http://dx.doi.org/10.1142/S0217732303011745}{{\em Mod. Phys. Lett. A}
  {\bfseries 18} (2003) 1793--1802},
  \href{http://arxiv.org/abs/hep-th/0306253}{{\ttfamily arXiv:hep-th/0306253}}.

\bibitem{Bhattacharya:2021jrn}
A.~Bhattacharya, A.~Bhattacharyya, P.~Nandy, and A.~K. Patra, ``{Islands and
  complexity of eternal black hole and radiation subsystems for a doubly
  holographic model},'' \href{http://dx.doi.org/10.1007/JHEP05(2021)135}{{\em
  JHEP} {\bfseries 05} (2021) 135},
  \href{http://arxiv.org/abs/2103.15852}{{\ttfamily arXiv:2103.15852
  [hep-th]}}.

\bibitem{Blommaert:2020yeo}
A.~Blommaert, T.~G. Mertens, and H.~Verschelde, ``{Unruh detectors and quantum
  chaos in JT gravity},'' \href{http://dx.doi.org/10.1007/JHEP03(2021)086}{{\em
  JHEP} {\bfseries 03} (2021) 086},
  \href{http://arxiv.org/abs/2005.13058}{{\ttfamily arXiv:2005.13058
  [hep-th]}}.

\bibitem{Weiss1999}
U.~Weiss, {\em {Quantum Dissipative Systems}}.
\newblock World Scientific, January, 1999.

\bibitem{Chamseddine:1991fg}
A.~H. Chamseddine, ``{Superstrings in arbitrary dimensions},''
  \href{http://dx.doi.org/10.1016/0370-2693(91)91215-H}{{\em Phys. Lett. B}
  {\bfseries 258} (1991) 97--103}.

\bibitem{Forste:2017kwy}
S.~Forste and I.~Golla, ``{Nearly AdS$_2$ sugra and the super-Schwarzian},''
  \href{http://dx.doi.org/10.1016/j.physletb.2017.05.039}{{\em Phys. Lett. B}
  {\bfseries 771} (2017) 157--161},
  \href{http://arxiv.org/abs/1703.10969}{{\ttfamily arXiv:1703.10969
  [hep-th]}}.

\bibitem{Arvis:1982tq}
J.~F. Arvis, ``{Classical Dynamics of the Supersymmetric Liouville Theory},''
  \href{http://dx.doi.org/10.1016/0550-3213(83)90602-8}{{\em Nucl. Phys. B}
  {\bfseries 212} (1983) 151--172}.

\bibitem{Fan:2021wsb}
Y.~Fan and T.~G. Mertens, ``{Supergroup structure of Jackiw-Teitelboim
  supergravity},'' \href{http://dx.doi.org/10.1007/JHEP08(2022)002}{{\em JHEP}
  {\bfseries 08} (2022) 002}, \href{http://arxiv.org/abs/2106.09353}{{\ttfamily
  arXiv:2106.09353 [hep-th]}}.

\bibitem{Cardy:talk}
J.~Cardy, ``{Entanglement in CFTs at Finite Chemical Potential}.''
  \url{http://www2.yukawa.kyoto-u.ac.jp/~entangle2016/YCardy.pdf}, 2016.

\end{thebibliography}\endgroup

\end{document}